\documentclass[12pt]{article}
\usepackage{epsfig}
\usepackage{graphicx}
\usepackage{version}
\usepackage{amsmath}
\usepackage{amssymb}
\usepackage{psfrag}
\usepackage{appendix}

\newcommand{\rme}{{\rm e}}
\newcommand{\rmi}{{\rm i}}
\newcommand{\ud}[1]{\hspace{-0em}\mathrm{d}{#1}\;}
\newcommand{\udd}[2]{\hspace{-0em}\mathrm{d}{#1}\,\mathrm{d}{#2}\;}
\newcommand{\Ud}[1]{\hspace{-0.5ex}\mathrm{d}{#1}\;}
\newcommand{\Udd}[2]{\hspace{-0.5ex}\hspace{-0em}\mathrm{d}{#1}\,\mathrm{d}{#2}\;}
\newcommand{\uD}{\mathcal{D}}

\makeatletter

\@addtoreset{equation}{section}
\makeatother

\begin{document}
\title{Coarsening of Disordered Quantum Rotors \\ under a Bias Voltage}
\author{Camille Aron$^*$, Giulio Biroli$^{\dagger}$ and Leticia F. Cugliandolo$^*$ \\
$^*${\small Laboratoire de Physique Th\'eorique et Hautes \'Energies,} \\
{\small Universit\'e Pierre et Marie Curie - Paris VI}, \\
{\small 4 Place Jussieu, Tour 13, 5\`eme \'etage, 75252 Paris Cedex 05, France.}\\
$^{\dagger}${\small Institut de Physique Th\'eorique, CEA Saclay, 91191 Gif-sur-Yvette, France.
}}
\date{}

%

\maketitle

\begin{abstract}
We solve the dynamics of an ensemble of interacting rotors coupled to
two leads at different chemical potential letting a current flow
through the system and driving it out of equilibrium. We show that at low temperature
the coarsening phase persists under the voltage drop up to a critical
value of the applied potential that depends on the characteristics of the electron reservoirs. We discuss the properties of the
critical surface in the temperature, voltage, strength of quantum
fluctuations, and coupling to the bath phase diagram.  We analyze the
coarsening regime finding, in particular, which features are
essentially quantum mechanical and which are basically classical in
nature. We demonstrate that the system evolves via the growth of a
coherence length with the same time dependence as in the classical
limit, $R(t) \simeq t^{1/2}$ -- the scalar curvature driven
universality class. We obtain the scaling function of the correlation 
function at late epochs in
the coarsening regime and we prove that it coincides with the classical
one once a prefactor that encodes the dependence on all the parameters is 
factorized.
We derive a generic formula for the current flowing through the system
and we show that, for this model, 
it rapidly approaches a constant that we compute.
\end{abstract}

\newpage
\tableofcontents

\pagebreak 

\section{Introduction}
Quantum mechanics determines the behavior of physical systems at
atomic and subatomic scales. The search for quantum effects at
macroscopic scales started soon after the development of quantum
mechanics. A number of quantum manifestations at such 
scales have been found including quantum tunneling of the phase in Josephson
junctions~\cite{Leggett} or resonant tunneling of magnetization in
spin cluster systems~\cite{Barbara}. 

Dynamic issues in isolated quantum many-body systems are the focus of active
research. Some of the problems that are currently being studied theoretically are:
the time evolution of the entropy of entanglement in spin
systems~\cite{Calabrese}, the nature of non-equilibrium steady states in small quantum systems driven
out of equilibrium~\cite{cut-offReichman,Seiji} due to  their relevance for nano-devices, 
quantum annealing techniques~\cite{quantum-annealing}, and the density of defects 
left over after a gradual change in a parameter~\cite{Zurek}.
The influence  of an environment on the dynamics of 
quantum systems was also dealt with in a number of cases
such as the spin-boson model~\cite{Leggett},
disordered spin chains coupled to bosonic baths~\cite{Greg}, 
or an electronic ring coupled to leads and further driven by a 
time-dependent field~\cite{Lili, LiliCu}.

Once the interest is set upon macroscopic systems, the question 
as to whether these undergo phase transitions naturally 
arises. The theory of equilibrium classical and
quantum phase transitions is well developed.
 {\it Non-equilibrium} phase transitions in which
quantum fluctuations can be neglected are also quite well understood. These are realized when a
system is forced in a non equilibrium steady state (by a shear rate,
an external current flowing through it,
etc.)~\cite{OnukiKawasaki,Hinrichsen,DombGreen,Tauber} or when it just fails to relax
(\textit{e.g}. after a quench) and displays aging
phenomena~\cite{Struick,LeticiaLesHouches}.  In contrast, the effect
of a drive on a {\it macroscopic} system close to a quantum phase
transition is a rather unexplored subject. Some works have focused on
non-linear transport properties close to an (equilibrium) quantum
phase
transition~\cite{DalidovichPhillips,GreenSondhi,HoganGreen}. Others
have studied how the critical properties are affected by
non-equilibrium drives \cite{MitraKimMillis,Feldman,MitraMillis}.
However, a global understanding of phase transitions in the control
parameter space $T,\ V,\ \Gamma$, with $T$ the temperature, $V$ the
driving strength, and $\Gamma$ the strength of quantum fluctuations,
is still lacking. Furthermore, to the best of our knowledge, the issue
of the {\it relaxation} toward the quantum non-equilibrium steady
state (QNESS) has not been addressed in the past.

In this paper we extend our study of driven quantum phase transitions and coarsening phenomena started in~\cite{AroBiroCuglio}.
We study a class of analytically tractable models, systems of $M$-component
$N$ quantum rotors that encompass an infinite range spin-glass and its
three dimensional pure counterpart modeling coarsening phenomena.  As
discussed in \cite{SachdevBook} models of quantum rotors are
non-trivial but still relatively simple and provide coarse-grained
descriptions of physical systems such as Bose-Hubbard models and
double layer antiferromagnets.  The system is coupled to two different
external electron reservoirs that lead to a current flowing through
it and driving it out of equilibrium. (For a two-dimensional model
the current flows perpendicular to it, see the sketch in Fig.~1 of
\cite{MitraKimMillis}.)  In the simplest setting~\cite{MitraKimMillis}
each rotor is coupled to independent reservoirs; more realistic
couplings are discussed in \cite{MitraMillis}.  Using the
Schwinger-Keldysh formalism \cite{SchwingerKeldysh, Kamenev} we obtain the complete out
of equilibrium dynamics of these models in the large $M$ limit. We
show that at sufficiently low $T,V,\Gamma$, see Fig.~1, the system
never reaches a QNESS and ages with remarkable universal properties.
\begin{figure}[t]
\centering
 \includegraphics[width=8cm]{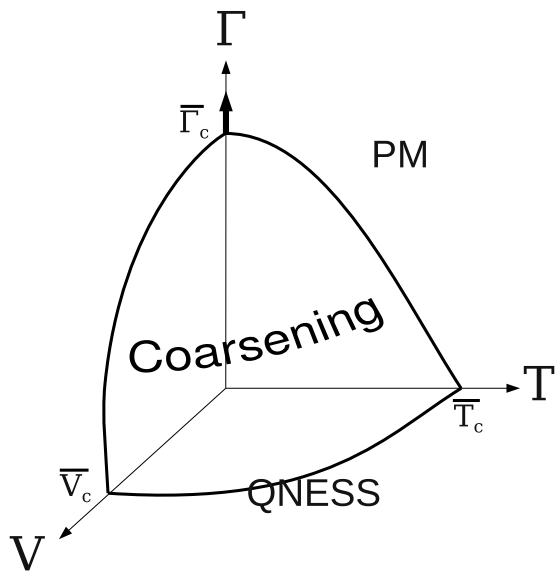}
\caption{\label{fig:epsart} Non-equilibrium phase diagram of the fully connected driven quantum rotor model with an infinite number of components.}
\end{figure}
We study the critical properties of
the phase transitions, in particular in the
vicinity of the (drive-induced) quantum out of equilibrium critical point
$\bar V_c$ at $\Gamma=0$, $T=0$ and the ``usual'' quantum critical point $\bar\Gamma_c$ at $V=0$, $T=0$.
We analyze in detail the relaxation in the coarsening regime and uncover the scaling properties of correlation functions and linear response. We derive a general formula for the current flowing through the system under such a voltage drop and we analyze its dependence on the 
dynamics of the system.
Some of these results were announced recently in~\cite{AroBiroCuglio}.

\section{The model}
\subsection{System of disordered quantum rotors}

The model we focus on is a quantum disordered system
made of $N$ $M$-component rotors interacting via random infinite-range
couplings \cite{SachdevYe}. We consider a fully-connected (mean-field) model where there is no
underlying geometry: each rotor is equivalently coupled to
all the others.
The Hamiltonian is given by
\begin{equation}
H_{S}= \frac{\Gamma}{2M}
\sum_{i=1}^N\mathbf{L}_i^2 - \frac{M}{\sqrt{N}} \sum_{i,j<i} J_{ij} \ \mathbf{n}_i
\cdot \mathbf{n}_j \;.
\label{eq:model}
\end{equation}
$n_{i}^\mu$ $(\mu=1\ldots M)$ are the $M$ components of the $i$-th rotor.
The coordinates $n_i^\mu$ constitute a complete set of commuting observables.
The scalar product $\mathbf{n}_i
\cdot \mathbf{n}_j $ is given by $\sum_{\mu=1}^M n_i^\mu n_j^\mu $.
The length of rotors is fixed to unity: $\mathbf{n}_i \cdot \mathbf{n}_i = 1,\, \forall \ i=1\dots N$.
The strengths $J_{ij}$'s are taken from a Gaussian distribution with
zero mean and variance $J^2$.
$J$ controls the strength of disorder.
$\mathbf{L}_i$ is the $i$-th generalized angular momentum operator
which $M(M-1)/2$ components are given by
\begin{equation}
L_i^{\mu\nu}=-\rmi\hbar\left( n_i^\mu \frac{\partial}{\partial n_i^\nu}-n_i^\nu \frac{\partial}{\partial n_i^\mu} \right) \qquad \mbox{ for } 
1\leq\mu<\nu\leq M \;,
\end{equation}
$\mathbf{L}_i^2 = \sum_{\mu<\nu} (L_i^{\mu\nu})^2$~\cite{SachdevBook,SachdevYe}.

$\Gamma$ acts like a moment of inertia and controls the strength of quantum fluctuations; when $\hbar^2\Gamma/J\to 0$ the
model approaches the classical Heisenberg fully-connected spin-glass.  In
the large $M$ limit it is equivalent to the quantum fully-connected
$p=2$ (or Sherrington-Kirkpatrick) spherical spin-glass~\cite{Shukla,Rokni}.
The classical mapping to
ferromagnetic coarsening in the O(${\cal N})$ model with 
${\cal N}\to\infty$~\cite{LeticiaLesHouches} holds, as we shall 
show in Sect.~\ref{subsubsec:long-time}, for the quantum model as well.

\subsection{Reservoirs of electrons}\label{sec:res_of_electrons}

The system is coupled to two, 
`left' ($L$) and `right' ($R$), reservoirs of electrons.
These independent reservoirs are both in equilibrium at inverse temperature
$\beta_L$ and $\beta_R$. The situation $\beta_L \neq \beta_R$ would create a
heat flow from one reservoir to the other. We are
interested in the simpler case in which $\beta_L = \beta_R \equiv \beta \equiv
T^{-1}$ ($k_B=1$). An electric
 current is forced by imposing different chemical potentials,
$\mu_L=\mu_0$ and $\mu_R=\mu_0+eV$ (where $-e$ is the electric charge of one 
electron). $eV$ is the strength of the drive.
As $eV/J \to 0$, the effect of the reservoirs on the system approaches the one
of an equilibrium bath at temperature $T$.
The details of the reservoir Hamiltonians $H_L$ and $H_R$ are not important
since only the electronic Green's functions matter in the small rotor-bath
coupling we concentrate on.
We consider the simple case in which left and right fermionic
reservoirs have the same density of states (DOS) $\rho_L = \rho_R = \rho$.
Moreover, we focus on simple cases in which the shape of the DOS is controlled by
only one typical energy scale $\epsilon_F$. In the rest of this paper, we
often consider the limit in which $\epsilon_F$ is much larger than all the
other energy scales involved. In this limit the results become independent of the detailed
functional form  of the DOS. We also give some results for finite $\epsilon_F$ 
using the specific DOS that we introduce below.

\subsubsection{DOS with a finite bandwidth}
\begin{figure}[t]
 \centering
 \includegraphics[height=4.00cm]{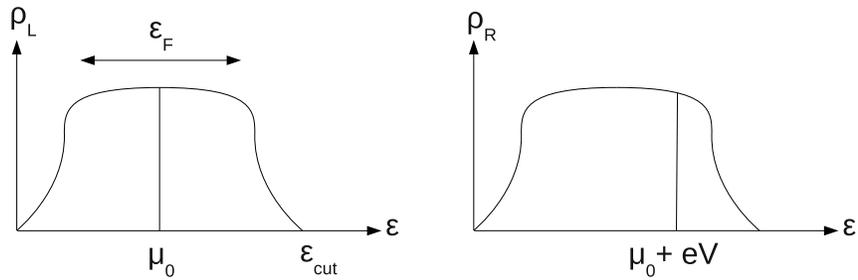}
 \caption{\footnotesize \label{fig:typeA} Density of states (DOS) of 
type A reservoirs. $\mu_0$ and $\mu_0 + eV$ are the left and right Fermi levels,
 respectively.  The left reservoir is half-filled.} 
\end{figure}

We first consider regular DOS which have a finite typical width (finite bandwidth) controlled by $\epsilon_F$ and $\mu_0$ is set around the maximum of the distribution. In the limit where $\epsilon_F$ is very large, they can be seen as almost flat distributions. We call $\epsilon_{cut}$ the finite energy cut-off beyond which the DOS vanishes, $\rho(|\epsilon|>\epsilon_{cut}) =0$. Since the DOS we consider have a single energy scale $\epsilon_F$, $\epsilon_{cut}$ should scale with $\epsilon_F$. Notice that a finite $\epsilon_{cut}$ constrains the voltage not to 
exceed $eV_{max}= \epsilon_{cut}-\mu_0$ since the right reservoir is then completely filled and 
therefore it cannot accept more fermions.

\begin{figure}
 \centering
 \includegraphics[width=6.00cm]{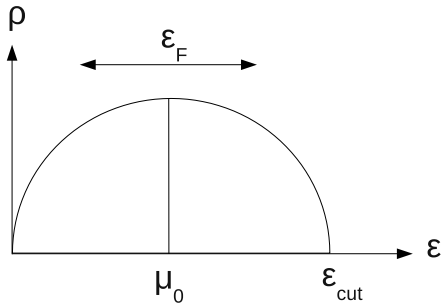}
 \caption{\footnotesize \label{fig:DOS1} An example of type A reservoir: the
semi-circle density of states (half-filled). } 
\end{figure}

We call reservoir of type A a half-filled\footnote{Half-filled means that half the total 
number of available states are occupied: $\int_{-\infty}^{\mu_0} \rm{d}\epsilon \rho(\epsilon) 
= \frac{1}{2} $ at $T=0$.} reservoir the DOS of which has a finite bandwidth controlled by $\epsilon_F$ and is symmetric and derivable in the vicinity of its maximum (see Fig.~\ref{fig:typeA}). 
The simplest example of a type A reservoir is given by the semi-circular DOS (see Fig.~\ref{fig:DOS1}), 
\begin{equation}
 \rho_A(\epsilon) \equiv \frac{2}{\pi\epsilon_F} \sqrt{1-\left(\frac{\epsilon-\epsilon_F}{\epsilon_F}\right)^2} \;,
\end{equation}
that is symmetric and centered around $\epsilon_F$. Here $\epsilon_{cut} = 2\epsilon_F$. We choose $\mu_0 = \epsilon_F$ so that the reservoirs are half-filled at zero drive ($eV=0$). In this case, at $T=0$, the voltage applied between both reservoirs cannot exceed $eV_{max} = \epsilon_{cut}-\mu_0 = \epsilon_F$.

\begin{figure}[h]
 \centering
 \includegraphics[height=4.00cm]{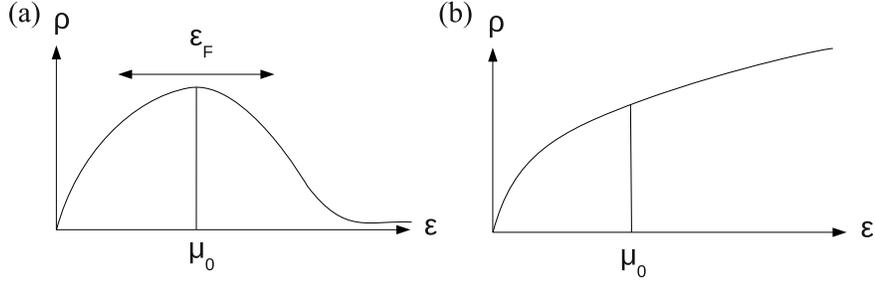}
 \caption{\footnotesize \label{fig:DOS3} Two examples of type $B$ reservoirs.
(a) The distribution $\rho_B$ vanishes asymptotically. (b) The square root distribution
diverges asymptotically.} 
\end{figure}

Type B reservoirs have finite bandwidth but no energy cut-off:  
$\epsilon_{cut} = eV_{max} \to\infty$.  A realization of these reservoirs is given 
by the following DOS [see Fig.~\ref{fig:DOS3}(a)]
\begin{equation}
 \rho_B(\epsilon) \equiv \frac{\alpha}{\epsilon_F} \sqrt{\frac{\epsilon}{\epsilon_F}} 
 \ \rme^{-\frac{1}{2}\left(\frac{\epsilon}{\epsilon_F}\right)^2} \;, \label{eq:DOSB}
\end{equation}
where $\alpha \approx 0.97$ is a numerical constant fixed by normalization.
The maximum of this distribution is located at $\epsilon_F / \sqrt{2}$. This reservoir is half-filled for 
$\mu_0 \approx 0.95 \ \epsilon_F$. This distribution resembles the semi-circular one in the sense that 
they both start with a square root behavior, have a maximum, and a bandwidth of order $\epsilon_F$. In contrast, the DOS in eq.~(\ref{eq:DOSB}) is different from zero at all finite $\epsilon$
and one can exploit this feature to apply 
strong voltages.

\subsubsection{DOS at low energy}
In the previous examples ($\rho_A$ and $\rho_B$), we focused on values of $\mu_0$ corresponding to 
high energy states where the DOS is regular. We are also interested in studying cases where 
$\mu_0$ is centered around low energy states. To analyze these cases, we
 focus on a DOS which reads [see Fig.~\ref{fig:DOS3}(b)]:
\begin{equation}
 \rho_{C3{\rm d}}(\epsilon) \equiv \frac{3}{4\sqrt 2\epsilon_F} \sqrt{\frac{\epsilon}{\epsilon_F}} \;.
\end{equation}
This square root behavior is actually the one of the $3d$ free fermions reservoir. In this case
$\epsilon_F $ is of the order of the hopping term for the free fermions. Since we shall only focus 
on the low energy states of the reservoir, we can neglect the non trivial high energy structure of the reservoir
and take the DOS equal to zero for $\epsilon>2\epsilon_F $. 

For the $2d$ free fermions, the density of states is given by
\begin{equation}
 \rho_{C2{\rm d}}(\epsilon) \equiv \frac{1}{2\epsilon_F} \;,
\end{equation}
whereas for the $1d$ free fermions, the density of states is given by
\begin{equation}
 \rho_{C1{\rm d}}(\epsilon) \equiv  \frac{1}{2 \sqrt 2\epsilon_F} \sqrt{\frac{\epsilon_F}{\epsilon}} 
\end{equation}
and, 
as for $\rho_{C3{\rm d}}$, we take these two densities of states to be equal to zero for $\epsilon>2\epsilon_F $.

\subsection{Coupling between the system and the reservoirs}

An electron hop from the $L$($R$) reservoir to the
$R$($L$) reservoir is linearly coupled to each rotor component:
\begin{equation}\label{eq:Hsb}
H_{SB}=  - \frac{\sqrt{M}}{N_s} \sum_{i=1}^N \sum_{\mu=1}^M \sum_{k,k'=1}^{N_s} \sum_{l,l'=1}^{\cal M} 
V_{kk'} \;
n_i^\mu \;
[\psi^{\dagger}_{Likl} \ \sigma_{ll'}^\mu \ \psi_{Rik'l'} + L
\leftrightarrow R] \;,
\end{equation}
where $\psi^{\dagger}_{Likl}$ is the $l$-th component of an $\cal{M}$-component spinor operator that
creates an additional fermion with energy $\hbar\omega_k$ in the
$L$ reservoir associated to the $i$-th
rotor.
$k$ labels the electron energy inside the reservoirs, $N_s$ is the total number of states in each reservoir.
$\mathbf{\sigma}^{\mu}$ are the generalized Pauli matrices for ${\rm SU}(\cal M)$ of
dimension ${\cal M} \times {\cal M}$ with ${\cal M}^2-1=M$. They are chosen to be normalized such that $\mbox{Tr}\,\sigma^\mu\sigma^\nu = \delta_{\mu\nu}$.
$V_{kk'}$ are the rotor-environment coupling parameters chosen to be constant:
$V_{kk'}=\hbar\omega_c$. $H_{SB}$ is ${\cal O}(MN)$ invariant.

\section{The dynamics}

\subsection{Quench setup}\label{sec:QuenchSetUp}

The system is initially prepared (at times $t<0$) in such a way that its initial
configuration (at time $t=0$) is neither correlated with disorder ($J_{ij}$'s)
nor with the reservoirs. This can be realized, for instance, by coupling the
system to an equilibrium bath at temperature $T_0 \gg J, \Gamma$ so 
that any correlation in the system is suppressed. At time $t=0$ the quench is performed by suddenly coupling the system to
the $L$ and $R$ reservoirs. These are supposed to be ``good reservoirs'' in
the sense that their properties are not affected by the state of the system.

This setup generates non-equilibrium dynamics at times $t>0$ for multiple reasons.
First of all, the rapid quenching procedure puts the system in a non-equilibrium
initial condition with respect to its new environment. Moreover, the latter
is not an equilibrium bath but a bias drive the role of which is to constantly
destabilize the system. 
Finally, as a consequence of its disordered interactions, the system of rotors experiences intrinsic difficulties to reach equilibrium.
Indeed, even if it were embedded within an equilibrium environment it would show a glassy
phase \cite{Rokni, CugliandoloLozano} in some parts of the phase diagram.

Since system and reservoirs are decoupled at times $t<0$, the
initial density matrix of the whole system is given by 
\begin{equation}
\varrho(t=0) = \varrho_S(t=0)
{ \begin{array}{c} {\scriptstyle N} \vspace{-0.7ex} \\ \otimes \vspace{-1ex} \\ {\scriptstyle  i=1} \end{array} }
 \varrho_{Li}
{ \begin{array}{c} {\scriptstyle N} \vspace{-0.7ex} \\ \otimes \vspace{-1ex} \\ {\scriptstyle  i=1} \end{array} }
\varrho_{Ri}\;.
\end{equation}
$\varrho_{Li/Ri}$ corresponds to the equilibrium density matrix of the
$L/R$ reservoir associated with the $i$-th rotor.
The system of rotors being prepared at very high temperature, its initial density matrix is the identity in the rotors space:
\begin{equation}
 \varrho_S(t=0) \propto I_S \;.
\end{equation}
All these density matrices are normalized to be of unit trace. The $t>0$ evolution of the whole system plus environment is encoded in
\begin{equation}
 \varrho(t) =  U(t,0)   \  \varrho(0) \ [U(t,0)]^\dagger \;,
\end{equation}
where the unitary evolution operator is given by
$U(t,0) \equiv \mathsf{T} \rme^{-\frac{\rmi}{\hbar} \int_0^t \ud{t'} H(t') }$
with $H = H_S + H_L + H_R + H_{SB}$ and $\mathsf{T}$ the time-ordering operator (see Appendix~\ref{app:Conventions}).
We analyze the non-equilibrium dynamics using  the Schwinger-Keldysh
formalism (see \cite{Kamenev} for a modern review) that we briefly introduce in the following lines.

\subsection{Schwinger-Keldysh formalism}

\begin{figure}
 \centering
 \includegraphics[width=9.00cm]{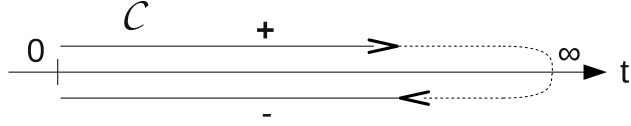}
 \caption{\footnotesize \label{fig:KeldyshContour} 
 The Keldysh contour $\mathcal{C}$ goes from $0$ to $+\infty$ and then back to $0$. 
 The Keldysh action involves forward fields (that live on the $+$branch of $\mathcal{C}$) 
 that are time-integrated from $0$ to $+\infty$ and backward fields (that live on the $-$branch of $\mathcal{C}$) and are time-integrated from $+\infty$ to $0$.}
\end{figure}

The Suzuki-Trotter decomposition of the two unitary evolution operators that appear in 
\begin{equation}\label{eq:Suzuki}
\mathcal{Z} \equiv  \lim\limits_{\tau\to \infty}\mbox{Tr }  U(\tau,0) \ \varrho(0) \ [U(\tau,0)]^\dagger  = 1\;,
\end{equation}
yields a path-integral involving two sets of fields with support on two
different branches. The first ones are time-integrated on a forward branch from
$t=0$ to $+\infty$. In the following, these fields carry a $+$ superscript. The
other ones are time-integrated on a backward branch from $+\infty$ to $0$ and carry a $-$ superscript. These two branches constitute the Keldysh contour $\mathcal{C}$, see Fig.~\ref{fig:KeldyshContour}. The identity~(\ref{eq:Suzuki}) can now be expressed as a path integral,
\begin{equation}\label{eq:PRER}
 \mathcal{Z} = \int_{\rm c} \mathcal{D}
[\boldsymbol{n}^{\pm}, \boldsymbol{\psi}^{\pm}, \boldsymbol{\bar\psi}^{\pm}]
\ 
\rme^{\frac{\rmi}{\hbar} S}
\
\langle \boldsymbol{n}^{+}(0),  \boldsymbol{\bar\psi}^{+}(0) |  \varrho(0) | \boldsymbol{n}^{-}(0), \boldsymbol{\psi}^{-}(0) \rangle 
 \;,
\end{equation}
where we collected all the  $n_i^{\mu a}$ fields into the notation $\boldsymbol{n}^a$, and all the fermionic fields $\psi_{\alpha i}^a$ and their Grassmannian conjugates into $\boldsymbol{\psi}^a$ and $\boldsymbol{\bar\psi}^a$ (with $a = \pm$). $\langle \boldsymbol{n}^{+}(0),  \boldsymbol{\bar\psi}^{+}(0) |  \varrho(0) | \boldsymbol{n}^{-}(0), \boldsymbol{\psi}^{-}(0) \rangle $ is the matrix element of the density matrix which has support at time $t=0$ only.
The action $S$ is a functional of all these fields:
\begin{eqnarray}
S = \sum_{a=\pm} a \int_0^{\infty} \ud{t} {\cal L}([\boldsymbol{n}^a, \boldsymbol{\psi}^a, \boldsymbol{\bar\psi}^a];t) \;.
\end{eqnarray}
The Lagrangian is given by ${\cal L} = {\cal L}_{S}  + {\cal L}_{SB} + {\cal L}_{L} + {\cal L}_{R}$ with
\begin{eqnarray}
  {\cal L}_{S}([\boldsymbol{n}^a] ;t)
\hspace{-1ex} &=& \hspace{-1ex}
\frac{M}{2\Gamma}
\sum_{i} {\dot {\bf n}_i^a(t)}^2 + \frac{M}{\sqrt{N}} \sum_{i,j<i} J_{ij} \ \mathbf{n}_i^a(t)
\cdot \mathbf{n}_j^a(t) \;, \\
 {\cal L}_{SB}([\boldsymbol{n}^a,\boldsymbol{\psi}^a, \boldsymbol{\bar\psi}^a] ;t) 
\hspace{-1ex} &=& \hspace{-1ex}
   \sqrt{M}\, \frac{\hbar\omega_c}{N_s} \sum_{i\mu kk' ll'}
n_i^{\mu a}(t) \;
[\bar \psi_{Likl}^a(t) \ \sigma^\mu_{ll'} \ \psi_{Rik'l'}^a(t) + L
\leftrightarrow R] \;.
\end{eqnarray}
${\cal L}_{L}$ and ${\cal L}_{R}$ are the Lagrangians of the free fermions in
the $L$ and $R$ reservoirs. The index `${\rm c}$' at the bottom of the integral
sign in eq.~(\ref{eq:PRER}) is here to remind us that the integration is performed
over fields satisfying the constraint that each rotor has a fixed unit length:
${{\bf n}_i^a(t)}^2 = 1 \ \forall \ a, i, t$. The
path-integral formalism gives a nice way to restore an unconstrained integration
over all fields ${\bf n}_i^a$ by the introduction of Lagrange multipliers
$z_i^a$:
\begin{eqnarray}
\int_{\rm c} \mathcal{D} [\boldsymbol{n}^a]
\hspace{-1ex} &=& \hspace{-1ex}
 \int \mathcal{D} [\boldsymbol{n}^a] \prod_{i, t} \delta( 1 - {{\bf n}_i^a(t)}^2) \\
\hspace{-1ex} &=& \hspace{-1ex}
\int \mathcal{D} [\boldsymbol{n}^a, \mathit{z}^a] \exp{\left( \frac{\rmi}{\hbar} \int_0^\infty \Ud{t} a \frac{M}{2} \sum_{i} z_i^a(t) \left( 1 -  {{\bf n}_i^a(t)}^2  \right) \right)} \;.
\end{eqnarray}
where we used the integral representation of the delta function (see Appendix~\ref{app:Conventions}) and collected the new auxiliary real fields $z_i^a$ into the notation $\mathit{z}^a$.
In terms of a Lagrangian, this gives rise to the new term
\begin{equation}
 {\cal L}_{\rm LM}([\boldsymbol{n}^a, \mathit{z}^a];t) = \frac{M}{2} \sum_{i} z_i^a(t) [1-{{\bf n}_i^a(t)}^2] \;.
\end{equation}

\subsection{Macroscopic observables}

We are interested in the macroscopic dynamics of the rotors after an infinitely rapid quench and we wish to give an answer to the following questions (among others). Does the system reach a steady state? 
Does a steady state current establish? What are the long-time dynamics? 
We first  obtain an effective generating functional for the rotors by 
expanding the system-drive interaction up to 
second order in the coupling,  integrating away the fermionic degrees of freedom, and  averaging over the 
disorder distribution.

Introducing the external real fields $\eta_{i\mu}^{a}(t)$ that we collect in the notation $\boldsymbol{\eta}^a(t)$ ($a=\pm$), the generating functional ${\cal Z}[\boldsymbol{\eta}^\pm]$ reads
\begin{eqnarray} \label{eq:generating_func}
 {\cal Z}[\boldsymbol{\eta}^\pm] \equiv \int \uD[\boldsymbol{n}^\pm,\mathit{z}^\pm,\boldsymbol{\psi}^\pm,\boldsymbol{\bar\psi}^\pm] 
 \
 \rme^{\frac{\rmi}{\hbar} S[\boldsymbol{n}^\pm,\mathit{z}^\pm,\boldsymbol{\psi}^\pm,\boldsymbol{\bar\psi}^\pm, {\boldsymbol\eta}^\pm]  } 
 \
 \langle \boldsymbol{n}^{+}(0),  \boldsymbol{\bar\psi}^{+}(0) |  \varrho(0) | \boldsymbol{n}^{-}(0), \boldsymbol{\psi}^{-}(0) \rangle  \;,
\end{eqnarray}
where we introduced the source term
\begin{eqnarray}
S  \longmapsto S + \hbar \sum_{a=\pm} \int \ud{t} \sum_{i}  \sum_{\mu}  n_i^{\mu a}(t) \eta_i^{\mu a}(t)\;.
\end{eqnarray}
The generating functional obeys the normalization property ${\cal Z}[\boldsymbol{\eta}^\pm={\bf 0}] = {\cal Z} = 1$ which is a fundamental feature of the Keldysh formalism in this setup (see eq.~(\ref{eq:Suzuki}) and Sect.~\ref{sec:average_disorder}).
One has
\begin{eqnarray}
 \langle n_i^{\mu a}(t)\rangle 
= - \frac{\rmi}{\cal Z} \left.\frac{\delta \ {\cal Z}[\boldsymbol{\eta}^\pm] }{\delta \eta_{i}^{\mu a}(t)}\right|_{\boldsymbol{\eta}^\pm=\boldsymbol{0}} \;,   
\end{eqnarray}
where we introduced the notation 
\begin{eqnarray}
\langle \ \cdots \ \rangle \equiv \int
\uD[\boldsymbol{n}^\pm,\mathit{z}^\pm,\boldsymbol{\psi}^\pm,\boldsymbol{
\bar\psi}^\pm] \ \cdots \ \rme^{\frac{\rmi}{\hbar} S  } \langle \boldsymbol{n}^{+}(0), 
\boldsymbol{\bar\psi}^{+}(0) |  \varrho(0) | \boldsymbol{n}^{-}(0),
\boldsymbol{\psi}^{-}(0) \rangle \;.
\end{eqnarray}
Notice that one can distinguish this bracket notation from the quantum
statistical average that we denote similarly by the occurrence of Keldysh
indices inside the brackets. However, they coincide in the case of one time
observables, {\it e.g.}
\begin{equation}
  \langle n_i^{\mu}(t) \rangle =  \langle n_i^{\mu a}(t) \rangle \;,
\end{equation}
with $a=+$ or $-$ equivalently if the observable is time-reversal invariant.

\subsubsection{Keldysh Green's functions}

We introduce the two-time Green's functions $G_{ij\mu\nu}^{\ ab}(t,t')$, defined on the Keldysh contour ($a,b=\pm$), as
\begin{eqnarray}
\langle  n_i^{\mu a}(t) n_j^{\nu b}(t') \rangle = - \frac{1}{\cal Z} \left.\frac{\delta^2 {\cal Z}[\boldsymbol{\eta}^\pm] }{\delta \eta_{i}^{\mu a}(t) \delta \eta_{j}^{\nu b}(t') }\right|_{\boldsymbol{\eta}^\pm=\boldsymbol{0}} \equiv \rmi\hbar G_{ij\mu\nu}^{\ ab}(t,t')\;.
\end{eqnarray}
$n_i^{\mu a}$ being real fields, one has the following time-reversal property
\begin{equation}
 G_{ij\mu\nu}^{\ ab}(t,t') = G_{ji\nu\mu}^{\ ba}(t',t) \;.
\end{equation}
In the operator formalism, the Keldysh Green's functions read
\begin{eqnarray}\label{eq:connexionOP}
 \rmi\hbar G_{ij\mu\nu}^{\ ab}(t,t') = \mbox{Tr} \left[ \mathsf{T}_{\cal C} \ n_{i {\rm H}}^{\mu}(t,a) \ n_{j {\rm H}}^{\nu}(t',b) \ \varrho(0) \right] \;,
\end{eqnarray}
where $n_{i {\rm H}}^{\mu}(t,a)$ denotes the Heisenberg representation of the operator $n_{i}^\mu$ at time $t$ and on the $a$-branch of the Keldysh contour. $\mathsf{T}_{\cal C}$ is the time-ordering operator acting with respect to the relative position of $(t,a)$ and $(t',b)$ on the Keldysh contour $ \cal C$ (see Appendix~\ref{app:Conventions}).

We define the macroscopic Keldysh Green's functions by summing over the $N$ rotors
and each of their $M$ components
\begin{eqnarray}
 G^{ab}(t,t') \equiv \frac{1}{N} \sum_{i=1}^N\sum_{\mu=1}^M G_{ii\mu\mu}^{\ ab}(t,t') \;.
\end{eqnarray}
From the identity (\ref{eq:connexionOP}), one establishes two relations between
the four Green's functions 
\begin{eqnarray}
\begin{array}{rcl}
 G^{++}(t,t') &=& G^{-+}(t,t')\Theta(t-t') + G^{+-}(t,t')\Theta(t'-t) \;, \label{eq:relat_Green1} \\
 G^{--}(t,t') &=& G^{+-}(t,t')\Theta(t-t') + G^{-+}(t,t')\Theta(t'-t) \;,
\end{array}
\end{eqnarray}
leading to
\begin{eqnarray}
\begin{array}{rcl}
 G^{++} + G^{--} &=& G^{+-} + G^{-+} \;, \\
 G^{++}(t,t') - G^{--}(t,t') &=& \mbox{sign}(t-t') \left[ G^{-+}(t,t') -
G^{+-}(t,t') \right]  \;.
\end{array}
\end{eqnarray}

\subsubsection{Self correlation}

We define the macroscopic two-time correlation as
\begin{eqnarray}
 C(t,t') &\equiv& \frac{1}{N}\sum_{i=1}^{N} \frac{1}{2} \langle {\bf n}_i^+(t) \cdot {\bf n}_i^-(t') + {\bf n}_i^-(t) \cdot {\bf n}_i^+(t) \rangle \label{eq:def_C} \\ 
&=& \frac{\rmi\hbar}{2} \left[ G^{+-}(t,t') +  G^{-+}(t,t') \right] = \frac{\rmi\hbar}{2} \left[ G^{--}(t,t') +  G^{++}(t,t') \right]\;.
\end{eqnarray}
It is symmetric in its time arguments $C(t,t') = C(t',t)$. Given the constraint ${\bf n}(t) \cdot {\bf n}(t) = 1$, it is one at equal times: $C(t,t) =1$. The two-time correlation function is the simplest
non-trivial quantity giving information on the dynamics of a system. In
particular, a loss of its time translational invariance (TTI) is a signature of aging.

\subsubsection{Self linear response}
The response at time $t$ of the observable $n_i^\mu$ to an infinitesimal perturbation performed at a previous time $t'$ on an observable $f_i^\mu$ linearly coupled to $n_i^\mu$ is defined as
\begin{equation} \label{eq:defR}
R_i^\mu(t,t') \equiv \left.  \frac{\delta \langle n_i^\mu(t) \rangle}{\delta  f_i^\mu(t')} \right|_{f_i^\mu=0} \;,
\end{equation}
with the modified Hamiltonian
\begin{equation}
 H \longmapsto  H - f_i^\mu n_i^\mu \;.
\end{equation}
Causality ensures that the response vanishes if $t < t'$.
We define the macroscopic linear response as
\begin{equation}
 R(t,t') = \frac{1}{N} \sum_{i=1}^N \sum_{\mu=1}^M R_i^\mu(t,t') \;.
\end{equation}
The functional derivative with respect to $f_i^\mu(t')$  in eq.~(\ref{eq:defR}) can be written in terms of the source fields $\eta_i^{\mu\pm}(t')$ since $f_i^\mu$ appears to play a similar role in the action functional:
\begin{equation}
 \frac{\delta}{\delta f_i^\mu(t')} \longleftrightarrow \frac{1}{\hbar} \left( \frac{\delta}{\delta \eta_i^{\mu +}(t')} - \frac{\delta}{\delta \eta_i^{\mu -}(t')} \right) \;.
\end{equation}
Therefore we obtain a Kubo relation, stating that the response can be expressed in terms of two-time Green's functions:
\begin{eqnarray}
 R(t,t') &=& -\frac{1}{N} \sum_{i=1}^N \sum_{\mu=1}^M \frac{\rmi}{\hbar} \frac{1}{\cal Z} \left( \left.\frac{\delta^2 {\cal Z}[\boldsymbol{\eta}^\pm] }{\delta \eta_{i}^{\mu a}(t) \delta \eta_i^{\mu +}(t')}\right|_{\boldsymbol{\eta}^\pm=\boldsymbol{0}} - \left.\frac{\delta^2 {\cal Z}[\boldsymbol{\eta}^\pm] }{\delta \eta_{i}^{\mu a}(t) \delta \eta_i^{\mu -}(t')}\right|_{\boldsymbol{\eta}^\pm=\boldsymbol{0}}\right) \nonumber \\
&=&   G^{a-}(t,t') - G^{a+}(t,t') \mbox{ with } a=+ \mbox{ or } - \mbox{equivalently} \nonumber  \\
&=&  \frac{1}{2} \left[ G^{--}(t,t') + G^{+-}(t,t') - G^{++}(t,t')- G^{-+}(t,t')  \right] \nonumber  \\
&=& \left[ G^{+-}(t,t') - G^{-+}(t,t') \right] \Theta(t-t')\;,
\end{eqnarray}
where we made use of the relations (\ref{eq:relat_Green1}).

\bigskip
Finally the four Keldysh Green's functions $G^{ab}(t,t')$ can be re-expressed in
terms of a couple of physical observables (namely correlation and response):
\begin{equation}
 \rmi\hbar G^{ab}(t,t') = C(t,t') - \frac{\rmi\hbar}{2} \left[ a R(t',t) + b
R(t,t') \right] \;.
\end{equation}

\subsubsection{Keldysh rotation}\label{sec:KeldyshRotaBos}
The Keldysh rotation of the fields is a change of basis that simplifies the expressions of the physical observables such as the correlation $C$ and the response $R$ in terms of Green's functions.
Moreover the connection with the Martin-Siggia-Rose generating functional in the classical limit is more straightforward in this representation \cite{Kamenev,CugliandoloLozano}.
One introduces new fields as
\begin{eqnarray}
\left\{
\begin{array}{rcl}
2 \; {\bf n}_i^{(1)} &\equiv& {\bf n}_i^+ + {\bf n}_i^- \;, \\
\hbar \; {\bf n}_i^{(2)} &\equiv& {\bf n}_i^+ - {\bf n}_i^- \;, 
\end{array}
\right.
\end{eqnarray}
and the inversion relation
\begin{eqnarray}
 {\bf n}_i^a = {\bf n}_i^{(1)} + a \frac{\hbar}{2} {\bf n}_i^{(2)}  \;.
\end{eqnarray}
We define the Green's functions of these new fields as $\rmi\hbar G^{rs}(t,t')
\equiv 1/N \ \sum_{i=1}^N \langle {\bf n}_i^{r}(t) \cdot {{\bf n}_i^{s}}(t')
\rangle$ with $r,s = (1), (2)$. We have
\begin{eqnarray} \label{sec:CRA}
\begin{array}{ll}
 \rmi\hbar G^{{(11)}}(t,t') =  C(t,t') \;, &
 \rmi\hbar G^{{(12)}}(t,t') =  -\rmi R(t,t') \;, \\
 \rmi\hbar G^{{(21)}}(t,t') =  -\rmi R(t',t) \;, &
 \rmi\hbar G^{{(22)}}(t,t') = 0 \;.
\end{array}
\end{eqnarray}
The fact that $G^{{(22)}}$ vanishes identically is very general and can be tracked back to be a consequence of causality.
The unit length constraint imposed on the rotor coordinates, ${\bf n}_i^a(t) \cdot {\bf n}_i^a(t) = 1$, becomes
an orthogonality constraint between the fields in the new basis, ${\bf n}_i^{(1)}(t) \cdot {\bf n}_i^{(2)}(t) = 0$, 
and a relation between their norms: $ {{\bf n}_i^{(1)}(t)}^2 + \frac{\hbar^2}{4} {{\bf n}_i^{(2)}(t)}^2 = 1$.

\subsubsection{Bosonic FDT}\label{subsec:canonical}

When the system of rotors is in equilibrium at a given temperature $\beta^{-1}$, the fluctuation-dissipation theorem (FDT) holds (in its bosonic version) giving an extra relation between the Green's functions. In Fourier space (see Appendix~\ref{app:Conventions} for our Fourier conventions) it reads
\begin{eqnarray}
 C(\omega) = \hbar \ \coth\left(\beta \hbar \omega/2 \right) \ \mbox{Im }  R(\omega) \;.
\end{eqnarray}
For completeness, we derive this theorem in Appendix~\ref{app:FDT2ndkind}.

\section{The influence of the fermion baths}\label{sec:Self-energy}

\subsection{Self-energy}

We treat the interactions with the environment in perturbation theory up to second order in the coupling. 
After the fermionic degrees of freedom are integrated out, the resulting effective action for 
the rotors acquires an extra term encoding the effects of the reservoirs. The detailed computation, given in Appendix~\ref{app:Sigma}, yields
\begin{eqnarray}
 S_{\rm eff} = S_{S} + S_{\rm LM}  + S^{(2)}_{SB} \;,
\end{eqnarray}
with
\begin{eqnarray}
  \frac{\rmi}{\hbar} S^{(2)}_{SB}[\boldsymbol{s}^{(1)}, \boldsymbol{s}^{(2)}] = \frac{1}{2} M \sum_{rs={(1)}, {(2)}} \iint_{0}^\infty \Udd{t}{t'}  \Sigma_{B}^{rs}(t,t') \sum_{i=1}^N {\bf n}_i^r(t) \cdot {\bf n}_i^s(t') \;,
\end{eqnarray}
and the four self-energy components
\begin{eqnarray}
 \Sigma_B^{{(22)}} &=& 2 (\hbar\omega_c)^2 \ \mbox{Re} \left[ G^{K}_L {G^{K}_R}^* - \hbar^2/4 \ \left( G^{A}_L {G^{A}_R}^* + G^{R}_L  {G^{R}_R}^* \right)   \right] \equiv - \Sigma_B^{K} \;, \\
 \Sigma_B^{{(21)}} &=& - 2 \rmi(\hbar\omega_c)^2 \ \mbox{Re} \left[  G^{R}_L {G^{K}_R}^* + G^{K}_L {G^{R}_R}^* \right] \equiv \rmi \Sigma_B^{R} \;, \\
\Sigma_B^{{(12)}} &=&  2 \rmi (\hbar\omega_c)^2 \ \mbox{Re} \left[  G^{A}_L {G^{K}_R}^* + G^{K}_L {G^{A}_R}^* \right]   \equiv - \rmi \Sigma_B^{A} \;, \\ 
 \Sigma_B^{{(11)}} &=&  0 \;.
\end{eqnarray}
The fact that $ \Sigma_B^{{(11)}}$ vanishes identically is a consequence of causality.
Similarly to what we have done in Sect.~\ref{sec:KeldyshRotaBos} we renamed $\Sigma_B^{{(22)}}$, $\Sigma_B^{{(21)}}$ and $\Sigma_B^{{(12)}}$ into $\Sigma_B^{K}$, $\Sigma_B^{R}$ and $\Sigma_B^{A}$. These real functions are usually referred to as the Keldysh, retarded and advanced components of the self-energy.
$G_\alpha^K$, $G_\alpha^R$ and $G_\alpha^A$ are the Keldysh, retarded and advanced Green's functions of the free electrons in the $\alpha$-reservoir respectively (see Appendix~\ref{app:KeldyRotFermions}).
Using their properties under time reversal (see Appendix~\ref{app:TimeReversal}), we establish
\begin{eqnarray}
 \Sigma_B^{K}(\tau) = \Sigma_B^{K}(-\tau) \;, \qquad
 \Sigma_B^{R}(\tau) = -\Sigma_B^{A}(-\tau)\;.
\end{eqnarray}
These relations reduce the number of independent self-energy components to two (namely $\Sigma_B^{K}$ and $\Sigma_B^{R}$). Plugging the expressions of the fermionic Green's functions given in Appendix~\ref{app:KeldyRotFermions}, we obtain
\begin{eqnarray}
 \Sigma_B^K(\tau)  = -\frac{1}{2} (\hbar\omega_c)^2 \langle \langle 	
\left[\tanh(\beta\frac{\epsilon_L-\mu_L}{2})
\tanh(\beta\frac{\epsilon_R-\mu_R}{2})  -1 \right] \cos\left( \frac{\epsilon_L
- \epsilon_R}{\hbar} \tau \right)
\rangle_L \rangle_R  \;, \\ 
\Sigma_B^R(\tau)  = \frac{1}{\hbar} (\hbar\omega_c)^2 \langle \langle 	
\left[\tanh(\beta\frac{\epsilon_L-\mu_L}{2}) 
-\tanh(\beta\frac{\epsilon_R-\mu_R}{2}) \right] \sin\left( \frac{\epsilon_L -
\epsilon_R}{\hbar} \tau\right)
\rangle_L \rangle_R  \Theta(\tau) \;. 
\end{eqnarray}
The notation $\langle\langle \ \cdots \ \rangle_{L}\rangle_{R}$
stands for $\int \Udd{\epsilon_L}{\epsilon_R} \rho_{L}(\epsilon_L)\rho_{R}(\epsilon_R) \ \cdots \ $.
The Fourier transforms read
\begin{eqnarray}
\Sigma_B^K(\omega) &=&  -\frac{1}{2} \pi \hbar (\hbar \omega_c)^2
\langle \langle
\left[
\tanh(\beta\frac{\epsilon_L-\mu_L}{2})  \,
\tanh(\beta\frac{\epsilon_R-\mu_R}{2}) -1
\right] \nonumber \\
& & \qquad\qquad\qquad \times
\left[
\delta(\hbar\omega - \epsilon_{LR}) + \delta(\hbar\omega + \epsilon_{LR})
\right]
\rangle_{L} \rangle_{R} \;, \label{eq:SKw} \\
\mbox{Re } \Sigma_B^R(\omega) \hspace{-0.7em} &=& \hspace{-0.7em} - (\hbar \omega_c)^2
\langle \langle 
\left[
\tanh(\beta\frac{\epsilon_L-\mu_L}{2})  -
\tanh(\beta\frac{\epsilon_R-\mu_R}{2}) \ 
\right]
\mbox{pv} \frac{\epsilon_{LR}}{(\hbar\omega)^2 - {\epsilon_{LR}}^2}
	\rangle_{L} \rangle_{R} \;,\nonumber   \\ 
\mbox{Im } \Sigma_B^R(\omega) &=& \frac{1}{2}  \pi \hbar (\hbar \omega_c)^2 
\langle \langle 
\left[
\tanh(\beta\frac{\epsilon_L-\mu_L}{2}) -
\tanh(\beta\frac{\epsilon_R-\mu_R}{2}) \ 
\right]
\nonumber \\
& & \qquad\qquad\qquad \times
\left[
\delta(\hbar\omega - \epsilon_{LR}) - \delta(\hbar\omega + \epsilon_{LR})
\right]
\rangle_{L} \rangle_{R} \label{eq:SRIw} \;,
\end{eqnarray}
where $\epsilon_{LR} \equiv \epsilon_L -\epsilon_R$.
Since $\Sigma_B^K(\tau)$ is a real and even function of $\tau$,
$\Sigma_B^K(\omega)$ is also a real and even function of
$\omega$. $\Sigma_B^R(\tau)$ being real, $\Sigma_B^R(\omega)$ is
Hermitian: $\Sigma_B^R(\omega) = {\Sigma_B^R(-\omega)}^*$.

\subsection{Some limits}\label{sec:Limits}
Expressions (\ref{eq:SKw}) and (\ref{eq:SRIw}) of the
Keldysh and retarded self-energies are somehow cumbersome. We simplify them 
here in some physical limits. These
expressions are heavily used in the rest of this work.

\subsubsection{Zero drive}

The $L$ and $R$ reservoirs constitute an equilibrium bath for the
rotors as soon as they share the same temperature and the strength of the drive
is set to zero ($\mu_L =
\mu_R$, $eV=0$). In this case, the fluctuation-dissipation theorem
 applies to the bath, and gives an extra relation between the bath
self-energy components. It reads
 \begin{equation}\label{eq:FDT2ndkind}
\Sigma_B^K (\omega) 
= \hbar \; \coth\left(\beta \frac{\hbar\omega}{2}\right) \; \mbox{Im }  \Sigma_B^R(\omega)
\;.
\end{equation}
Ultimately the number of independent self-energy components reduces to one.
We checked in Appendix~\ref{app:FDT2ndkind} that the expressions (\ref{eq:SKw}) and 
(\ref{eq:SRIw}) comply with the FDT in the equilibrium case.

\subsubsection{Low frequency}
\label{sec:low-freq}

Let us consider the low frequency limit ($\omega\to0$), or long time-difference in real time, of the self-energy components of a generic non-equilibrium bath ($eV\neq0$  \textit{a priori}). Parity considerations on $\Sigma_B^K$ and $\Sigma_B^R$ show that $\Sigma_B^K(\omega)$ approaches $\Sigma_B^K(\omega=0)$ which depends on $T$, $eV$ and $\epsilon_F$ whereas $\mbox{Im } \Sigma_B^R(\omega) \propto \omega$.
The low frequency limit, which can also be seen as the classical limit ($\hbar\omega\ll T$) of the quantum fluctuation-dissipation theorem in eq.~(\ref{eq:FDT2ndkind}) gives a way to express the temperature of an equilibrium bath as
\begin{equation}
 T =  \lim\limits_{\omega\to0}  \frac{1}{2} \
 \frac{\Sigma_B^K(\omega)}{\partial_\omega \mbox{Im } \Sigma_B^R(\omega)} 
  \;.
\end{equation}
By analogy with the equilibrium case, we introduce for non-equilibrium situations
\begin{equation} \label{eq:defTstar}
 T^* \equiv  \lim\limits_{\omega\to0}  \frac{1}{2} \
  \frac{\Sigma_B^K(\omega)}{\partial_\omega \mbox{Im } \Sigma_B^R(\omega)} \;.
\end{equation}
We expect that the effect of the reservoirs on the long time-difference dynamics of the rotors is the one of an equilibrium bath at temperature $T^*$.

\subsubsection{$\epsilon_F$ much larger than all other energy scales}\label{sec:Ohmic}

The reservoirs act as an \textit{Ohmic} bath in the limit in which
$\epsilon_F$ is much larger than the temperature, the drive and
$\hbar\omega$ ($eV, T, \hbar\omega \ll \epsilon_F$). 
Equation~(\ref{eq:SRIw}) with $\Delta\epsilon \equiv\epsilon_L - \epsilon_R
$ reads
\begin{eqnarray}
 \mbox{Im }  \Sigma_B^R(\omega)
&=& \frac{1}{2} \pi (\hbar\omega_c)^2  \int \Ud{\epsilon'} \int \Ud{\Delta \epsilon} \rho(\epsilon') \rho(\epsilon'-\Delta \epsilon)
\left[\delta(\hbar\omega-\Delta \epsilon)-\delta(\hbar\omega+\Delta \epsilon)\right] \nonumber  \\
& & \qquad \times \left[\tanh(\beta\frac{\epsilon'-\mu_0}{2}) - \tanh(\beta\frac{\hbar (\epsilon'-\Delta \epsilon) - \mu_0 - eV}{2})\right] \;.
\end{eqnarray}
In the limit $\hbar\omega \ll \epsilon_F$, we use $\rho(\epsilon'\pm \hbar \omega) \simeq \rho(\epsilon')$ and we derive
\begin{eqnarray*}
 \mbox{Im }  \Sigma_B^R(\omega) \simeq  \frac{1}{2} \pi (\hbar\omega_c)^2 \int \Ud{\epsilon'} \rho^2(\epsilon') \left[\tanh(\beta\frac{\epsilon'+\hbar\omega-\mu_0-eV}{2}) - \tanh(\beta\frac{\epsilon'-\hbar\omega-\mu_0-eV}{2}) \right] \;.
\end{eqnarray*}
The factor within the square brackets in
the integrand 
is peaked at $\epsilon'=\mu_0+eV$. Hence we can approximate 
$\rho^2(\epsilon') \simeq
\rho^2(\mu_0)$ and then compute the remaining integral exactly to 
obtain an Ohmic (in the sense that it is proportional to $\omega$)
behavior for the imaginary part of the retarded self-energy:
\begin{eqnarray}
 \mbox{Im } \Sigma_B^R(\omega) \simeq  2\pi\hbar (\hbar\omega_c)^2 \rho^2(\mu_0) \ \omega \;. \label{eq:Sig_R} \label{eq:ohmic}
\end{eqnarray}
Interesting enough, this expression is independent of $T$ and $V$.
Similar calculations give
\begin{equation}
 \Sigma_B^K(\omega) \simeq  2\pi\hbar (\hbar\omega_c)^2   \rho^2(\mu_0) \frac{eV\sinh(\beta eV) - \hbar\omega\sinh(\beta\hbar\omega)}{\cosh(\beta eV)-\cosh(\beta\hbar\omega)} \;. \label{eq:Sig_K}
\end{equation}
In order to determine $T^*$, we investigate the low frequency limit of $\Sigma_B^K(\omega)$ given in eq.~(\ref{eq:Sig_K}).

\paragraph{Zero drive.} \label{sec:V=0} For $eV \ll T \ll
\epsilon_F$, eq.~(\ref{eq:Sig_K}) yields
\begin{eqnarray} \label{eq:ZeroDriveLimit}
 \Sigma_B^K(\omega) \simeq  2\pi\hbar^2 (\hbar\omega_c)^2  \rho^2(\mu_0) \ \omega \coth{\left(\beta\hbar\omega/2\right)} \;.
\end{eqnarray}
Equations~(\ref{eq:ohmic}) and (\ref{eq:ZeroDriveLimit}) are linked through FDT.
In the low frequency limit 
($\hbar\omega, eV \ll T  \ll \epsilon_F$) it reads
\begin{equation}
 \Sigma_B^K(\omega) \simeq  
4 \pi  \hbar (\hbar \omega_c)^2  \rho^2(\mu_0) \ T \;,
\label{eq:equilow}
\end{equation}
yielding $T^* = T$ as expected in this equilibrium situation.

\paragraph{Finite drive.}\label{sec:FiniteV}
As soon as the drive is not negligible compared to temperature, in the low frequency regime ($\hbar \omega \ll T \ll \epsilon_F$ and $eV \ll \epsilon_F$)
\begin{equation}
 \Sigma_B^K(\omega) \simeq 2 \pi  \hbar (\hbar \omega_c)^2 \rho^2(\mu_0) \ eV \coth{\left( \beta eV/2 \right)} \;, \label{eq:SklowW}
\end{equation}
yielding 
\begin{equation}\label{eq:Tstar_epsinf}
T^* = \frac{eV}{2} \ \coth{\left( \beta eV/2 \right)}\;.
\end{equation}
An ``FDT like'' relation is verified in these limits
\begin{eqnarray}
 \Sigma_B^K(\omega)= \hbar \coth\left({\hbar\omega}/{2T^*}\right) \mbox{Im }\Sigma_B^R(\omega) \;.
\end{eqnarray}
A similar interpretation of the effect of a two-leads bath in these limits on the dynamics of a single localized spin was given in \cite{NunezDuine} and \cite{Basko}.

Furthermore, in the low temperature limit ($\hbar\omega \ll T \ll eV \ll \epsilon_F$)
\begin{equation}
 \Sigma_B^K(\omega) \simeq 2 \pi  \hbar (\hbar \omega_c)^2 \rho^2(\mu_0) \ |eV| \;, \label{eq:lowTlowF}
\end{equation}
yielding $T^* \equiv |eV|/2$.

Finally in the zero temperature limit ($0=T \ll \hbar\omega, eV \ll \epsilon_F$)
\begin{eqnarray}
 \Sigma_B^K(\omega) =  2\pi \hbar (\hbar \omega_c)^2  \rho^2(\mu_0) \left\{  
\begin{array}{l}
 |eV| \mbox{ if } |\hbar\omega| \leq |eV| \;, \\
 |\hbar\omega|  \mbox{ if } |\hbar\omega| > |eV| \;.
\end{array}
\right. 
\end{eqnarray}
In the low frequency regime, we recover expression~(\ref{eq:lowTlowF}). In the zero temperature and zero drive limit ($0=T= eV \ll \hbar\omega \ll \epsilon_F$) 
the Keldysh component of the bath self-energy 
reads
$\Sigma_B^K(\omega) =  2\pi \hbar (\hbar \omega_c)^2  \rho^2(\mu_0) \ |\hbar \omega| $ that goes linearly to zero in the $\hbar\omega\to0$ limit.

\subsubsection{Zero temperature}\label{sec:T=0}
In the $T=0$ limit, we obtain for finite values of the other parameters ($eV, \hbar\omega, \epsilon_F$)
\begin{eqnarray}
\Sigma_B^K(\omega) &=&   \pi \hbar (\hbar \omega_c)^2  \left[  \mbox{sign}(eV + \hbar \omega) \int_{\mu_0}^{\mu_0 + eV + \hbar\omega} \Ud{\epsilon} \rho(\epsilon) \rho(\epsilon - \hbar\omega) \right. \nonumber \\
 & & \qquad\qquad\quad  + \left. \mbox{sign}(eV - \hbar\omega) \int_{\mu_0}^{\mu_0 + eV - \hbar\omega} \Ud{\epsilon} \rho(\epsilon) \rho(\epsilon + \hbar\omega) \right] \;, \label{eq:Sig_KT} \\
\mbox{Im } \Sigma_B^R(\omega)\hspace{-0.7em}  &=& \hspace{-0.7em}  \pi (\hbar \omega_c)^2  \left[ \int_{\mu_0}^{\mu_0 + eV + \hbar\omega} \Ud{\epsilon} \rho(\epsilon) \rho(\epsilon - \hbar\omega) - \int_{\mu_0}^{\mu_0 + eV - \hbar\omega} \Ud{\epsilon} \rho(\epsilon) \rho(\epsilon + \hbar\omega) \right]. \label{eq:Sig_RT}
\end{eqnarray}
In the low frequency limit ($0=T\ll \hbar\omega \ll eV, \epsilon_F$) they yield
\begin{eqnarray}
 \Sigma_B^K(\omega) &\simeq&   2 \pi \hbar (\hbar \omega_c)^2  \, \mbox{sign}(eV)  \int_{\mu_0}^{\mu_0 + eV} \Ud{\epsilon} \rho^2(\epsilon)  \;,  \\
\mbox{Im } \Sigma_B^R(\omega)\hspace{-0.7em}  &\simeq&  \pi \hbar(\hbar \omega_c)^2  \left[ \rho^2(\mu_0) +  \rho^2(\mu_0+eV) \right] \omega\;,
\end{eqnarray}
so that
\begin{eqnarray} \label{eq:Tstar_T0}
 T^*(T=0) = \mbox{sign}(eV)   \frac{\int_{\mu_0}^{\mu_0 + eV} \Ud{\epsilon} \rho^2(\epsilon)}{\rho^2(\mu_0) +  \rho^2(\mu_0+eV)}\;.
\end{eqnarray}

\subsubsection{Some specific reservoirs}
\begin{figure}
 \includegraphics[height=8.00cm, angle=-90]{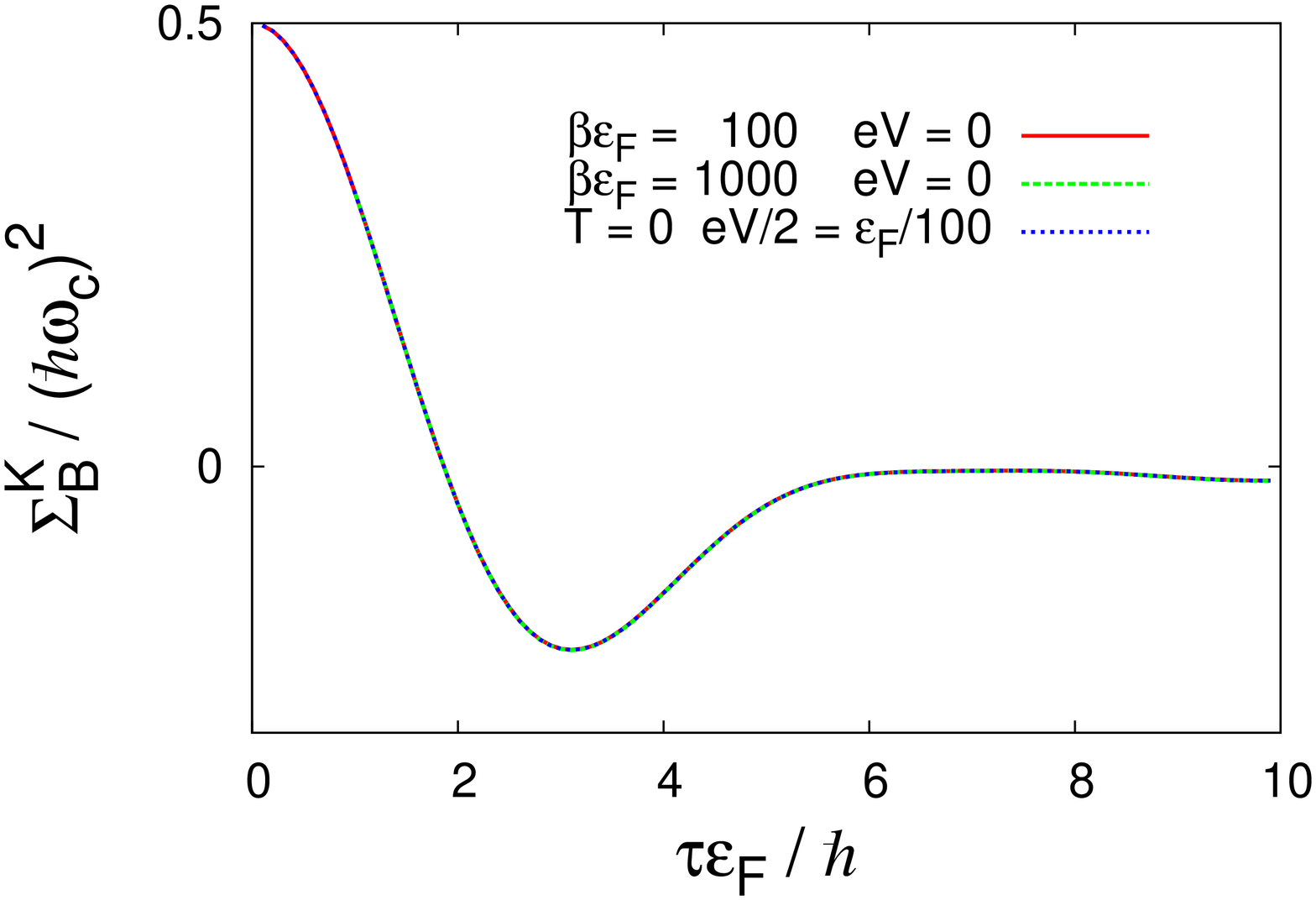}
 \includegraphics[height=8.00cm, angle=-90]{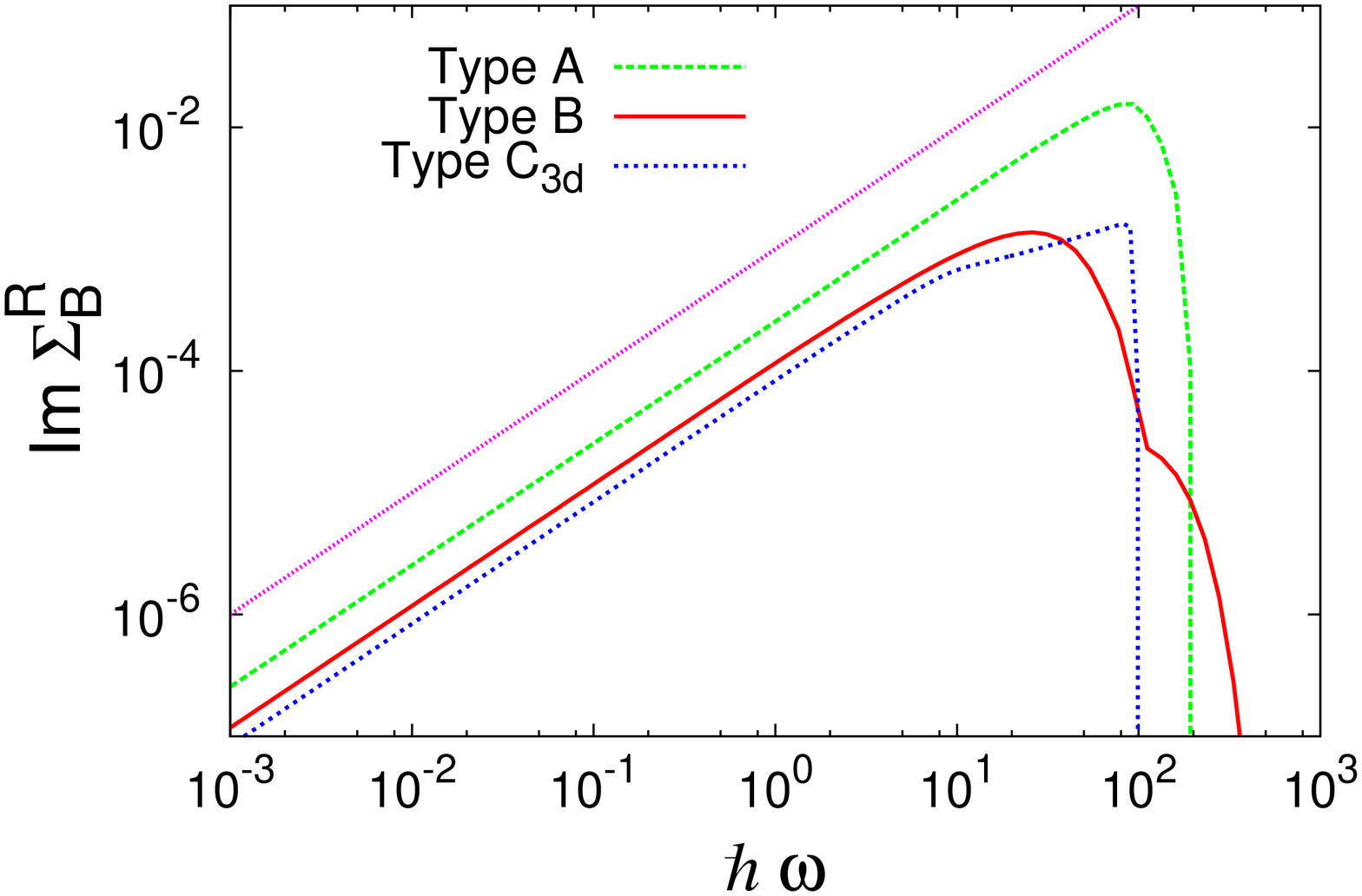}
 \caption{\footnotesize \label{fig:SigmaK_tau} (Color online.) (a) $\Sigma_B^K$ (for
   the half-filled semi-circle DOS) as a function of
   $\tau\epsilon_F$ in the regime where $\epsilon_F$ is much
   larger than any other energy scale: for $\beta\epsilon_F = 100$ and
   $\beta\epsilon_F = 1000$ at $eV=0$ and also for
   $eV/2=\epsilon_F/100$ at $T = 0$. The three curves are indistinguishable. This shows that
   $\Sigma_B^K$ is indeed a function of $\tau\epsilon_F$ in this
   regime and shows furthermore that $eV/2$ plays the same role as
   $T$.  \label{fig:Ohmic} (b) $\mbox{Im } \Sigma_B^R(\omega)$ is
   represented in a double logarithmic scale for the three following DOSs with
   $\beta \epsilon_F = \epsilon_F / eV = 100$: the half-filled
   semi-circle $\rho_A(\epsilon)$, the half-filled type $B$ with
   $\rho_B(\epsilon)$ and the $3{\rm d}$ free electrons DOS
   $\rho_{C3{\rm d}}(\epsilon)$. The straight line above all is a
   guide to the eyes for a pure Ohmic ($\propto \omega$) behavior. 
  The rapid decay above $\hbar\omega \sim \epsilon_F$ is a signature of the energy cut-off, $\epsilon_{cut} \propto \epsilon_F$, of the DOS.
}
\end{figure}

For the half-filled semi-circular DOS (type A), 
at zero drive and zero temperature, we establish the following analytical results at finite $\epsilon_F$:
\begin{eqnarray}
\Sigma_B^K(\tau) &=& 
 2 \left(\frac{\hbar\omega_c}{\epsilon_F}\right)^2 \frac{J_1^2(\tau\epsilon_F/\hbar)-S_1^2(\tau\epsilon_F/\hbar)}{(\tau/\hbar)^2} \label{eq:Struve1} \;, \\
 \Sigma_B^R(\tau) &=&  
\frac{8}{\hbar} \left(\frac{\hbar\omega_c}{\epsilon_F}\right)^2  \frac{J_1(\tau\epsilon_F/\hbar)S_1(\tau\epsilon_F/\hbar)}{(\tau/\hbar)^2} \Theta(\tau) \;, \label{eq:Struve2}
\end{eqnarray}
with $\Sigma_B^R(\tau=0) = 0$, $\Sigma_B^K(\tau=0) =  \frac{1}{2}
(\hbar\omega_c)^2$. $J_1$ and $S_1$ are  the
Bessel and the Struve functions of first kind and first order, respectively. From
eqs.~(\ref{eq:Struve1}) and (\ref{eq:Struve2}), we see that the
temporal extent of both $\Sigma_B^R$ and $\Sigma_B^K$ is of order
$\hbar/\epsilon_F$.
In the limit in which $\epsilon_F$ is much larger
than any other energy scale, a numerical analysis shows that this
property holds for finite values of the temperature and the
drive as well. As a way of summary, in Fig.~\ref{fig:SigmaK_tau} (a) we plot $\Sigma_B^K$
as a function of $\tau \epsilon_F$ for $\epsilon_F = 10 J, 100J$
and at $(T=J, V=0)$ and $(T=0, V=J)$.
In the case in which $\epsilon_F$ is finite, one can compute $T^*$ for the half-filled semi-circular DOS at zero temperature:
\begin{equation}
 T^*(T=0) = \frac{|eV|}{2} \frac{1-1/3 \
(eV/\epsilon_F)^2}{1-1/2\ (eV/\epsilon_F)^2}
 \mbox{  for }  |eV|<eV_{max} = \epsilon_F \;.
\end{equation}

In Fig.~\ref{fig:Ohmic} (b) we give a numerical integration of
$\mbox{Im } \Sigma_B^{R}(\omega)$ for the three types of reservoirs we
introduced in Sect.~\ref{sec:res_of_electrons} and in the case in which
$\epsilon_F$ is the largest energy scale. This shows that the
self-energy is indeed the one of an Ohmic bath. The fact that their
Ohmic behavior is approximately valid until $\hbar \omega = \epsilon_F$
supports the property that the temporal extent of the self-energies
(in real time) is of the order of $\hbar/\epsilon_F$.

\section{Results}

In this section we present our results. We first complete the calculation of disorder
averaged generating function and, from it, we derive Schwinger-Dyson 
equations for the two-time correlation and linear response valid for all values of the 
parameters. We next  derive the dynamical phase diagram as a function of the temperature of the
reservoirs ($T$), the strength of quantum
  fluctuations ($\Gamma$), the voltage ($eV$) and the coupling to the leads for
which we introduce the new dimensionless parameter $g\equiv \hbar\omega_c /
\epsilon_F$. We distinguish two phases separated by a second order phase
transition.
For high values of the temperature and/or strong drive and/or strong quantum
fluctuations, we find a non-equilibrium steady state that approaches the usual
paramagnet when $eV\to0$. Whereas for low temperatures and/or low drive and/or
quantum fluctuations we find a coarsening phase.

\subsection{Average over disorder} \label{sec:average_disorder}

At this stage, after tracing out all fermionic degrees of freedom, the
effective action of our system is quadratic in the fields and reads
\begin{eqnarray}
 \frac{\rmi}{\hbar} S_{\rm eff} 
 \hspace{-1ex} &=& \hspace{-1ex}
M \sum_{i=1}^N \int  \ud{t}   \left\{  
 \frac{\rmi}{\Gamma}   \ \dot{\bf n}_i^{(1)}(t) \cdot  \dot{\bf n}_i^{(2)}(t)
  +   \frac{\rmi}{\sqrt{N}} \sum_{j<i}^N J_{ij}  \left[ {\bf n}_i^{(1)}(t) 
\cdot {\bf n}_j^{(2)}(t) +  {\bf n}_i^{(2)}(t) \cdot {\bf n}_j^{(1)}(t) \right] 
\right. \nonumber\\
 & & \quad\qquad -   \frac{1}{2} \int \ud{t'} \Sigma_B^K(t-t') \ {\bf n}_i^{(2)}(t) \cdot
{\bf n}_i^{(2)}(t') 
  + \rmi \int \ud{t'} \Sigma_B^R(t-t') \ {\bf n}_i^{(2)}(t) \cdot {\bf
n}_i^{(1)}(t') \nonumber  \\
 & & \quad \left. +    \frac{\rmi}{2\hbar}  \sum_{a=\pm} a z_i^a(t) \left[ 1 - \frac{1}{2}
\left({\bf n}_i^{(1)}(t)\right)^2 - a  \hbar \ {\bf n}_i^{(1)}(t) \cdot 
{\bf n}_i^{(2)}(t) - \frac{\hbar^2}{4}  \left({\bf n}_i^{(2)}(t) \right)^2
\right] 
 \right\} \;. 
\end{eqnarray}
Given that the initial condition for the rotors is taken to be
uncorrelated with the disorder configuration (the $J_{ij}$'s), neither the
initial density matrix
$\varrho(0)$ nor the
generating functional without sources (${\cal
  Z}[\boldsymbol{\eta}^\pm=0]=1$) 
depend upon disorder. This property allows us to write dynamic
equations by averaging over disorder 
the generating functional itself
hence without resorting to the use of replicas~\cite{CugliandoloLozano}. 
As in other quantum systems with quenched disorder \cite{Rokni,CugliandoloLozano,effect-bath,other-quantum}, we
are therefore interested in 
\begin{equation}
\overline{\cal Z[\boldsymbol{\eta}^\pm]}^J \equiv 
  \int \left( \prod_{i,j<i}   \ud{J_{ij}} P(J_{ij}) \right) \cal
Z[\boldsymbol{\eta}^\pm]
\;,
\end{equation}
where $P(J_{ij})$ is the Gaussian density distribution for the rotor couplings with zero mean and variance $J^2$.
The disorder average over a random Gaussian potential
can be readily done and the effective action of the 
system is quartic in the fields and reads
\begin{eqnarray}
 \frac{\rmi}{\hbar} S_{\rm eff} 
\hspace{-1ex} &=& \hspace{-1ex}
M \sum_{i=1}^N \int  \ud{t}   \left\{  
 \frac{\rmi}{\Gamma}   \ \dot{\bf n}_i^{(1)}(t) \cdot \dot{\bf n}_i^{(2)}(t)
 \right. 
\\
&& \qquad 
 -  \frac{J^2 M}{2N} \int \ud{t'} \sum_{j}
 \left[ {\bf n}_i^{(1)}(t) \cdot {\bf n}_j^{(2)}(t) \right] \left[ {\bf
n}_i^{(1)}(t') \cdot {\bf n}_j^{(2)}(t')
+ {\bf n}_i^{(2)}(t') \cdot {\bf n}_j^{(1)}(t') \right]
 \nonumber \\
 & &  
\qquad 
- \frac{1}{2} \int \ud{t'} \Sigma_B^K(t-t') \ {\bf n}_i^{(2)}(t) \cdot {\bf
n}_i^{(2)}(t') 
  +\rmi \int \ud{t'} \Sigma_B^R(t-t') \ {\bf n}_i^{(2)}(t) \cdot {\bf
n}_i^{(1)}(t') \nonumber
\\
&& \qquad \left.
 +  \frac{\rmi}{2\hbar}  \sum_{a=\pm} a z_i^a(t) \left[ 1 - \frac{1}{2}
\left({\bf n}_i^{(1)}(t)\right)^2 - a  \hbar \ {\bf n}_i^{(1)}(t) \cdot {\bf
n}_i^{(2)}(t)  - \frac{\hbar^2}{4}  \left( {\bf n}_i^{(2)}(t) \right)^2 \right] 
 \right\} \;. \nonumber
\end{eqnarray}

\subsection{Schwinger-Dyson equations}
\label{sec:Schwinger-Dyson}

In the large $M$ limit, we show that the Lagrange multipliers are homogeneous,
\begin{eqnarray}
 z_i^+(t) = z_i^-(t) \equiv z(t) \ \forall \ i,\,t\;.
\end{eqnarray}
See Appendix~\ref{app:Dynamics} for a detailed computation. Moreover,
introducing
\begin{eqnarray}
 \Sigma^K  \equiv  J^2 C + \Sigma_B^K  \;, \qquad
  \Sigma^R   \equiv J^2 R + \Sigma_B^R  \;,
\end{eqnarray}
we obtain the Schwinger-Dyson equations which fully determine the dynamics of
the system:
\begin{eqnarray} 
\left[\frac{1}{\Gamma}\frac{\partial^2}{\partial t^2} + z(t)\right] C(t,t') 
         &=& \int_{0}^{t'} \ud{t''}  \Sigma^K (t,t'')R(t',t'') + \int_{0}^{t}
\ud{t''}  \Sigma^R (t,t'')C(t'',t') \;,  \label{eq:Dyson1} \\
\left[\frac{1}{\Gamma}\frac{\partial^2}{\partial t^2} + z(t)\right] R(t,t') 
         &=& \delta(t-t') + \int_{t'}^{t} \ud{t''}  \Sigma^R (t,t'')R(t'',t')
\;, \label{eq:Dyson2} 
\end{eqnarray}
\vspace{-1em}
\begin{eqnarray} 
  z(t) =  \int_{0}^{t} \ud{t''}  \Sigma^K (t,t'')R(t,t'') +  \Sigma^R
(t,t'')C(t,t'') - \frac{1}{\Gamma} \frac{\partial^2 C}{\partial t^2}(t,t'\to
t^-)\;. \label{eq:z}\label{eq:Dyson3}
\end{eqnarray}
We remark that the expression for the response is decoupled from the self
correlation apart from a residual coupling through the Lagrange multiplier. This
is actually a consequence of two features of the model: the disordered potential
is quadratic in the rotors and the coupling to the reservoirs is linear in the
rotors. The ``initial'' conditions are given by
\begin{equation}
 C(t,t) = 1, \qquad R(t,t) = 0 \quad \forall \ t\;.
\end{equation}
Moreover, integrating eqs.~(\ref{eq:Dyson1}) and (\ref{eq:Dyson2}) over an
infinitesimal interval around $t'=t$,  one sees that the first derivative of the
correlation is continuous at equal times
\begin{equation}
 \lim\limits_{t'\to  t^-} \partial_t C(t,t') = \lim\limits_{t'\to  t^+}
\partial_t C(t,t') = 0 \;,
\end{equation}
whereas the one of the response function is discontinuous
\begin{equation}
 \lim\limits_{t'\to  t^-} \partial_t R(t,t') = \Gamma, \qquad
 \lim\limits_{t'\to  t^+} \partial_t R(t,t') = 0 \;.
\end{equation}
The structure of these equations is the same as the one in other out of equilibrium problems studied in \cite{Rokni,CugliandoloLozano,effect-bath,other-quantum,CuBi}.

\subsection{Quantum non-equilibrium steady state (QNESS) phase}\label{sec:para}
One expects that if the system is quenched into the high temperature phase, after a short transient it
should relax toward a quantum non-equilibrium steady state (QNESS). The system of rotors cannot be in equilibrium since, for $V\neq
0$, an electronic current is passing through it. Nevertheless the dynamics are still
stationary  (time translationally invariant). This implies that $C(t,t')$ and
$R(t,t')$ are only functions of $t-t'$. Guided by a numerical analysis (see
Sect.~\ref{sec:LagrangeMcheck}), we make the assumption (that we later check to be consistent) that the quantity $z(t)$
is a one-time 
observable that
converges toward a finite value $z^\infty$.
In this situation, one can Fourier transform the  Schwinger-Dyson equations
(\ref{eq:Dyson1}) and (\ref{eq:Dyson2}) with respect to $t-t'$ to find
\begin{eqnarray}
 R(\omega) &=&  \frac{1}{-\Gamma^{-1}\omega^2 + z^{\infty} - { \Sigma^R(\omega)}
} \;, \label{eq:Rw0} \\
 C(\omega) &=&  {\Sigma^K(\omega)} |R(\omega)|^2 \;,\\ 
 C(\omega) &=& \frac{\Sigma_B^K(\omega)}{\mbox{Im }  \Sigma_B^R(\omega)} \mbox{Im
}  R(\omega) \label{eq:CK} \;, 
\end{eqnarray}
Using the fact that $\lim\limits_{\omega \rightarrow \infty} R(\omega)$ has to vanish,
eq.~(\ref{eq:Rw0}) implies
\begin{equation}\label{eq:Rsol}
 R(\omega) = \frac{1}{2J^2} \left(-\Gamma^{-1}\omega^2 + z^{\infty}
-\Sigma_B^R(\omega) + \sqrt{\left(-\Gamma^{-1}\omega^2 + z^{\infty}  -
\Sigma_B^R(\omega)\right)^2- 4J^2} \right) \;.
\end{equation}
We note that in the cases in which the DOS of the reservoirs have an energy
cut-off $\epsilon_{cut}$,
\begin{equation}
 C(\omega) = \mbox{Im } R(\omega) = \Sigma_B^K(\omega) =  \mbox{Im }
\Sigma_B^R(\omega) = 0 \mbox{ for } \hbar \omega > \epsilon_{cut} \;.
\end{equation}

\subsection{Critical manifold}
\subsubsection{Equation for criticality}

Approaching the putative critical manifold from the disordered phase, see Fig.~\ref{fig:epsart}, 
where after a short transient the system should be time translationally invariant, we
look for a singularity in the Fourier transformed Schwinger-Dyson equations that
would be the signature of the loss of time translational invariance and
ultimately of a phase transition toward an out of equilibrium behavior. Anticipating a
second order phase transition scenario where the onset of criticality is
characterized by long-wavelength instabilities, we inspect these equations at
$\omega=0$.

The constraint that rotors have a unit length $C(t,t) = 1$ implies
\begin{equation}
 \int_0^{\infty} \frac{\rm{d}\omega}{2 \pi} \ C(\omega) = \frac{1}{2} \;,
\end{equation}
and replacing $C(\omega)$ with its expression in eq.~(\ref{eq:CK}):
\begin{equation} \label{eq:critical}
 \int_0^{\infty} \frac{\rm{d}\omega}{2 \pi}  \frac{\Sigma_B^K(\omega)}{\mbox{Im
}  \Sigma_B^R(\omega)} \mbox{Im }  R(\omega) = \frac{1}{2} \;.
\end{equation}
Equation~(\ref{eq:Rsol}) at $\omega = 0$ reads
\begin{equation}\label{eq:R0}
 R(\omega = 0) = \frac{1}{2J^2} \left(z^{\infty} - \Sigma_B^R(\omega = 0) +
\sqrt{\left(z^{\infty} - \Sigma_B^R(\omega=0)\right)^2- 4J^2} \right) \;.
\end{equation}
$R(\omega = 0) = \int_{0}^\infty \Ud{\tau} R(\tau)$ has to be real since
$R(\tau)$ is
real\footnote{$\Sigma_B^R(\omega=0)$ is real for the same reason.}.
However, it is clear from eq.~(\ref{eq:R0})  that  $z^{\infty} = z^\infty_c \equiv 2J +
\Sigma_B^R(\omega=0)$ is a singular point (a minus sign would be incoherent with the approach in
Sect.~\ref{lab:CugDean}). This is the signature of the phase transition we
were looking for. At criticality,
\begin{equation}\label{eq:R0c}
\left.R(\omega = 0)\right\lvert_{z^\infty=z^\infty_c} = 1/J \;.
\end{equation}
Concomitantly, the value of $C(\omega=0)$ blows up. 
Inserting $z^\infty_c$ in eq.~(\ref{eq:critical}), we obtain the equation for
the critical manifold,
\begin{equation} \label{eq:critical2}
 \int_0^{\infty} \frac{\rm{d}\omega}{2 \pi}  \frac{\Sigma_B^K(\omega)}{\mbox{Im
}  \Sigma_B^R(\omega)} \mbox{Im }  R(\omega)\lvert_{z^\infty_c} = \frac{1}{2}
\;.
\end{equation}
The parameters are the strength of
quantum fluctuations $\Gamma$, the temperature $T$, the voltage applied between
the two reservoirs $V$. We recall that $J$ is the typical interaction
between two rotors. The energy variation scale of the reservoirs is characterized
by $\epsilon_F$ and $\hbar \omega_c$ quantifies the coupling strength of the
rotors to their environment through the dimensionless small parameter
$g\equiv{\hbar\omega_c}/{\epsilon_F}$.

In the rest of this Section, we use eq.~(\ref{eq:critical2}) to uncover the phase
diagram of Fig. ~\ref{fig:epsart}. The critical surface is parametrized in the $T$, $\Gamma$ $V$ space by $T_c$,
$\Gamma_c$, $V_c$ ($g$ is kept constant). We introduce the critical points $\bar
T_c \equiv
T_c(\Gamma=V=0)$, $\bar V_c \equiv
V_c(T=\Gamma=0)$, $\bar \Gamma_c \equiv
\Gamma_c(T=V=0)$.
Anticipating the coming results, we introduce the dimensionless reduced
parameters $\theta \equiv T / J$, and $\upsilon \equiv eV / 2J$, $\gamma \equiv  (4\hbar/3\pi)^2
\ \Gamma/ J$.
In the plane $V=0$, where the reservoirs act like an
equilibrium bath, we recover the results in \cite{Rokni}.
In the classical limit $V=\Gamma=0$, we recover the ones in \cite{CugliandoDean}.

In the limit in which $\epsilon_F$ is much larger than any other energy scale,
using eqs.~(\ref{eq:Sig_R}) and (\ref{eq:Sig_K}), the equation for the critical surface reads
\begin{equation} \label{eq:critical_2}
 \int_0^{\infty} \frac{\rm{d}\omega}{2 \pi}  \frac{1}{\omega} 
\frac{eV\sinh(\beta eV) - \hbar\omega\sinh(\beta\hbar\omega)}{\cosh(\beta
eV)-\cosh(\beta\hbar\omega)} \mbox{Im }  R(\omega)\lvert_{z^\infty_c} =
\frac{1}{2} \;.
\end{equation}

\subsubsection{Critical points on the $\Gamma=0$ plane}\label{sec:CPG0}
\begin{figure}[t]
 \includegraphics[width=5.50cm,angle=-90]{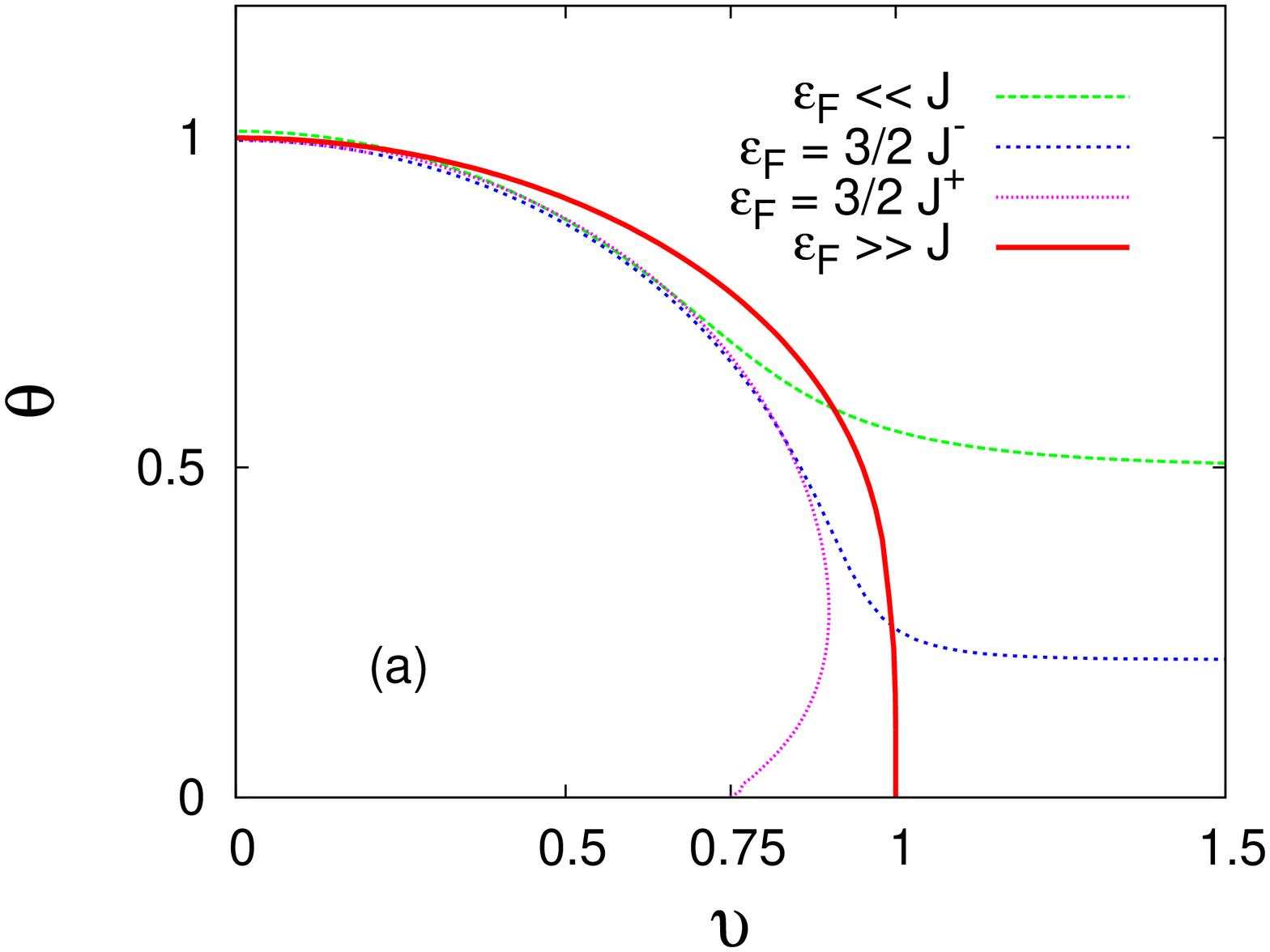}
 \includegraphics[width=6.00cm,angle=-90]{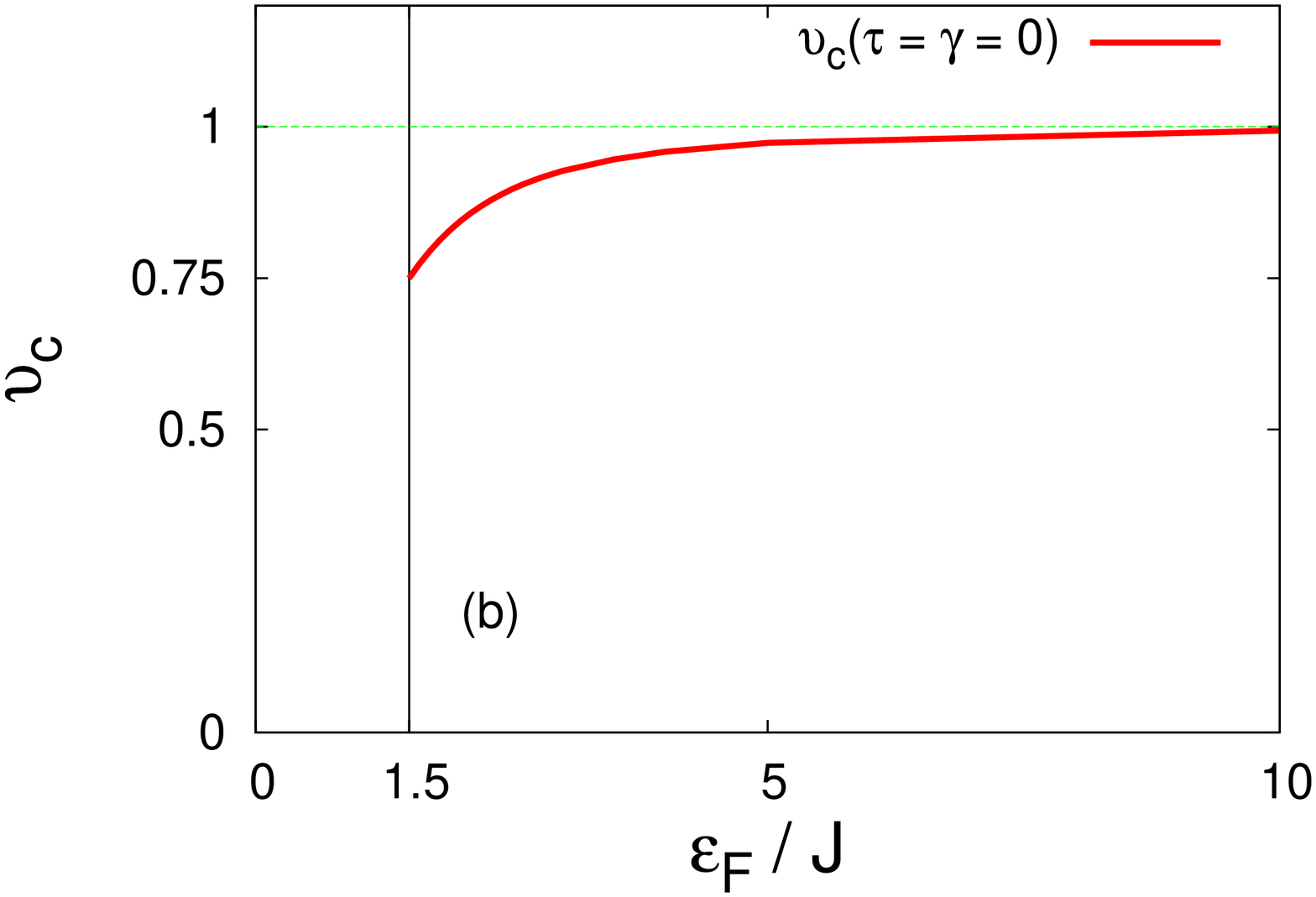}
 \caption{\footnotesize \label{fig:Vc-semicircle} Study of the behavior of the
$\gamma = 0$ critical line with the ratio $\epsilon_F / J$ for the half-filled
semi-circle DOS. (a) The $\gamma=0$ critical line $\theta_c(\upsilon)$ is given for four
different values of the ratio $\epsilon_F / J$. The analytical expression of the
$\epsilon_F / J \to \infty$ curve is given in eq.~(\ref{eq:Vc_exact}). For
$\epsilon_F / J < 3/2$ the critical point $\bar \upsilon_c$ is rejected to infinity.
(b)  $\bar \upsilon_c \equiv \upsilon_c(\theta=\gamma=0)$ is plotted against $\epsilon_F / J$. All
these $\gamma=0$ results are independent of the value $g$.} 
\end{figure}
Taking the $\Gamma\to 0$ limit of expression (\ref{eq:Rsol}) one has
\begin{eqnarray}
 \mbox{Im } R(\omega')\lvert_{z^\infty_c} &=& \left\{ 
\begin{array}{lcl}
\frac{1}{J} \sqrt{1-(1-\omega'^2)^2} &\mbox{ for }& \omega' \in [0,\sqrt{2}] \;,
\\ 
0 &\mbox{ for }& \omega' \geq \sqrt{2}\;,
\end{array}
\right.
\end{eqnarray}
where we introduced  $\omega' \equiv  \omega / \sqrt{2J\Gamma}$. 
The expression of $\mbox{Im } R(\omega)$ does not involve the reservoirs:
the time scale of the rotors (controlled by $\Gamma$) totally decouples from the
one of the reservoirs in such a way that the rotors only couple with the zero
mode (the slowest) of the reservoirs. Using eq.~(\ref{eq:critical2}), we write
the equation of the critical manifold in the $\Gamma=0$ plane
\begin{eqnarray}
 \lim\limits_{\Gamma\to0} \sqrt{\frac{2 \Gamma}{J}} \int_0^{\sqrt{2}}
\frac{\ud{\omega'}}{2 \pi} \sqrt{1-(1-\omega'^2)^2} \ 
\frac{\Sigma_B^K(\sqrt{2J\Gamma}\omega')}{\mbox{Im } 
\Sigma_B^R(\sqrt{2J\Gamma}\omega')}  = \frac{1}{2} \;.
\end{eqnarray}
Using the definition~(\ref{eq:defTstar}) of $T^*(T,eV)$ introduced in Sect.~\ref{sec:low-freq}, this simply reads
\begin{eqnarray} \label{eq:Critik}
 T^*(T_c, eV_c) = J \;.
\end{eqnarray}

At $eV=0$, for which the reservoirs constitute an equilibrium bath, the ratio
${\Sigma_B^K}/{\mbox{Im }  \Sigma_B^R}$ is given by the FDT and we find a
temperature-induced classical critical point $\bar T_c \equiv T_c(\Gamma=V=0) =
J$. In terms of the reduced temperature this reads $\bar \theta_c = 1$. In the next
two paragraphs we look at how this critical point is affected by a finite drive
($eV\neq0$).

\begin{figure}[t]
 \centering
 \includegraphics[width=5.00cm,angle=-90]{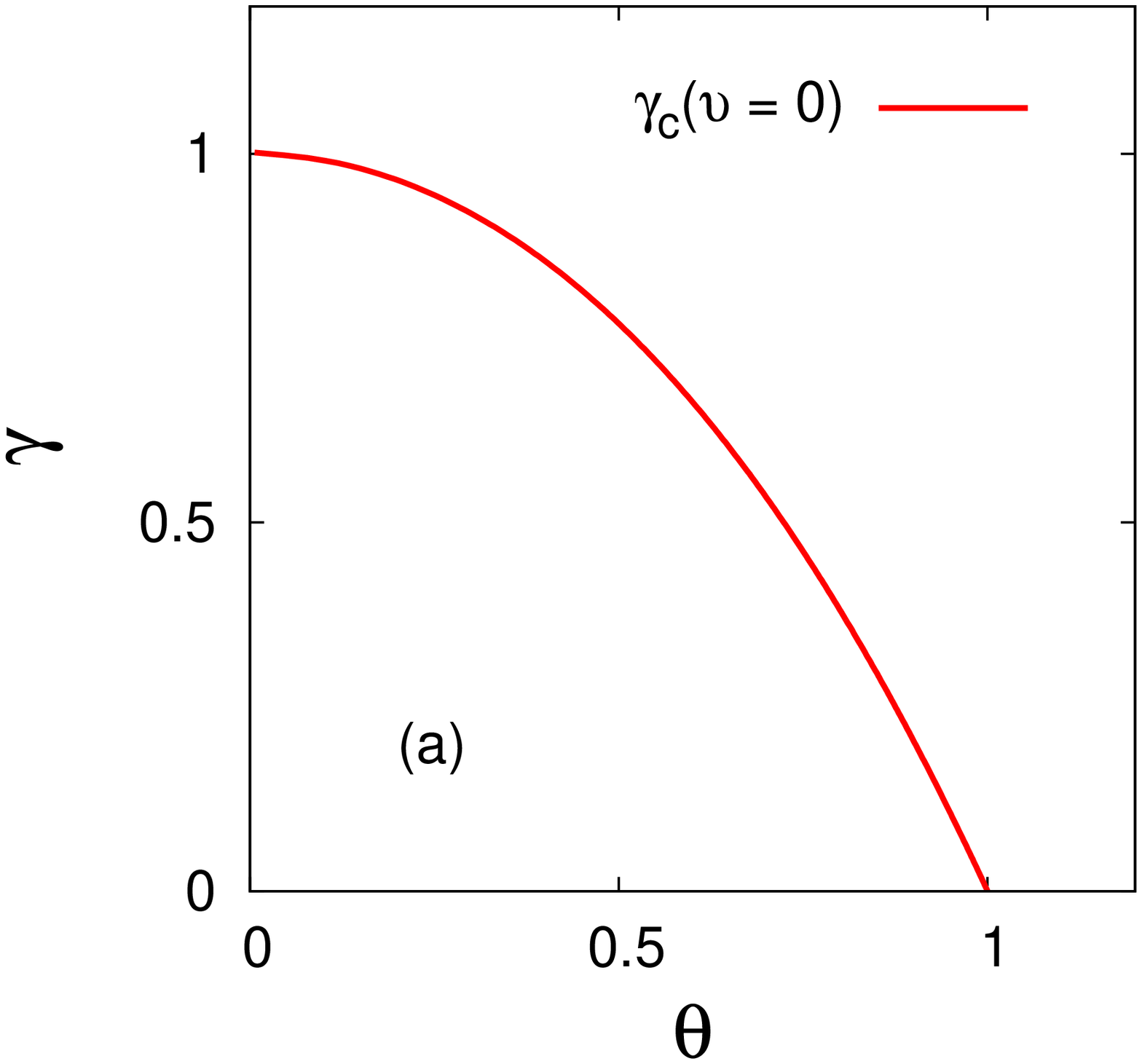}
 \includegraphics[width=5.00cm,angle=-90]{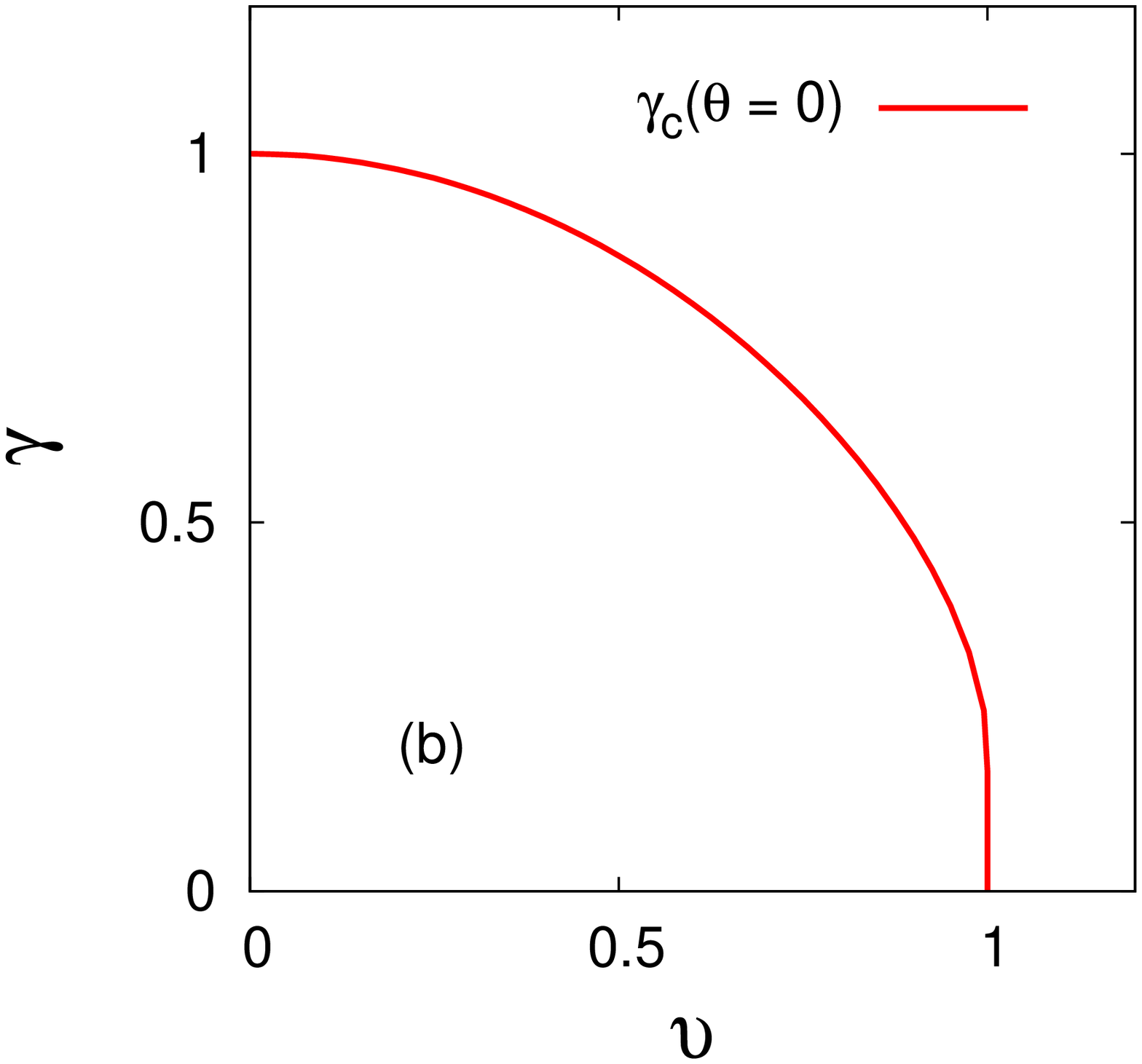}
 \caption{\footnotesize \label{fig:diag_g} Phase diagram in terms of the reduced
parameters analytically determined in the limit $g \to 0$. (a)~Critical line for
$V=0$. (b)~Critical line for $T = 0$ in the limit $\epsilon_F \to \infty$.}
\end{figure}

\paragraph{Infinite $\epsilon_F$.}
We first consider the limit $\epsilon_F \to \infty$, using the explicit expression~(\ref{eq:Tstar_epsinf}) for $T^*$ one finds:
\begin{eqnarray}
 T_c(eV) =  {\left. \frac{eV}{2}  \middle/  \mathrm{arccoth}\left(
\frac{2J}{eV} \right) \right.} \;. \label{eq:Vc_exact}
\end{eqnarray}
From this equation we find a drive-induced critical point at
${e\bar V_c}/2 = J$. In terms of the reduced voltage this reads $\bar \upsilon_c = 1$.
The departure from the classical critical temperature on the $\gamma=0$
plane is quadratic: $\theta_c \simeq 1 - \left(1/3\right) \upsilon^2$ for $\upsilon \ll 1$.  Instead, on the
zero-drive plane, $\upsilon=0$, the critical line leaves $\bar \theta_c$ linearly: $\theta_c
\simeq 1 - \left({3 \pi^2}/{16}\right) \gamma$ for $\gamma \ll 1$. More details on the critical
line $\gamma_c(t)$ at $\upsilon=0$ are given in \cite{SachdevYe}.
Close to $\bar \upsilon_c$ on the $\theta=0$ and $\gamma=0$ planes the departure of the
critical lines $\gamma_c(\upsilon)$ and $\theta_c(\upsilon)$, respectively, are non-analytical and
thus very steep [see Figs.~\ref{fig:Vc-semicircle}~(a)
and~\ref{fig:diag_g}~(b)].

\paragraph{Finite $\epsilon_F$.}
Let us now investigate the $T=0$ critical point $\bar V_c$ for finite values of
$\epsilon_F$. For our simple DOS depending on a unique parameter $\epsilon_F$, $\bar \upsilon_c$ is
controlled by $\epsilon_F / J$. Plugging the expression~(\ref{eq:Tstar_T0}) for $T^*(T=0)$ into the expression (\ref{eq:Critik}) we obtain
\begin{equation} \label{eq:critikV}
\mbox{sign}(e \bar V_c) \frac{1}{J} \frac{\int_{\mu_0}^{\mu_0+e \bar V_c} 
\ud{\epsilon'} \rho^2(\epsilon')}{\rho^2(\mu_0) + \rho^2(\mu_0 + e\bar V_c)} = 1
\;.
\end{equation}
The existence and the value of the solution $\bar V_c$ depend on the details of
the DOS $\rho(\epsilon)$.
If the DOS has an energy cut-off $\epsilon_{cut}$, the existence of a solution
is guaranteed if the cut-off is larger than the solution $\epsilon_{cut}^{min}$
of
\begin{equation}
 \int_{\mu_0}^{\epsilon_{cut}^{min}} \Ud{\epsilon} \rho^2(\epsilon) = J
\rho^2(\mu_0) \;.
\end{equation}
For the type A half-filled semi-circle distribution ($\mu_0 = \epsilon_F$,
$\epsilon_{cut} = 2 \epsilon_{F}$), it turns out that eq.~(\ref{eq:critikV})
admits
a finite solution as soon as $\epsilon_F / J  \geq 3/2$. For
$\epsilon_F / J=3/2$, one finds $e \bar V_c = 3/2 \ J$ ($\bar \upsilon_c  = 3/4$) . 
For $\epsilon_F / J > 3/2$, the finite solution $\bar \upsilon_c$ goes to one as one
increases the ratio $\epsilon_F/J$. For $\epsilon_F / J < 3/2$ the critical
point is rejected to infinity and the critical line in the $\Gamma = 0$ plane
converges to the asymptotic value $\theta_c(\upsilon\gg1)  = 1/2$ as $\epsilon_F/J \to 0$. See
Fig.~\ref{fig:Vc-semicircle}.

For the distribution B, if $\mu_0 \neq 0$,  the scenario is the same as for the
semi-circle distribution
 there is a finite value of the ratio $\epsilon_F/J$ under which, the critical
point $\bar \upsilon_c$ is rejected to infinity, and above which, $\bar \upsilon_c$ has a
finite value that goes to $1$ in the limit $\epsilon_F \to \infty$. If $\mu_0 =
0$ then $\bar \upsilon_c$ remains finite.

For the distribution of type C, eq.~(\ref{eq:critikV}) always admits a finite
solution $\bar \upsilon_c$ independent of $\epsilon_F$. For the distribution $C3{\rm
d}$, $\bar \upsilon_c = 1$ regardless of $\mu_0$, $\epsilon_F$ and $J$.
For the distribution $C2{\rm d}$, we also get $\bar \upsilon_c = 1$.
For the distribution $C1{\rm d}$, one can show that as long as $\mu_0 >0$, there
is a finite $\bar \upsilon_c$, function only of $u\equiv J/\mu_0$: $\bar \upsilon_c = \left[
\exp{\left( u + \rm{L}(u \rme^{-u}) \right)} -1 \right]/2u$, where $\rm{L}(x)$ is
the only solution of the equation $L \rme^L = x$ that is analytic in $0$. For
$\mu_0(\epsilon_F\to \infty) \to \infty$, we recover $\bar \upsilon_c = 1$.

\subsubsection{Quantum critical point}
\paragraph{Weak coupling limit.}
We first consider the limit of the weak coupling to the reservoirs $g \to  0$
after the long-time limit such that the asymptotic regime has been established.
It is actually in this $g \to  0$ limit that the self-energy was computed (we
expanded the total action up to second order in $g$) in Sect.~\ref{sec:Self-energy}. $g \equiv \hbar\omega_c/\epsilon_F$ can be sent
to zero by sending the coupling parameters to zero, but for our simple DOS, it
can also be realized by sending $\epsilon_F$ to infinity.

In equilibrium ($V=0$) at $T=0$,  the FDT gives 
\begin{equation}
\frac{\Sigma_B^K(\omega)}{\mbox{Im }  \Sigma_B^R(\omega)} = \hbar  \mbox{ for }
0<\hbar\omega < \epsilon_{cut} \;. \label{eq:FDTT0}
\end{equation}
By turning off the coupling to the reservoirs ($g\to0$) in eq.~(\ref{eq:R0}) on
has
\begin{eqnarray}
 \mbox{Im }  R(\omega')\lvert_{z^\infty_c}  &=& \left\{
\begin{array}{lcl}
\frac{1}{J}
\sqrt{1-(1-\omega'^2)^2} &\mbox{ for }& \omega' \in [0,\sqrt{2}] \;, \\
 0 &\mbox{ for }& \omega' \geq \sqrt{2} \;,
\end{array}
\right.
\label{eq:ImRg0}
\end{eqnarray}
where we introduced  $\omega' \equiv \omega /  {\sqrt{2J\Gamma}} $. Plugging
eqs.~(\ref{eq:FDTT0}) and (\ref{eq:ImRg0}) in the equation for the critical
manifold ($\ref{eq:critical2}$) gives the quantum critical point
\begin{eqnarray}
 \hbar^2 \bar \Gamma_c \equiv \left(\frac{3 \pi}{4}\right)^2 J \; \mbox{  if } 
\epsilon_{cut} > \frac{3 \pi}{2} J \mbox{ and no solution otherwise.}
\end{eqnarray}
For type A reservoirs in the $\epsilon_F \to  \infty$ limit, one can prove that
the critical surface is parabolic close to the quantum critical point
$\bar \gamma_c$, \textit{i.e.}, $\gamma_c \simeq 1 - \left({16}/{3 \pi^2}\right) \theta^2$ at $\theta\ll 1$ and
$\upsilon=0$, and $\gamma_c \simeq 1 -  \left({16}/{3 \pi^2}\right) \upsilon^2$ for $\upsilon \ll 1 $ and $\theta=0$.

\paragraph{Finite coupling.}
\begin{figure}[t]\label{fig:diags_2}
 \centering
\includegraphics[width=5.00cm,angle=-90]{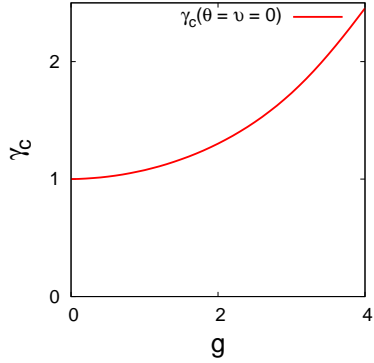}  
\caption{\footnotesize \label{fig:diag_T} Numerical study of the evolution of
the critical point $\bar \gamma_c \equiv \gamma_c(\theta=0,v=0)$ with the coupling
parameter $g$ (here for $\epsilon_F/J = 10$). }
\end{figure}
When the coupling to the electronic reservoirs $g$ is finite  this quantum
critical point (actually the whole critical surface) moves upward when
increasing the coupling constant (see Fig.~\ref{fig:diag_T}). The coarsening
phase is thus stabilized when increasing the
coupling to the reservoirs. In the $\epsilon_F\to\infty$ limit, one has for $g
\ll 1 $
\begin{equation}
\bar \gamma_c \simeq 1 + 2  \left( \frac{3\pi}{4}\right)^2 (\hbar\omega_c)^2
\rho^2(\mu_0)\;.
\end{equation}
In the case of the type A half-filled semi-circle distribution this reads $\bar
\gamma_c \simeq 1 + \left(9/2\right) g^2$.
This is similar to what was found for other quantum spin models embedded
in an Ohmic harmonic oscillator bath and is due to a spin-localization-like
effect~\cite{Rokni, effect-bath}. This similitude is not surprising since we
showed in Sect.~\ref{sec:Limits}, eq.~(\ref{eq:Sig_R}), that the mixed
electronic reservoirs behave like an Ohmic bath in the $\epsilon_F \to  \infty$
limit.

\subsubsection{Summary of the phase diagram}
\begin{table}
\begin{center}
\begin{tabular}{|lll|}\hline
\begin{tabular}{ccl}
$\scriptstyle \Gamma_c(V=0)$ & $\scriptstyle\sim$ & $\scriptstyle\overline{T}_c
- T$\\
$\scriptstyle \Gamma_c(T=0)$ & $\scriptstyle\sim$ & $\scriptstyle{\rm NA}$
\end{tabular} & \begin{tabular}{ccl}
$\scriptstyle T_c(\Gamma=0)$ & $\scriptstyle\sim$ & $\scriptstyle{\rm NA}$\\
$\scriptstyle T_c(V=0)$ & $\scriptstyle\sim$ & $\scriptstyle\left(
\overline{\Gamma}_c - \Gamma \right)^{1/2}$
\end{tabular} & \begin{tabular}{ccl}
$\scriptstyle V_c(T=0)$ & $\scriptstyle\sim$ &$\scriptstyle\left(
\overline{\Gamma}_c - \Gamma \right)^{1/2}$\\
$\scriptstyle V_c(\Gamma=0)$ & $\scriptstyle\sim$ & $\scriptstyle\left(
\overline{T}_c - T \right)^{1/2}$ \\ 
\end{tabular} \\\hline
\end{tabular}
\end{center}
 \caption{ \footnotesize \label{table:behav} Behavior of the critical manifold
close to the critical points for $g\to0$ and $\epsilon_F\to\infty$. Close to the critical point $\overline V_c=V_c(T=\Gamma=0)$ the critical lines are non-analytical (${\rm NA}$).}
\end{table}

Let us summarize the key features of the critical manifold in the case of a DOS
with $\epsilon_F\to\infty$. When the coupling to the reservoir  $g$ is set to
zero, the values of three critical points ($\bar T_c$, $\bar \Gamma_c$, $e \bar
V_c$) are only controlled by $J$ that measures the disorder strength.
Figure~\ref{fig:epsart} gathers all the $g\to0$ results in the $T$, $\Gamma$,
$V$ space. The increase in either the thermal or quantum fluctuations, by
raising $\Gamma$ or the temperature $T$, respectively, leads to the
destabilization of the coarsening phase. The same occurs for an increase in the
bias voltage $V$. The summary of the behavior of the critical manifold close to the critical
points $\bar T_c$ , $\bar \Gamma_c$ and $\bar V_c$ is given in
Table~\ref{table:behav}. Furthermore, an increase in the rotors-reservoirs coupling
$g$ pulls the quantum critical point $\bar \Gamma_c$ upward (as indicated in
 Fig.~\ref{fig:epsart} by a vertical arrow) enlarging the low temperature phase.

\subsection{Coarsening phase} \label{lab:CugDean}

We study the dynamics in the low $T$, weak $\Gamma$, weak $V$ region of the
phase diagram by solving the Schwinger-Keldysh equations in two ways: with an
exact numerical approach and using analytic approximation in the long-time
dynamics. We prove that in this region of the phase diagram there is
coarsening and that the aging dynamics that occur are universal and
equivalent to the ones of the classical (and undriven) limit of our model
(\textit{a.k.a.} the $p=2$ spherical model with quenched disorder).

\subsubsection{Numerical solution}
Our numerical analysis consists in solving the Schwinger-Dyson equations
(\ref{eq:Dyson1}), (\ref{eq:Dyson2}) and (\ref{eq:Dyson3}) after a quench into
the low temperature, weak quantumness, weak drive phase. Thanks to their causal
structure, the equations on $C$, $R$ and $z$ can be
integrated step by step in time, with a Runge-Kutta method. Apart from
arbitrarily small numerical errors, this approach is exact.

We concentrate on reservoirs at temperature $T$ that have a type A semi-circle
DOS (both $L$ and $R$ reservoirs). $L$ reservoirs are kept half-filled while a
voltage $V$ is applied between $L$ and $R$ reservoirs. $\epsilon_F$ is chosen to
be the largest energy scale. Typically, we consider the following values for the
parameters: $T \sim \Gamma \sim eV \sim 0.1 J$ and $\epsilon_F \sim 10 J$. 

The analysis shows (analytical arguments are given in 
Sect.~\ref{sec:Cugliandolo-Dean}) that the dynamics after the quench below the critical surface do not
reach a QNESS. There is a separation of two-time scales typical of aging
phenomena \cite{LeticiaLesHouches}. The 
data in Figs.~\ref{fig:C-chi}-\ref{fig:zeta} were obtained using the algorithm briefly described.

\subsubsection{Mapping to Langevin dynamics}

The goal of this subsection is to map our quantum field theory description of
the rotors dynamics, which involves the two fields $\boldsymbol{n}^{(1)}$ and
$\boldsymbol{n}^{(2)}$ (see Sect.~\ref{sec:KeldyshRotaBos}), to an equivalent description 
in terms of Langevin dynamics.
In the long-time limit of the coarsening dynamics, we establish that the
equation of motion for the field $\boldsymbol{n}^{(1)}$ is actually a Langevin
equation driven by a colored noise $\boldsymbol{\xi}$ the statistical
characteristics of which are controlled by the self-energies of the fermion reservoirs.

Let us take a step back and rewrite the effective action as it was before
averaging over disorder. Making the assumption (we later check
its consistency) that the Lagrange multipliers satisfy
$z_i^+(t) = z_i^-(t) = z_i(t) \ \forall \ i, t$, the effective action reads
\begin{eqnarray}
 \frac{\rmi}{\hbar} S_{\rm eff} [\boldsymbol{n}^{(1)}, \boldsymbol{n}^{(2)},
z]
\hspace{-1ex} &=& \hspace{-1ex}
M \sum_{i=1}^N \int  \ud{t}  \left\{  
\frac{\rmi}{\Gamma}   \ \dot{\bf n}_i^{(1)}(t) \cdot \dot{\bf n}_i^{(2)}(t)
 \right.
 +\rmi  \sum_{j=1}^N  \mathcal{J}_{ij} \ {\bf n}_i^{(1)}(t) \cdot {\bf
n}_j^{(2)}(t)  \nonumber\\
& & -   \frac{1}{2} \int \ud{t'} \Sigma_B^K(t-t') \ {\bf n}_i^{(2)}(t) \cdot
{\bf n}_i^{(2)}(t') 
   +\rmi \int \ud{t'} \Sigma_B^R(t-t') \ {\bf n}_i^{(2)}(t) \cdot {\bf
n}_i^{(1)}(t') \nonumber \\
& & \left.  - \rmi  z_i(t)  \ {\bf n}_i^{(1)}(t) \cdot {\bf n}_i^{(2)}(t)   
 \right\} \;,
\end{eqnarray}
where introduced the real and symmetric matrix ${\bf \mathcal{J}}$ defined by
$\mathcal{J}_{ij} \equiv J_{ji}/\sqrt{N}$ if $j<i$, $\mathcal{J}_{ij} \equiv
\mathcal{J}_{ji}$ if $j>i$. Like the other components of this matrix, we set
$\mathcal{J}_{ii} $ to be taken from a Gaussian distribution with
zero mean and variance $J^2/N$ [we saw that the constraint  ${{\bf n}_i(t)}^2 = 1$ yields ${\bf n}_i^{(1)}(t) \cdot {\bf
n}_i^{(2)}(t) = 0$].
The total effective action adopts the quadratic form
\begin{eqnarray}
 \frac{\rmi}{\hbar} S_{\rm eff}  = -M \sum_{i=1}^N \int \ud{t}  \left\{ \rmi {\bf
n}_i^{(2)}(t) \cdot \boldsymbol{\xi}_i(t) + \frac{1}{2} \int \ud{t'}
{\bf n}_i^{(2)}(t) \cdot \Sigma_B^K(t-t') \ {\bf n}_i^{(2)}(t') \right\} \;,
\end{eqnarray}
where we  introduced the $N$ auxiliary fields $\boldsymbol{\xi}_i$:
\begin{eqnarray}\label{eq:avantprojo}
\boldsymbol{\xi}_i(t) \equiv \sum_{j=1}^N \int \ud{t'} \left\{   \left[  \left(
\frac{1}{\Gamma} \partial^2_t + z_i(t)\right) \delta_{ij} - \mathcal{J}_{ij} 
\right] \delta(t-t') -\Sigma_B^R(t-t') \delta_{ij} \right\} {\bf n}_j^{(1)}(t')
\;.
\end{eqnarray}
By integrating over $ {\bf n}_i^{(2)}$, we are left with
\begin{eqnarray}\label{eq:Lang}
 \frac{\rmi}{\hbar} S_{\rm eff}  =  - M \sum_{i=1}^N \iint \udd{t}{t'} 
\boldsymbol{\xi}_i(t) \cdot \frac{1}{2} {{\Sigma_B^K}^{-1}(t-t')} \
\boldsymbol{\xi}_i(t') \;.
\end{eqnarray}
From this Gaussian action, the quantity $\boldsymbol{\xi}_i(t)$ can be
interpreted
as a Gaussian random process with a zero average and variance
$\langle\boldsymbol{\xi}_i(t) \cdot \boldsymbol{\xi}_j(t')\rangle_\xi =
\delta_{ij} \Sigma_B^K(t-t')$ and eq.~(\ref{eq:avantprojo}) as a set of coupled 
Langevin equations. This mapping is possible since the action of the 
rotor system, once the constraint on each rotor has been imposed 
through $z_i(t)$ and $z_i(t)$ is treated independently, is quadratic.
In more general models the mapping is not exact, see {\it e.g.} the 
discussion in~\cite{Ingold}.

Under the further assumption $z_i(t)=z(t)$, justified in the large $M$ limit, 
the stochastic equations (\ref{eq:avantprojo}) are rendered independent --
apart from a
residual coupling through the Lagrange multiplier -- by a rotation
onto the basis that diagonalizes the interaction matrix $\mathcal{J}$. Indeed,
${\bf \mathcal{J}}$ being real and symmetric, it has $N$ real eigenvalues
$J_\sigma$ with corresponding eigenvectors $\boldsymbol{\sigma}$ that
constitute a complete and orthonormal basis of the space of rotor sites:
$\boldsymbol{\sigma} \bullet \boldsymbol{\sigma}' = \delta_{\sigma \sigma'}$
where $\bullet$ is the usual scalar product in this space. Let us collect all
the rotors  in the vector $\boldsymbol{n} \equiv \{ {\bf n}_i^{(1)} 
\}_{i\in[1,N]}$ and introduce its projections on the eigenvectors:   ${\bf
n}_\sigma \equiv \boldsymbol{n} \bullet \boldsymbol{\sigma}$.
If we project eq.~(\ref{eq:avantprojo}) onto $\boldsymbol{\sigma}$, we are left
with $N$ uncoupled Langevin equations reading
\begin{equation} \label{eq:edpS01}
\left( \frac{1}{\Gamma} \partial_t^2 - 
J_\sigma+z(t) \right) {\bf n}_\sigma(t) - \int \ud{t'}\Sigma_B^R(t-t') {\bf
n}_\sigma(t')=  {\boldsymbol{\xi}}_\sigma(t)   \;,
\end{equation}
with
\begin{eqnarray}
 \langle \boldsymbol{\xi}_\sigma(t) \rangle_{\xi}  =  0 \;, \qquad
 \langle \boldsymbol{\xi}_\sigma(t) \cdot \boldsymbol{\xi}_{\sigma'}(t')
\rangle_{\xi}  =  \delta_{\sigma\sigma'}  \, \Sigma_B^K(t-t')  \;.
\end{eqnarray}
The noise statistics is peculiar because of the quantum origin of the environment: it has memory (colored), and depends on $T, eV, \hbar$.  

\paragraph{Two-time self correlation.}

Within the effective Langevin formalism, the two-time self correlation function defined in eq.~(\ref{eq:def_C}) reads
\begin{equation}
	C(t,t') = \overline{ \langle {\bf n}_\sigma(t) \cdot {\bf n}_\sigma(t') \rangle }^J\;,
\label{eq:corr-n}
\end{equation}
where the average over disorder is realized by
\begin{equation}
 \overline{ \ \cdots \ }^{J} \equiv \int \ud{J_\sigma} \rho_J(J_\sigma) \
\cdots \ \;,
\end{equation}
and $\rho_J(J_\sigma)$ is the probability density of the eigenvalues of the interaction matrix ${\cal J}$.
Following the analysis in \cite{CugliandoDean}, the correlation function (\ref{eq:corr-n})
is expected to show a separation of time scales (at least in some parts of the phase diagram). This 
is usual in coarsening phenomena and corresponds to a stationary regime at short time-difference 
and an aging one at long time-difference with respect to a waiting-time dependent characteristic 
time. The stationary part of the correlation approaches a plateau at the Edwards-Anderson order 
parameter, $q_{\rm EA} \equiv \overline{\langle {\bf n}_\sigma  \rangle_\xi^2 }^J$, that measures 
the fraction of frozen rotor fluctuations on time scales much smaller than this characteristic time. 
The value of $q_{\rm EA}$ depends on all parameters ($T,eV, \Gamma, g$). It is non-vanishing in 
the spontaneously symmetry-broken phase and continuously goes to $0$ on the critical surface. 
In  certain cases it can be computed exactly. 

It is reasonable to expect that the long-time aging dynamics is determined by the low frequency (or long time) 
form of the Langevin equations only. The simplification arising in this asymptotic limit are discussed below.   

\subsubsection{Long-time dynamics}
\label{subsubsec:long-time}

In the low-frequency, long time-difference limit, $\hbar\omega\ll T$, the Keldysh self-energy can be approximated by a constant 
[see, \textit{e.g.}, eq.~(\ref{eq:SklowW}) in Sect.~\ref{sec:FiniteV} for its exact expression in the $\epsilon_F\to\infty$ limit]
\begin{equation}
 \Sigma_B^K(\tau) \simeq \delta(\tau) \Sigma_B^K(\omega=0)\geq 0 \;.
\end{equation}
Similarly, we keep the leading contributions in the derivative expansion of $\Sigma_B^R$:
\begin{equation}
\Sigma_B^R(\tau) \simeq \Sigma_B^R(\omega=0) \delta(\tau)  + \eta \delta(\tau)\partial_\tau\;,
\end{equation}
with $\eta \equiv \partial_\omega \mbox{Im } \Sigma_B^R(\omega=0) > 0$. The Langevin equations read in this limit
\begin{equation} \label{eq:edpS0}
 \frac{1}{\Gamma} \partial_t^2   {\bf n}_\sigma(t)  + \eta \partial_t   {\bf n}_\sigma(t)
 = \left( J_\sigma - z(t) + \Sigma_B^R(\omega=0) \right) {\bf n}_\sigma(t) +  {\boldsymbol{\xi}}_\sigma(t) \;,
\end{equation}
where $\eta$ plays the role of a friction coefficient and ${\boldsymbol{\xi}}_\sigma(t)$
has white noise statistics:
\begin{equation}
 \langle \boldsymbol{\xi}_\sigma(t) \cdot \boldsymbol{\xi}_{\sigma'}(t') \rangle_\xi =  \delta_{\sigma\sigma'} \delta(t-t')\,  \Sigma_B^K(\omega=0) \;.
\end{equation}
In the Langevin formalism, the kernel of an equilibrium white bath is given by the Einstein relation (known as the FDT of the second kind): $\langle \xi(t) \xi(t') \rangle_{\xi} = 2 \eta T \delta(t-t')$. Thus, the temperature $T$ of the bath can be seen as the ratio of the diffusion coefficient of a particle embedded in that
bath with the friction coefficient $\eta$ of the bath on the particle.
For our reservoirs, in the low-frequency long time-difference limit, one can associate this ratio to an equivalent temperature $T^*$
\begin{equation}
 T^* \equiv  \lim\limits_{\omega\to0} \frac{1}{2} \ \frac{\Sigma_B^K
(\omega)}{\partial_\omega \mbox{Im } \Sigma_B^R (\omega) }
 \;,
\end{equation}
the properties of which were discussed in Sect.~\ref{sec:low-freq}. Thus, we confirm here that $T^*$
acts like a temperature in the sense that the effect of the
(out of equilibrium) reservoirs on the long-time dynamics is the one of an
\textit{equilibrium} dissipative (Ohmic) bath at a temperature $T^*$.
This has been reported in different works
and is at the root of the derivation of the stochastic Gilbert
equation for a spin under bias \cite{NunezDuine}.

We expect that as far as the long time dynamical behavior is concerned, the inertial term in eq.  (\ref{eq:edpS0}) 
can also be dropped, thus leading to the equations:
\begin{equation} \label{eq:edpS}
\partial_t {\bf n}_\sigma(t) = \lambda_\sigma(t) \ {\bf n}_\sigma(t) +
\frac{1}{\eta} \ {\boldsymbol{\xi}}_\sigma(t)  \;,
\end{equation}
where we introduced  the shorthand notation $\lambda_\sigma(t) \equiv  \left[ J_\sigma -
\Delta z(t) \right] / \eta$ and $\Delta z(t) \equiv z(t)-\Sigma_B^R(\omega=0)$ and
the spherical constraint is enforced by $z(t)$. 

This particular Langevin equation has been analyzed intensively in the study of
the classical spherical Sherrington-Kirkpatrick model (or spherical $p=2$
spin-glass  model) and the results in \cite{CugliandoDean} apply to our problem with $T \mapsto T^*$. The solution to eq.~(\ref{eq:edpS}) for a given disorder realization and noise history is
\begin{equation}\label{eq:solutionTOfirstOrder}
  {\bf n}_\sigma(t) = {\bf n}_\sigma(0) \exp{\left(  \int_0^t \Ud{\tau}
\lambda_\sigma(\tau)  \right)}
 	+  \frac{1}{\eta}  \int_{0}^t \Ud{\tau}
{\boldsymbol{\xi}}_\sigma(\tau) \exp{\left( \int_{\tau}^{t} \Ud{\tau'}
\lambda_\sigma(\tau')  \right)}  \;.
\end{equation}
Copying results in \cite{CugliandoDean}, the aging part of the correlation (in the limit $t' \gg t \to \infty$)
shows a simple aging scaling behavior
\begin{equation} \label{eq:simpleaging}
 C(t,t')  \simeq 2\sqrt{2} \ q_{\rm EA}  \  \frac{(t/t')^{3/4}}{(1+t/t')^{3/2}}
= C(t/t') 
\;.
\end{equation}
The solution to eqs. (\ref{eq:edpS}) leads to $q_{\rm EA}=1-T^*(eV,T)/J$.  However, this 
result is obtained by taking the limit of relatively close times -- with respect to 
$t'$ --  whereas, as we stressed, eq. (\ref{eq:edpS}) 
is valid for the long time $t'$ and long time-difference $t-t'$ properties only.
 As a consequence, we expect the scaling result, eq. (\ref{eq:simpleaging}), to 
 hold at long times with  
 the value of the Edwards-Anderson parameter not necessarily  given by $1-T^*(eV,T)/J$.
 Its computation requires a full solution of the equations of motion. 

We now focus on the aging dynamics in different parts of the phase diagram and argue 
that the Langevin dynamics of eq.~(\ref{eq:edpS0}) 
indeed provide a correct description of the dynamical evolution. 

\paragraph{Dynamics in the $eV=0$ plane.}
In this case, the Edwards-Anderson order parameter $q_{\rm EA}$ measures the static order parameter.
 The dynamic calculations based on the use of the quantum FDT to relate the correlation to the linear response in the stationary regime detailed in \cite{Rokni}, or the replica equilibrium computation in \cite{effect-bath}, can be easily extended to deal with a generic electronic bath in equilibrium. One confirms that $q_{\rm EA}=1$ at $T=\Gamma=eV=0$ and continuously approaches $0$ on the critical line $\Gamma_c(T)$ for all values of $g$. The precise variation of $q_{\rm EA}$ within the coarsening phase depends on the bath kernels. In the $\epsilon_F\to\infty$ limit, the results in \cite{Rokni} apply also to our problem. The solution of the Schwinger-Dyson equations in the aging regime confirms that the scaling result, eq. 
 (\ref{eq:simpleaging}), holds.

\paragraph{Dynamics in the $\Gamma=0$ plane.} \label{sec:Cugliandolo-Dean}
Another interesting case is the effective overdamped Langevin limit obtained for $\Gamma\to0$ and $(eV, T)$ in the coarsening phase. In this case dropping the inertial term in eq.  (\ref{eq:edpS0}) is exact and not an approximation. 

Here the result $q_{\rm EA}=1-T^*(eV,T)/J$ can be shown to hold. The Edwards Anderson parameter approaches one for $T=V=\Gamma=0$ and goes continuously to zero on the critical line, as in a second order phase transition. Consistently with the analysis of the critical surface derived from the QNESS phase (see Sect.~\ref{sec:CPG0}), one finds $T^*(T_c, eV_c) = J$. Numerical integration of the integro-differential equations of motion  
confirms that the scaling result, eq. 
 (\ref{eq:simpleaging}), holds in the aging regime.
 
Despite the fact that dropping the inertial term is exact, the equations (\ref{eq:edpS}) are still not exact at all times. 
In particular, the initial conditions for this
approximated equation of motion should be given by the state of the system a
short while after the quench when the long-timescale description starts to be valid. 
Apparently, this delay seems to be not sufficient to significantly correlate the rotors with the interaction matrix $\mathcal{J}$ and, to any practical purpose ${\bf n}_\sigma(0)$ can still be considered ``random'', 
at least as far as the Edwards-Anderson parameter is concerned. 

\begin{figure}[t]
\includegraphics[width=5.5cm,angle=-90]{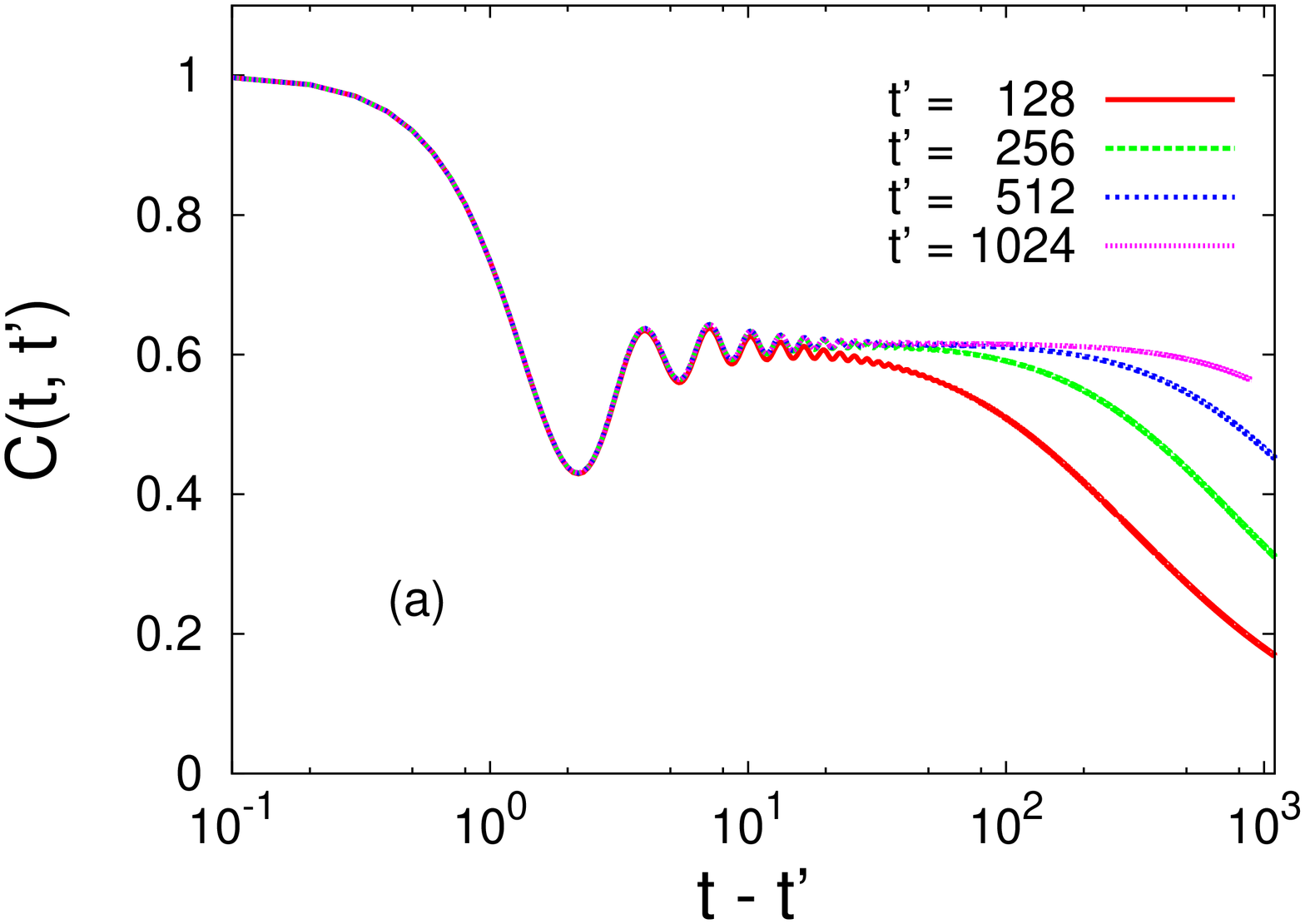}
\includegraphics[width=5.5cm,angle=-90]{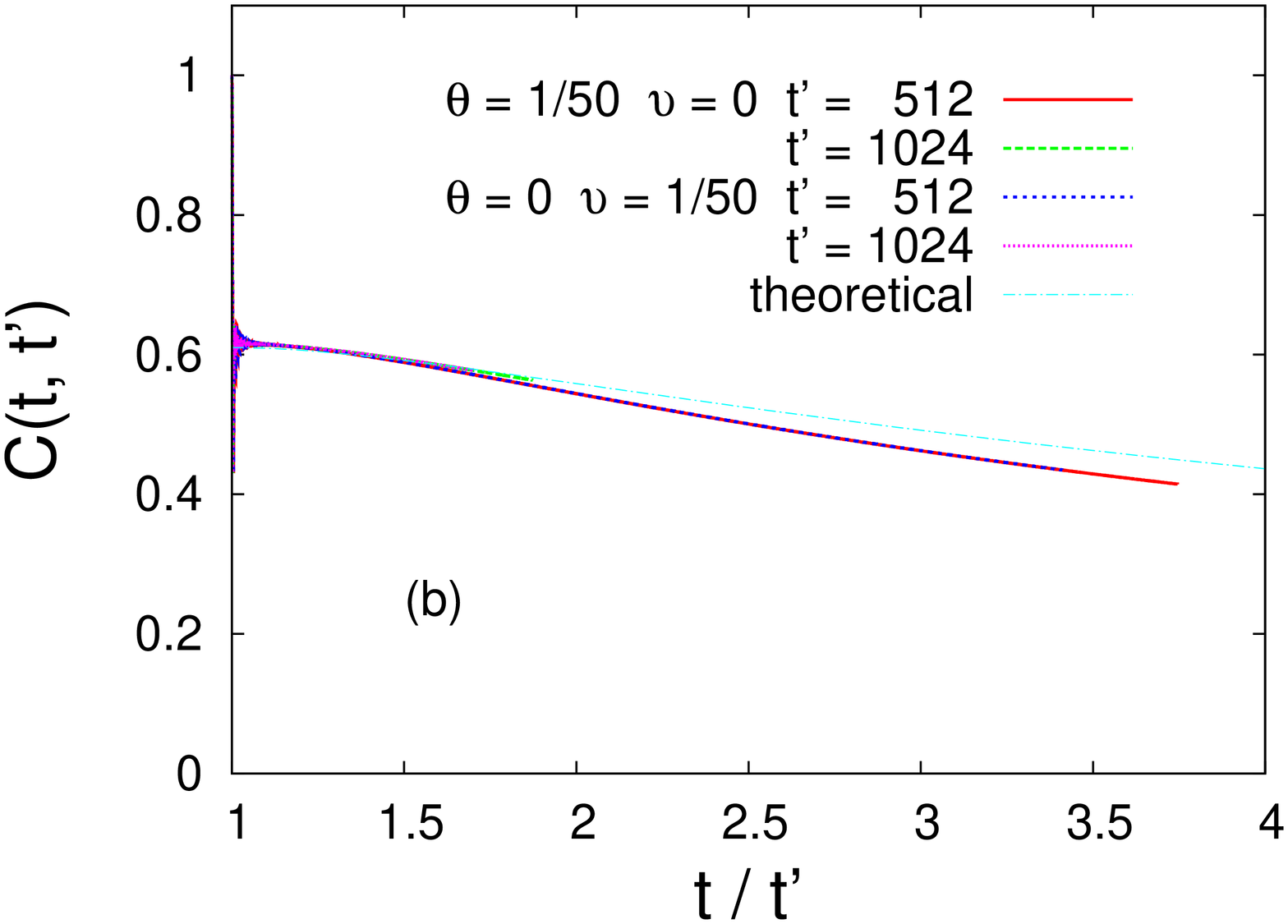}
\caption{\footnotesize \label{fig:C-chi} (Color online.) Dynamics in the driven coarsening
regime:
numerical solution to Schwinger-Dyson eqs.~(\ref{eq:Dyson1}) and
(\ref{eq:Dyson2})  where the reservoirs have a half-filled semi-circle DOS with
$\epsilon_F = 10 J$.  (a) The self
correlation $C(t,t')$ after a quench to $\theta=0.02, v=0.02
$, $\gamma=0.2$, $g=1$ (in reduced quantities) shows first a stationary regime for short $t-t'$, then a
slow aging regime where the time translational invariance is lost. (b)
\label{fig:C-scaled} The self correlation $C$ is plotted versus $t/t'$ for two
waiting times after two quenches into the coarsening region: $\theta=0.02 , v=0, \gamma=0.2$ and
$\theta=0, v=0.02, \gamma=0.2$. There is a double collapse of the curves.
The collapse for the different $t'$ proves the simple aging scaling $C(t'/t)$
and the collapse for the two different quenches shows that $T^* \simeq eV/2$
plays the role of a temperature. The theoretical curve is the solution
eq.~(\ref{eq:simpleaging}) with $q_{\rm EA} \approx 0.6$.}
\end{figure}

\paragraph{Dynamics in the $T=0$ plane.}
The zero-temperature plane is more difficult to deal with analytically. One is not entitled to use FDT since the system is driven by $eV$ nor dropping the second time-derivative is exact. Furthermore, this is the case where the simplification leading 
to eq. (\ref{eq:edpS})  are more dangerous because of the power law tails appearing at $T=0$ in correlation and response
functions. 
 
In order to check that the scaling result, eq. 
 (\ref{eq:simpleaging}), holds we numerically integrate the full set of Schwinger-Dyson equations.
 
 In Fig.~\ref{fig:C-chi}~(a) we show the
decay of the two-time correlation function. For short time differences $t-t'$
with respect to the waiting time $t'$, there is a stationary regime depending on
all control parameters in which the correlation approaches a plateau
asymptotically in the time-difference. The plateau value is $q_{\rm EA}$ and
measures the fraction of frozen rotor fluctuations on
timescales much smaller than $t'$. Afterwards, there is an aging regime in which
$C$ depends on the two times explicitly. In Fig.~\ref{fig:C-scaled}~(b), we
plot $C$ against $t/t'$ to prove that the simple aging scaling predicted
analytically with eq.~(\ref{eq:simpleaging}) holds at these long times. Moreover,
we show that the dynamics after a quench to $\theta=0.2$, $v=0$ are the same that the
ones after a quench to $\theta=0$, $v=0.2$,  illustrating the fact that $T^* \simeq
eV/2$ acts here like a temperature.

\paragraph{Super-universality.} It is remarkable that in the large $M$ limit, the long-time dynamics of our
model are exactly the ones of the classical fully connected $p=2$ spherical spin
glass. The latter being a classical model in contact with an equilibrium bath
($\Gamma=0, eV=0$), the former being its quantum version in contact with a
non-equilibrium bath ($\Gamma\neq0, eV\neq0$). The fact that the scaling
functions are \textit{super-universal}, in the sense that they do not depend on
the external parameters $T, eV, \Gamma$ once $q_{\rm EA}$ is extracted as a factor, can be understood as follows.
First the fact that the non-equilibrium environment of our model gives rise to
the same long-time dynamics than an equilibrium environment can be seen as a
consequence of the Ohmic behavior of the reservoirs self-energy kernels at small
frequencies (see Sect.~\ref{sec:low-freq}). Second, the fact that our quantum
model shows a classical behavior at late times can be understood as a
consequence of decoherence due to the dissipative (and Ohmic) bath. Furthermore,
the effect of the temperature $T$ on the long-time dynamics being irrelevant (in
a RG sense) in the classical limit, one can expect the same to hold in the
quantum case with respect to all parameters.

This result has an interesting consequence. In the case of (large $M$) quantum
$3d$
coarsening the classical-quantum mapping extends to space-time
correlations and proves the existence of a growing
coherence length $R(t_w)\propto t_w^{1/2}$ over which the rotors are
oriented in the same direction. This real-space interpretation of aging unveils
the connection with coarsening that was announced all along this manuscript.

We found quite naturally that
the long-time dynamics correspond to a Bose-Einstein-type condensation
process of the $N$ $M$-dimensional ``vectors'' ${\bf n}_\sigma$ on the
direction of the edge eigenvector. The relaxation is controlled
by the decay of $\rho(J_\sigma)$ close to its edge.  For Gaussian
i.i.d. couplings $\rho(J_\sigma) \propto [(2J)^2-J^2_\sigma]^{1/2}$. This coincides
with the distribution of the modulus of the Laplacian eigenvalues,
$\rho(k^2) \propto \left(k^2\right)^{d/2-1}$ in $d=3$.  For this reason
all models with a square root singularity of the distribution of
``masses'' $J_\sigma$, as the ferromagnetic rotor model in $d=3$ and the
completely connected spin glass rotor model, are characterized by the
same long-time dynamics.

\subsubsection{Linear response}
\begin{figure}[t]
\centering
\includegraphics[width=6cm,angle=-90]{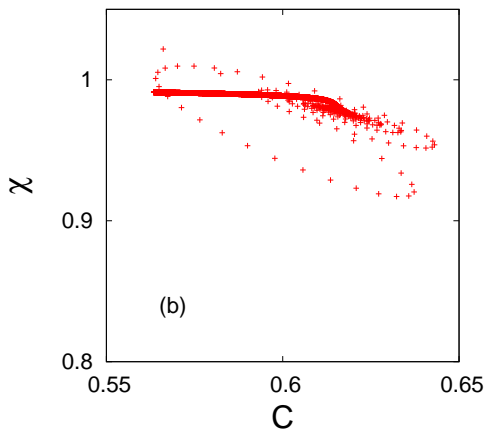}
\caption{\footnotesize \label{fig:OFDR} The integrated linear response,
$\chi(t,t') =\int_{t'}^t {\rm d}\tau R(t,\tau)$ against $C(t,t')$, for
$t'=1024$ and using $t$ as a parameter. The curved part corresponds
to the stationary and oscillatory regime with $(t-t')/t'\to 0$ while
the straight line is for times in the monotonic aging decay of $C$.
}
\end{figure}
It has already been noticed in Sect.~\ref{sec:Schwinger-Dyson} that the response
function was somehow peculiar since its equation of motion is decoupled from the
one of the self correlation. Having argued that the long-time dynamics are
governed by their classical counterparts, the linear response should also scale
as in the classical limit. Therefore, the quantum fluctuation-dissipation
relation between integrated linear response, $\chi(t,t') \equiv \int_{t'}^{t}
\ud{t''} R(t,t'')$ and self correlation $C(t,t')$ approaches the
classical one, $\chi \sim \mbox{ct} + (q_{\rm EA}-C)/T_{\rm eff}$,
with an {\it infinite} effective temperature~\cite{Cukupe}, $T_{\rm
  eff}\to\infty$, as shown in Fig.~\ref{fig:OFDR}. The relations between
integrated responses and correlation functions in other quantum problems that
also approach classical-like form in the aging regime were shown
in~\cite{CugliandoloLozano,other-quantum}.

\subsubsection{The Lagrange multiplier} \label{sec:LagrangeMcheck}

One should check the validity of a key assumption that was used to derive the
phase diagram: the convergence of $z(t)$ to an asymptotic value on the critical
manifold. We first derive analytically the asymptotic behavior (within our
long-time approximation) of $z(t)$ in the $\Gamma=0$ coarsening phase showing that this is
indeed the case. Then we give numerical evidence that $z(t)$ converges in the
whole phase space.

The condition $C(t,t) = \int \ud{J_\sigma} \rho_J(J_\sigma) \ \langle {\bf
n}_\sigma(t) \cdot {\bf n}_\sigma(t) \rangle_{\xi} = 1$ reads after taking its
time derivative and assuming furthermore that ${\bf n}_\sigma(0)$ is
uncorrelated with $\sigma$ (${\bf n}_\sigma(0) =   {\bf n}_0, \forall \ \sigma$),
that is valid for random initial conditions (coming from infinite temperature
for instance)
\begin{eqnarray}
 0 &=&  \int \ud{J_\sigma} \rho_J(J_\sigma) \  \langle  \partial_t {\bf
n}_\sigma(t) \cdot {\bf n}_\sigma(t) \rangle_{\xi}  \\
  &=& \int \ud{J_\sigma} \rho_J(J_\sigma)  \left\{
	 {\bf n}_0^2 \lambda_\sigma(t) \rme^{2\int_0^t \ud{\tau}
\lambda_\sigma(\tau)}
	+ \frac{T^*}{\eta} \left[
1+2 \lambda_\sigma(t) \int_0^t \ud{\tau'} \rme^{2
\int_{\tau'}^t \ud{\tau''} \lambda_\sigma(\tau'')}  \right]
 \right\} \;.
\end{eqnarray}
Taking the derivative with respect to ${\bf n}_0^2$  yields
\begin{eqnarray}\label{eq:condunity}
 0 =  \int \ud{J_\sigma} \rho_J(J_\sigma)  \lambda_\sigma(t)  
 \ 
\rme^{2\int_0^t \ud{\tau} \lambda_\sigma(\tau) } \;,
\end{eqnarray}
that can be recast into
\begin{equation}\label{eq:zlimitgeneral}
 \Delta z(t) =  \frac{\eta}{2} \
 \partial_t \ln \int \ud{J_\sigma}
\rho_J(J_\sigma) \ \rme^{2 J_\sigma t / \eta} \,.
\end{equation}

\begin{figure}[t]
\includegraphics[width=6cm,angle=-90]{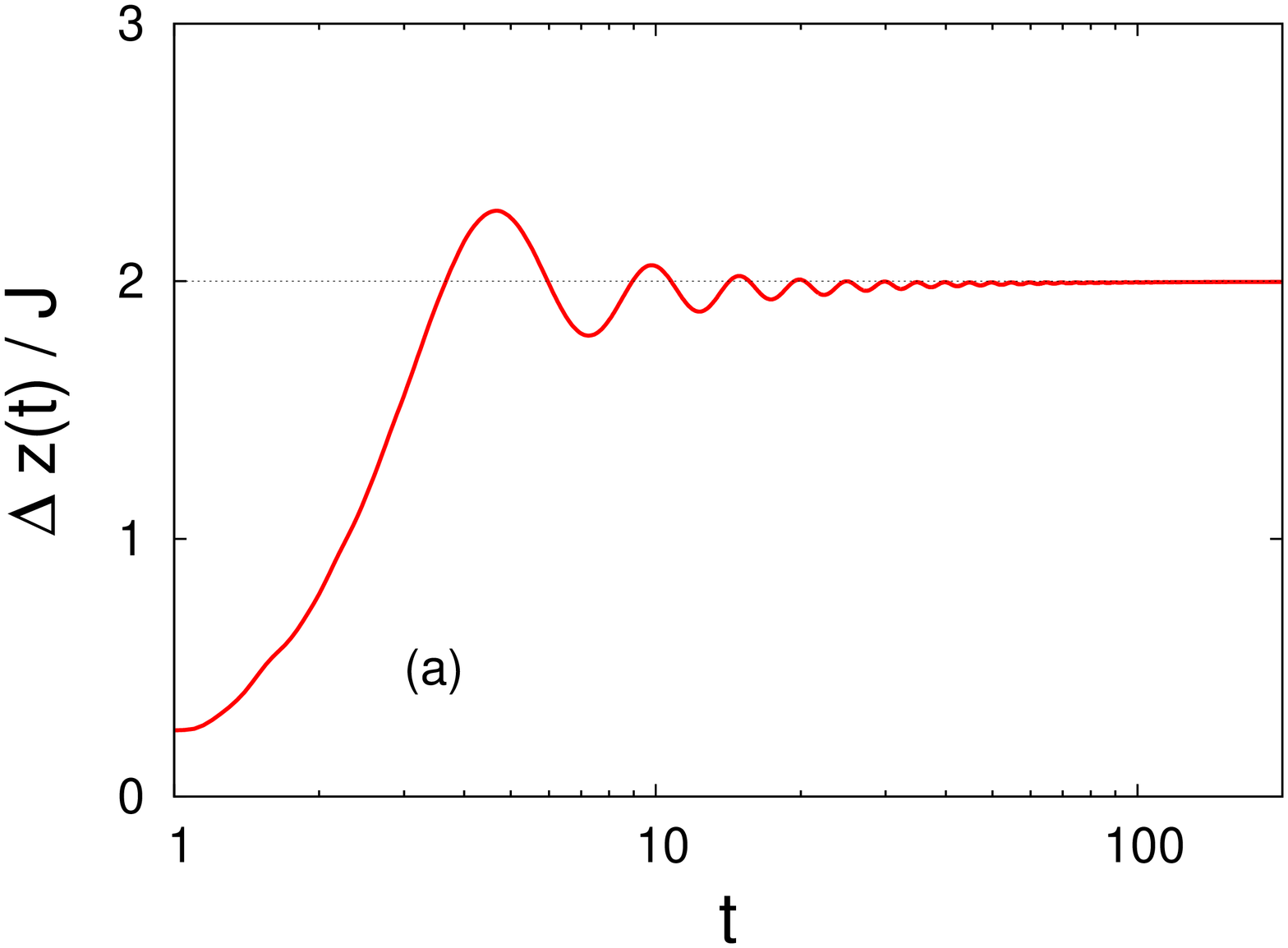}
\includegraphics[width=6cm,angle=-90]{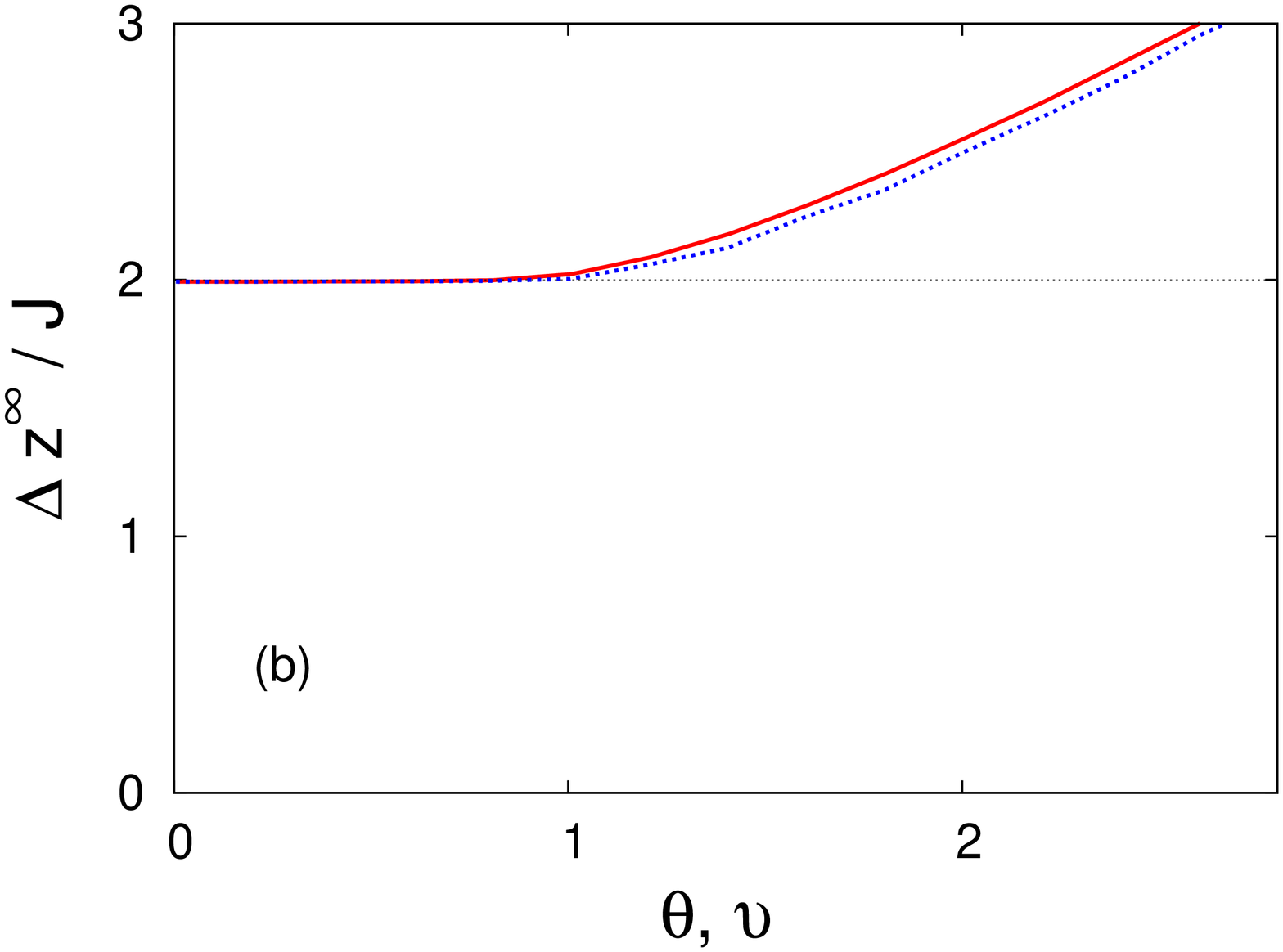}
\caption{\footnotesize \label{fig:zeta} (Color online.) (a)  $\Delta z(t) \equiv z(t) - \Sigma_B^R(\omega=0)$ quickly converges
toward $2J$, the largest eigenvalue of the $\mathcal{J}_{ij}$ matrix (here
$\Gamma = eV = T =0.1 J$, $g=1$, and $\epsilon_F = 10 J$).
\label{fig:zdep} (b) Dependence of $z^\infty$ with $T$ (plain curve) and $eV$
(dashed curve).
}
\end{figure}

\paragraph{Asymptotic behavior of $z(t)$.} \label{sec:asy_z}
By plugging the density of eigenvalues of an infinite ($N\rightarrow\infty$) and
symmetric random matrix with Gaussian elements of variance $J^2 / N$
\begin{equation}
\rho_J(J_\sigma) \equiv \frac{1}{\pi J}
\sqrt{1-\left(\frac{J_\sigma}{2J}\right)^2} \qquad \mbox{ for } \qquad J_\sigma \in
[-2J;+2J] \; , 
\end{equation}
and zero elsewhere, we obtain
\begin{equation}\label{eq:zlimitgeneral2}
 \Delta z(t) =  \frac{\eta}{2} \partial_t \ln \frac{\eta}{2J}
\frac{1}{t} I_1\left(\frac{4J}{\eta} t \right) \,,
\end{equation}
where $I_1$ is the modified Bessel function of the first kind and first order.
We obtain, the pre-asymptotic behavior for $t \gg \eta/J$
\begin{equation}
z(t) \simeq  2J + \Sigma_B^R(\omega=0) - \eta \ \frac{3}{4t} \;.
\end{equation}
We just showed that inside the coarsening phase, the Lagrange multiplier $z(t)$
reaches an asymptotic value which is actually the critical value, $z_c^\infty = 
 2J + \Sigma_B^R(\omega=0)$, calculated in Sect.~\ref{sec:para} from the QNESS
phase TTI equations without neglecting any term. The coherence between those two
results somehow justifies the approximations made previously. In the $\epsilon_F\to\infty$ limit (reservoirs acting like an Ohmic bath) $\Sigma_R(\omega=0)$ vanishes and we recover the same mechanism as in the classical case \cite{CugliandoDean}.

These analytical results are supported by the numerical analysis. Computed after
the quench, the Lagrange multiplier $z(t)$ quickly converges to an asymptotic
value $z^\infty$. As an example, we plot in Fig.~\ref{fig:zeta}~(a)  the behavior of
$z(t)$ after a quench into the QNESS phase. 
The oscillations and the zero initial slope are signatures of the second and higher order derivatives in eq.~(\ref{eq:edpS01}). These terms were dropped in the analytical study of the long-time limit, see eq.~(\ref{eq:edpS}), but the numerical integration does not neglect them.
We give in Fig.~\ref{fig:zdep}~(b) the
dependence of $z^\infty$ with $T$ and $eV$. It is quite clear that $z^\infty$ is
constant (and equal to $z_c^\infty$) inside the critical surface and increases
with $T$, $\Gamma$, and $eV$ as soon as entering the QNESS phase. This justifies
the assumptions made in Sect.~\ref{sec:para}.

To summarize the results, in the whole phase diagram $z(t)$ always rapidly
reaches an asymptotic value $z^\infty$. Inside the QNESS phase, $z^\infty$ is a
growing function of the parameters $T, \Gamma, V$ whereas on the critical
surface and inside the coarsening region, it is fixed to $z_c^\infty$.

\paragraph{Link between $z(t)$ and the potential energy density}\label{sec:link_z}
One is interested in computing the energy density $\epsilon(t)$ of the effective
Brownian particle. It is given by
\begin{eqnarray}
\epsilon(t) = - \frac{1}{2}\sum_{i,j=1}^N  \overline{ J_{ij} {\bf n}_i(t) {\bf
n}_j(t) }^J
= -\frac{1}{2} \int \ud{J_\sigma} \rho_J(J_\sigma) J_\sigma {\bf n}_\sigma^2(t)
\;.
\end{eqnarray}
Using the solution (\ref{eq:solutionTOfirstOrder}) for ${\bf n}_\sigma(t)$ at
$T^*=0$, one has
\begin{eqnarray}
2 \epsilon(t) = - {\bf n}^2_0 \
e^{- \frac{2}{\eta}
\
 \int_0^t \ud{\tau} \Delta z (\tau)}
 \
\int \ud{J_\sigma} J_\sigma \rho(J_\sigma)
\ \rme^{{2 J_\sigma t}/{\eta}} \;.
\end{eqnarray}
By use of eq.~(\ref{eq:zlimitgeneral}), we obtain
\begin{eqnarray}
2 \epsilon(t) = - \frac{\eta}{2} \partial_t  \ln \int \ud{J_\sigma}
\rho(J_\sigma) \
\rme^{{2 J_\sigma t}/{\eta}}  \;.
\end{eqnarray}
We recognize eq.~(\ref{eq:zlimitgeneral}) in the right-hand-side ({\sc rhs}) of this last expression,
giving finally
\begin{eqnarray}
 \epsilon(t) = -\frac{1}{2} \Delta z(t) \;.
\end{eqnarray}
This result is valid for any disorder density $\rho(J_\sigma)$. For a non-zero
$T^*$, similar calculations give, see \cite{CugliandoDean}, 
\begin{eqnarray}
 \epsilon(t) = \frac{1}{2} \left[T^* - \Delta z(t) \right] \;.
\end{eqnarray}

\section{The current}
The physics of electric currents through mesoscopic quantum impurities in out-of-equilibrium settings has attracted a lot of attention in the recent years. The Kondo impurity is the canonical example of a strongly correlated system that has both been tackled experimentally~\cite{KondoExp} and theoretically by non-perturbative methods~\cite{Kondo}. It is, to our knowledge, the first time that some fermionic reservoirs are coupled to a macroscopic disordered quantum system.
In the previous Sections we analyzed the effects of the voltage drop on the system dynamics. In this Section we study the properties of the current that establishes between the two reservoirs. In particular we are interested in the possible influence of the rotors on the current. Is the current, that is rather easy to measure experimentally, able to give information about the dynamics of the rotors ?

We recall the expression of the interaction Hamiltonian given in eq.~(\ref{eq:Hsb}):
\begin{equation}
H_{SB}=  -\sqrt{M}\,  \frac{\hbar\omega_c}{N_s}  \sum_{i=1}^N \sum_{\mu=1}^M \sum_{k,k'=1}^{N_s} \sum_{l,l'=1}^{\cal M} 
 \;
n_i^\mu \;
[\psi^{\dagger}_{Likl} \ \sigma_{ll'}^\mu \ \psi_{Rik'l'} + L
\leftrightarrow R] \;.
\end{equation}
From the point of view of the electric current, our model consists in two reservoirs coupled through time-dependent tunneling constants $n_i^\mu(t)$. It is different from the usual quantum impurity problems in the fact that the electrons cannot stay on the rotor system but only hop directly from one reservoir to the other. Furthermore, the quantum character of the system is not expected to play any significant role since its level spacings are smaller than any other energy scale in the large $M N$ limit. The computation of the current will therefore lead to Landauer formula~\cite{Landauer} \textit{a priori} dependent on the rotors states.

The electric current carried by the fermions flowing from the right to the left
reservoirs is
\begin{equation}
 I_{R\to L}(t) = -e \ \langle \frac{{\rm d} N_L}{{\rm d t}}\rangle 
	=  -\frac{\rmi e}{\hbar} \langle \left[H, N_L \right] \rangle
	=  -\frac{\rmi e}{\hbar} \langle \left[H_{SB}, N_L \right] \rangle \;,
\end{equation}
where $-e$ is the electric charge of a fermion and $N_L \equiv \sum_{ikl}
\psi_{Likl}^\dagger \psi_{Likl}$  is the number operator of the left reservoirs. $H_{SB}$ is the part of the total Hamiltonian $H$ that couples the system and the reservoirs,  see eq.~(\ref{eq:Hsb}).
After straightforward algebra, we obtain
\begin{equation}
 I_{R\to L}(t) =   -\frac{\rmi e}{\hbar} \langle 
\sqrt{M} \, \frac{\hbar\omega_c}{N_s}  \sum_{i\mu kk' ll}   \sigma^\mu_{ll'} n_i^\mu  \left[ \psi_{Likl}^\dagger
\psi_{Rjk'l'} - L \leftrightarrow R \right]\rangle \;.
\end{equation}
In the Keldysh field theory formalism, this corresponds to the quantity
\begin{eqnarray}
 I_{R\to L}(t) = \frac{1}{2} \left( I_{R\to L}^+(t) + I_{R\to L}^-(t) \right)
\;,
\end{eqnarray}
with
\begin{equation}
 I_{R\to L}^a(t) \equiv -\frac{\rmi e}{\hbar}   
\langle  \sqrt{M} \, \frac{\hbar\omega_c}{N_s} \sum_{i\mu kk' ll}  \sigma^\mu_{ll'} n_i^{\mu a}(t)  \left[
\bar\psi_{Likl}^{a}(t) \psi_{Rjk'l'}^{a}(t) - L \leftrightarrow R \right]
\rangle \;.
\end{equation}
Expanding the action up to first order in the coupling constant $g$, we obtain an average
over the rotors and the free fermions that are now uncoupled, that we note
$\langle \ \cdots \
\rangle_{SB}$
\begin{eqnarray}
I_{R\to L}(t)  &=& \frac{1}{2} \langle \left( I_{L\to R}^+(t) + I_{L\to R}^-(t)
\right) \frac{\rmi}{\hbar} S_{SB} \rangle_{SB}
\nonumber\\
 &=& \frac{e}{2\hbar^2} M \left(\frac{\hbar\omega_c}{N_s}\right)^2   \sum_{ab}  \sum_{i \mu kk'
ll'} \sum_{j \nu qq' mm'} b \int \ud{t'}  \sigma^\mu_{ll'} \sigma^\nu_{mm'} 
\langle n_i^{\mu a}(t) n_j^{\nu b}(t')  \\
 & & \qquad \times  \left[ \bar\psi_{Likl}^{a}(t) \psi_{Rjk'l'}^a(t) - L
\leftrightarrow R \right]\left[ \bar\psi_{Ljqm}^{b}(t') \psi_{Rjq'm'}^{b}(t') +
L \leftrightarrow R \right] \rangle_{SB} \;. \nonumber
\end{eqnarray}
Averaging over the free fermions, we obtain
\begin{eqnarray}
 I_{R\to L}(t)  = \frac{e}{2\hbar^2} M N  (\hbar\omega_c)^2 \sum_{ab=\pm}
 b \int \ud{t'} \rmi\hbar G^{ab}(t,t') \left[ \rmi\hbar G_L^{ab}(t,t') \rmi\hbar G_R^{ba}(t',t) - L
\leftrightarrow R \right] \,.
\end{eqnarray}
$G^{ab}$ are the macroscopic Keldysh Green's functions for the rotors and
$G^{ab}_{L/R}$ are the Green's functions of the free fermions in the
$L/R$-reservoirs.
This reads, after Keldysh rotations, 
\begin{eqnarray}
 I_{R\to L}(t)  = -\frac{e}{\hbar} M N  \int_{0}^{t}
\ud{\tau}  
C(t,t-\tau) \ \Pi_B^R(\tau)  + R(t,t-\tau) \ \Pi_B^K(\tau)
\;,
\label{eq:Current}
\end{eqnarray}
with
\begin{eqnarray}
\begin{array}{rcl}
  \Pi_B^K &\equiv& \displaystyle  -2 (\hbar\omega_c)^2 \ \mbox{Im}   \left[ G^K_L {G^K_R}^* -
\frac{\hbar^2}{4} \left( G^R_L {G^R_R}^* + G^A_L
{G^A_R}^* \right) \right] \;, \\
\Pi_B^R &\equiv& \displaystyle -2 (\hbar\omega_c)^2  \ \mbox{Im}  \left[ G^R_L {G^K_R}^* +
G^K_L
{G^R_R}^*  \right] \;.
\end{array}
\end{eqnarray}
The expression for the current given in eq.~(\ref{eq:Current}) is quite generic. It is valid as soon as the system and the fermionic leads are coupled with an interaction $H_{SB}$. The details of the system and the leads enter in the formula through their respective Green's functions. The formula was obtain after a first order expansion in the coupling constant $g$. The second order term like all the even order terms are zero by use of Wick's theorem. The third and higher odd order terms would have involved higher order correlation functions of the system.
Plugging the expressions of the fermionic Green's functions $G_\alpha^K$,
$G_\alpha^R$, and $G_\alpha^A$ ($\alpha = L, R$) that are given in
Appendix~\ref{app:KeldyRotFermions}, we get
\begin{eqnarray}
 \Pi_B^K(\tau) =  \frac{1}{2} (\hbar\omega_c)^2  \langle\langle \left[
\tanh{(\beta \frac{\epsilon_L-\mu_L}{2} )} \tanh{(\beta
\frac{\epsilon_R-\mu_R}{2})} -1 \right] \sin \left(\frac{\epsilon_L -
\epsilon_R}{\hbar} \tau \right) \rangle_L\rangle_R \;, \\
\Pi_B^R(\tau) =  \frac{1}{\hbar} (\hbar\omega_c)^2  \langle\langle \left[
\tanh{(\beta \frac{\epsilon_L-\mu_L}{2} )} - \tanh{(\beta
\frac{\epsilon_R-\mu_R}{2})} \right] \cos \left(\frac{\epsilon_L -
\epsilon_R}{\hbar} \tau \right) \rangle_L\rangle_R  \Theta(\tau) \;,
\end{eqnarray}
where the notation $\langle\langle \ \cdots \ \rangle_{L}\rangle_{R}$ stands for $\iint
\Udd{\epsilon}{\epsilon'} \rho_{L}(\epsilon) \rho_{R}(\epsilon') \ \cdots \ $.
One can check that the current vanishes when the bias voltage ($eV \equiv \mu_R
- \mu_L$) is set to zero.

\paragraph{Linear conductance.}
We develop the current formula (\ref{eq:Current}) to the first order in $eV$ and
compute the linear conductance
\begin{eqnarray}
 I_{R\to L}(t)  \hspace{-0.2em}= \hspace{-0.2em} -\frac{e}{\hbar} M N \,eV\,
\int_{0}^{t}
\Ud{\tau} \left. C(t,t-\tau)\right\rvert_{eV=0} 
\left.\frac{\Ud{\Pi_B^R}(\tau)}{\ud{eV}}\right\rvert_{eV=0}  
\hspace{-1em} + \left. R(t,t-\tau)\right\rvert_{eV=0}
\left.\frac{\ud{\Pi_B^K}(\tau) }{\ud{eV}}\right\rvert_{eV=0}\hspace{-2em},
\end{eqnarray}
One can derive for a flat half-filled DOS, $\rho(\epsilon) \propto \Theta(\epsilon_F - |\epsilon-\epsilon_F|)$, in the limit $\epsilon_F \to \infty$ (in that limit we expect the results to depend very little on the precise shape of the DOS)
\begin{eqnarray}
\left.\frac{\ud{\Pi_B^R}(\tau)}{\ud{eV}}\right\rvert_{eV=0} &=&   -\pi
g^2 \delta(\tau) \;, \\
\left.\frac{\ud{\Pi_B^K}(\tau)}{\ud{eV}}\right\rvert_{eV=0} &=& -\hbar
g^2 \frac{1}{2\tau}    
\;.
\end{eqnarray}
Therefore the linear current very quickly goes from zero to
\begin{eqnarray}
 I_{R\to L}(t)  = \frac{e}{2\hbar} M N g^2 \, eV
\left(
{\pi} + \hbar \int_{0}^{t} \ud{\tau} \frac{R(t,t-\tau)}{\tau}
\right)\;.
\label{eq:conductance}
\end{eqnarray}
The dependence on the history of the two-time correlation function has disappeared and the second term in eq.~(\ref{eq:conductance}) goes to zero
due to the rapid decay of the response function. Finally the current quickly takes an asymptotic value 
\begin{eqnarray}
 I_{R\to L}^\infty = \frac{e}{2\hbar} \pi M N g^2 \,  eV\;.
\end{eqnarray}
From this computation, it appears that the current only probes the very fast dynamics of the system it passes through and does not give information on the long-time dynamics. Since the short-time dynamics of the system are equilibrium ones even in the coarsening regime, the current cannot be used to tell in which regime the system is. An exact numerical integration of eq.~(\ref{eq:Current}) supports these findings for other types of DOS, for finite values of $\epsilon_F$ and far from the linear regime.

\section{Conclusions and discussion}

In this paper we presented a detailed study of the quantum fully-connected 
rotor model driven out of equilibrium by a fermionic drive.
We determined analytically the phase diagram of the model and we showed that a
critical manifold, controlled by the value of the disorder strength, separates a
QNESS with zero order parameter from an ordering phase with
non-zero order parameter.
We solved the equations that describe the dynamics in the different phases 
with a  numerical integration and analytically by using
various approximation schemes that give valuable physical insights.
In particular, we showed that this (quasi) quadratic model maps to a set of Langevin 
equations with additive colored noise that describes the dynamics of the rotors.
The nature of the noise is determined by the type of electron baths used 
and, in the driven case, the friction kernel and noise-noise correlation 
are not linked by any fluctuation-dissipation relation. By using this 
effective Langevin description we established the connection with the 3$d$ coarsening 
dynamics of the ${\cal O}(M)$ model and we showed that the long-time ordering dynamics are in the class of
the classical limit of our model without a drive, \textit{i.e.}, with the typical
length growing as $t^{1/2}$.

Finally, we derived a generic expression for the current flowing through the system
that involves a time-convolution between the characteristics of the system (through
its correlation and linear response) and the ones the leads (through their 
retarded and Keldysh kernels). 
Interestingly enough, for the type of density of states used 
in the large $\epsilon_F$ limit the current depends only on the short-time 
difference (stationary) regime in which coarsening is not relevant. 

Future studies along these lines include the analysis of the fate of first order phase transitions, common in disordered quantum spin systems with multi-spin interactions when driven out of equilibrium.

\section{Acknowledgments}

We thank C. Chamon, L. Chaput, A. Millis and A. Mitra for useful 
discussions. This work was financially supported by ANR-BLAN-0346 (FAMOUS).

\appendix
\appendixpage
\section{Conventions}\label{app:Conventions}
$\Theta$ is the Heaviside step function. We choose $\Theta(0)= 1/2$, so that
$\Theta(x)+\Theta(-x) = 1 \ \forall \ x \in \mathbb{R}$.  We recall the
identities
\begin{equation}
 \int_{-\infty}^{\infty} \frac{\ud{x}}{2\pi} \rme^{ixy}  = \delta(y)
\qquad \qquad 
\mbox{and}
\qquad \qquad 
 \int_{-\infty}^y \Ud{x} \delta(x)  = \Theta(y)
\;,
\end{equation}
where $\delta$ is the Dirac delta function. In particular $\int_{-\infty}^0 \ud{x} \delta(x)  = 1/2$.

\subsection{Fourier transform}
The convention for the Fourier transform ${\cal F}$ that we use is
\begin{eqnarray}
\begin{array}{rcl}
  {\cal F}[f(\tau)](\omega) \equiv f(\omega) &\equiv&\displaystyle    \int_{-\infty}^{\infty}
\ud{\tau} \; \rme^{+\rmi\omega \tau} \, f(\tau) 
\;, 
\\
{\cal F}^{-1}[f(\omega)](\tau) \equiv f(\tau) &=& \displaystyle  \int_{-\infty}^{\infty}
\frac{{\rm d}{\omega}}{2\pi} \; \rme^{-\rmi\omega \tau} \, f(\omega)
\;, 
\end{array}
\end{eqnarray}
The Fourier transform of the step function is
\begin{eqnarray}
{\cal F}[\Theta(\tau)](\omega) = \rmi \ \mbox{pv} \frac{1}{\omega} + \pi \delta(\omega) \;,
\end{eqnarray}
where `$\mbox{pv}$' denotes the principal value.
Convolutions in real and Fourier spaces are defined by
\begin{eqnarray}
\begin{array}{rcl}
(f \circ  g) (\tau)  &\equiv& \displaystyle \int \ud{\tau'}   f(\tau')   g(\tau-\tau')   =
{\cal F}^{-1}[(f \, g)(\omega)](\tau)  \;, \\
(f \circ g) (\omega) &\equiv& \displaystyle \int \frac{{\rm d}{\omega'}}{2\pi} \; f(\omega')
g(\omega-\omega') =  {\cal F}[(f \,  g)(\tau)](\omega)\;.
\end{array}
\end{eqnarray}

\subsection{Heisenberg representation}
In the Heisenberg representation the operators evolve as 
\begin{equation}
A_{\rm H}(t) = U^\dagger(t)  A(t) U(t) \;.
\end{equation}
with the unitary operator 
\begin{equation}
U(t) \equiv \mathsf{T} \rme^{-\frac{\rmi}{\hbar}  \int_{0}^{t} \ud{t'} H(t')}\;,
\end{equation}
and thus $U^\dagger(t) = \mathsf{\tilde T} \rme^{-\frac{\rmi}{\hbar} \int_{t}^{0}
\ud{t'} H(t')}$.  $\mathsf{T}$ and $\mathsf{\tilde T}$ are respectively the time and
anti-time-ordering operators (see Appendix~\ref{app:TimeOrder}). For
Hamiltonians $H$ that do not explicitly depend on time we get
\begin{equation}
A_{\rm H}(t) = \rme^{\rmi Ht/\hbar} A(t) \rme^{-\rmi Ht/\hbar}
\; . 
\end{equation}

\subsection{Time-ordering operator}\label{app:TimeOrder}
On the real time axis, the time-ordering operator $\mathsf{T}$ rearranges operators with
ascending times to the left:
\begin{equation}
 \mathsf{T} \ A_{\rm H}(t) B_{\rm H}(t') = A_{\rm H}(t) B_{\rm H}(t') \Theta(t-t') +
\zeta\ B_{\rm H}(t') A_{\rm H}(t) \Theta(t'-t) \;,
\end{equation}
with $\zeta=-1$ if both $A$ and $B$ are fermionic operators, $\zeta=1$
otherwise.
The anti-time-ordering operator $ \mathsf{\tilde T}$ rearranges operators the other way
round:
\begin{equation}
 \mathsf{\tilde T} \ A_{\rm H}(t) B_{\rm H}(t') = A_{\rm H}(t) B_{\rm H}(t') \Theta(t'-t)
+ \zeta\ B_{\rm H}(t') A_{\rm H}(t) \Theta(t-t') \;,
\end{equation}
On the Keldysh contour $\cal C$, the position of an operator is specified by
both the time and the branch index. By the notation $A_{\rm H}(t,a)$, we denote
the operator $A$ in the Heisenberg representation at time $t$
($t\in[0,+\infty[$) on the branch $a$ ($a=\pm$).
One can similarly define a time-ordering operator $\mathsf{T}_{\cal C}$ that rearranges
operators along the contour $\cal C$ represented in Fig.~\ref{fig:KeldyshContour}. The rules are
\begin{eqnarray}
\begin{array}{rcl}
  \mathsf{T}_{\cal C} \ A_{\rm H}(t,-) B_{\rm H}(t',+) &=& A_{\rm H}(t) B_{\rm H}(t') 
\;, \\
  \mathsf{T}_{\cal C} \ A_{\rm H}(t,+) B_{\rm H}(t',-) &=& \zeta\ B_{\rm H}(t')  A_{\rm
H}(t) \;, \\
  \mathsf{T}_{\cal C} \ A_{\rm H}(t,+) B_{\rm H}(t',+) &=& A_{\rm H}(t) B_{\rm H}(t')
\Theta(t-t') + \zeta\ B_{\rm H}(t')  A_{\rm H}(t) \Theta(t'-t) \;, \\
  \mathsf{T}_{\cal C} \ A_{\rm H}(t,-) B_{\rm H}(t',-) &=& A_{\rm H}(t) B_{\rm H}(t')
\Theta(t'-t) + \zeta\ B_{\rm H}(t')  A_{\rm H}(t) \Theta(t-t') \;.
\end{array}
\end{eqnarray}

\subsection{Green's functions}\label{app:GreenFonctions}
Let $\phi$ and $\phi^\dagger$ be respectively 
annihilation and creation operators (bosonic or fermionic). 
In the field theory formalism of the Keldysh approach, we define the Green's
functions as
\begin{eqnarray}
 \rmi\hbar G^{ab}(t,t') \equiv \langle  \phi^a(t)  \bar\phi^b(t')  \rangle\;.
\end{eqnarray}
$a,b=\pm$, $\bar\phi$ is either the complex conjugate (for bosons) or the
Grassmannian conjugate (for fermions) of $\phi$ and the average is understood as
\begin{eqnarray}
 \langle \ \cdots \ \rangle \equiv \int \uD{[\phi^\pm,\bar\phi^\pm]} \ \cdots \
\exp \left( \frac{\rmi}{\hbar} S[\phi^\pm,\bar\phi^\pm] \right)\;.
\end{eqnarray}
In the operator formalism the Green's function read
\begin{eqnarray}
 \rmi\hbar G^{ab}(t,t') \equiv \mbox{Tr} \left[ \mathsf{T}_{\cal C} \ \phi_{\rm H}(t,a) \
\phi_{\rm H}^{\dagger}(t',b) \ \varrho_{\rm H}(0,\pm) \right]\;,
\end{eqnarray}
where $\phi_{\rm H}(t,a)$ denotes the Heisenberg representation of the operator
$\phi$ at time $t$ on the $a$-branch of the Keldysh contour. $\varrho_{\rm H}(0,\pm) = \varrho(0)$ is the
initial density matrix (normalized to be of unit trace) and its location on the $+$ or $-$-branch does not matter thanks to the cyclicity of the trace. $\mathsf{T}_{\cal C}$ is the time-ordering operator acting with
respect to the relative position of $(t,a)$ and $(t',b)$ on the Keldysh contour
(see Appendix~\ref{app:TimeOrder}).

One has, independently of the bosonicity or fermonicity of the field
\begin{eqnarray} \label{eq:G_properties}
 G^{ab}(t',t) = -G^{\bar b \bar a}(t,t')^*\;,
\end{eqnarray}
where the star indicates complex conjugate and $\bar a \equiv -a$.

\section{Fermionic bath}
We define the fermionic Keldysh Green's functions
\begin{equation}
\rmi\hbar G^{ab}(t,t') \equiv  \langle \psi^{a}(t) \bar\psi^{b}(t') \rangle\;,
\end{equation}
where $a,b = \pm$. Like for bosons [see eqs.~(\ref{eq:relat_Green1}) , one has
\begin{eqnarray}
\begin{array}{rcl}
 G^{++}(t,t') &=& G^{-+}(t,t')\Theta(t-t') + G^{+-}(t,t')\Theta(t'-t)\;, \\
 G^{--}(t,t') &=& G^{+-}(t,t')\Theta(t-t') + G^{-+}(t,t')\Theta(t'-t)\;,
\end{array}
\end{eqnarray}
leading to the relation between Keldysh Green's functions
\begin{equation} \label{eq:FQ=0}
 G^{++} + G^{--} = G^{+-} + G^{-+} \;.
\end{equation}

\subsection{Keldysh rotation}\label{app:KeldyRotFermions}
We introduce the new fermionic fields
\begin{eqnarray}
\left\{
\begin{array}{rclrcl}
2 \; \psi^{(1)} &\equiv& \psi^+ + \psi^- \;,
&
2 \; \bar\psi^{(1)} &\equiv& \bar\psi^{+} + \bar\psi^{-} \;,
\\
\hbar \; \psi^{(2)} &\equiv& \psi^+ - \psi^- \;,
&
\hbar \; \bar\psi^{(2)} &\equiv& \bar\psi^{+} - \bar\psi^{-} \;.
\end{array}
\right.
\end{eqnarray}
These definitions leads to
\begin{eqnarray}
\begin{array}{rcl}
\rmi\hbar G^{{(11)}}(t,t') &\equiv& \displaystyle \langle  \psi^{(1)}(t) \bar\psi^{(1)}(t')
\rangle 
= \rmi\hbar/4 \, \left[G^{++}+G^{--}+G^{-+} +G^{+-}\right]
\equiv  G^K \;, 
\\
\rmi\hbar G^{{(12)}}(t,t') &\equiv& \displaystyle \langle  \psi^{(1)}(t) \bar{\psi}^{(2)}(t')
\rangle 
= \rmi/2 \, \left[G^{++}-G^{--}+G^{-+} -G^{+-}\right]
\equiv -\rmi G^R \;, 
\\
\rmi\hbar G^{{(21)}}(t,t') &\equiv& \displaystyle \langle  \psi^{(2)}(t) \bar{\psi}^{(1)}(t')
\rangle 
= \rmi/2 \, \left[G^{++}-G^{--}-G^{-+} +G^{+-}\right]
\equiv \rmi G^A \;, 
\\
\rmi\hbar G^{{(22)}}(t,t') &\equiv& \displaystyle \langle  \psi^{(2)}(t) \bar{\psi}^{(2)}(t')
\rangle 
= \rmi/\hbar\,  \left[G^{++}+G^{--}-G^{-+} -G^{+-}\right]
= 0 \;.
\end{array}
\end{eqnarray}
Where we defined, \textit{en passant}, the Keldysh $G^K$, the retarded $G^R$ and
the advanced $G^A$ Green's functions in the same manner that we did for $C$ and
$R$ in Sect.~\ref{sec:KeldyshRotaBos}.
Using relation (\ref{eq:FQ=0}) we get
\begin{eqnarray}
 G^K &=& \rmi\hbar/2 \, \left[G^{++} + G^{--} \right] = \rmi\hbar/2 \, \left[G^{+-} +
G^{-+}\right] \;, \label{eq:RelationPMBasisToLOBasis1} \\
G^R &=& -\left[G^{++} - G^{+-}\right] = \left[G^{+-}-G^{-+}\right]\Theta(\tau)
\;, \label{eq:RelationPMBasisToLOBasis2} \\
G^A &=& \left[G^{++} - G^{-+}\right] =  \left[G^{+-}-G^{-+}\right]\Theta(-\tau)
\;, \label{eq:RelationPMBasisToLOBasis3}
\end{eqnarray}
which are inverted as
\begin{equation}
 \rmi\hbar G^{ab} = G^{K} + \frac{\rmi\hbar}{2}( a \ G^{A} - b \ G^{R} ) \;.
\end{equation}

\subsection{Symmetry properties under $t \leftrightarrow
t'$}\label{app:TimeReversal}
Using eq.~(\ref{eq:G_properties}), one establishes
\begin{eqnarray}
G^R(\tau) = -{G^A(-\tau)}^* \;, \qquad
G^K(\tau) = {G^K(-\tau)}^* \;.
\end{eqnarray}
And hence in Fourier space
\begin{eqnarray}
G^R(\omega) = -{G^A(\omega)}^* \;, \qquad
G^K(\omega) \in \mathbb{R} \;.
\end{eqnarray}

\subsection{Free fermions}

\subsubsection{Single free fermion}\label{app:FreeFermions}
The free fermion Hamiltonian is
\begin{equation}
 H = \epsilon \ \psi^\dagger \psi\;.
\end{equation}
Starting from the expression in terms of operators of the Keldysh Green's
functions,
\begin{equation}
\rmi\hbar G^{ab}(t,t') = \mbox{Tr}\left[ \mathsf{T}_{\cal C}\ \psi_H(t,a)
\psi_H^{\dagger}(t',b) \varrho(0) \right]\;,
\end{equation}
with $a,b=\pm$ and the grand-canonical density matrix $\varrho(0) \propto
e^{-\beta (H - \mu N)}$, one computes
\begin{eqnarray}
\begin{array}{rcl}
\rmi\hbar G^{+-}(\epsilon; \tau)
&=&
 - n_F \rme^{-\frac{\rmi}{\hbar} \epsilon \tau} \;, \\
\rmi\hbar G^{-+}(\epsilon; \tau)
&=&
 (1-n_F) \rme^{-\frac{\rmi}{\hbar} \epsilon \tau} \;.
\end{array}
\end{eqnarray}
 $n_F$ is the Fermi factor given by $ n_F(\epsilon) \equiv \left( 1+ \rme^{\beta (\epsilon - \mu) }\right)^{-1}$.
After the Keldysh rotation we get
\begin{eqnarray}
G^K(\epsilon; \tau) &=&  \frac{1}{2} \tanh\left(\beta\frac{\epsilon-\mu}{2}\right)
\rme^{-\frac{\rmi}{\hbar} \epsilon\tau} \;, \nonumber \\
G^R(\epsilon; \tau) &=&   \frac{\rmi}{\hbar} \rme^{- \frac{\rmi}{\hbar}
\epsilon\tau}\Theta(\tau) \;, \\
G^A(\epsilon; \tau) &=&   \frac{\rmi}{\hbar} \rme^{-\frac{\rmi}{\hbar}
\epsilon\tau}\Theta(-\tau) \;. \nonumber
\end{eqnarray}

\subsubsection{Collection of free fermions}
For our left and right reservoirs, we consider continuous distribution
(density of states) $\rho_L(\epsilon)$ and $\rho_R(\epsilon)$  of these free
fermions. This yields to the Keldysh Green's functions
\begin{eqnarray}
  G^{ab}_{\alpha}(\tau) = \int \ud{\epsilon} \rho_{\alpha}(\epsilon)
G_{\alpha}^{ab}(\epsilon;\tau) \;,
\end{eqnarray}
with $\alpha = L,R$. After a Keldysh rotation it yields
\begin{eqnarray}\label{eq:G_compo}
\begin{array}{rcl}
G^K(\tau) &=& \displaystyle
\int \Ud{\epsilon} \; \rho(\epsilon) \, \frac{1}{2}
\tanh[\beta(\epsilon-\mu)/2] \, 
\rme^{-\frac{\rmi}{\hbar} \epsilon \tau}
= \frac{1}{2} \; 
\langle \, 
\tanh[\beta(\epsilon-\mu)/2] \, 
\rme^{-\frac{\rmi}{\hbar} \epsilon \tau}
\, \rangle_{\epsilon} \;,
\\
G^R(\tau) &=& \displaystyle \int \Ud{\epsilon} \; \rho(\epsilon) \, \frac{\rmi}{\hbar}
\rme^{-\frac{\rmi}{\hbar} \epsilon \tau}
\Theta(\tau)
= 
\frac{\rmi}{\hbar} \; 
\langle \, 
\rme^{-\frac{\rmi}{\hbar} \epsilon \tau}
\, \rangle_{\epsilon}
\; 
\Theta(\tau) \;,
\\
G^A(\tau) &=&  \displaystyle
\int \Ud{\epsilon} \; \rho(\epsilon) \, \frac{\rmi}{\hbar}
\rme^{-\frac{\rmi}{\hbar} \epsilon \tau}\, 
\Theta(-\tau)
= \frac{\rmi}{\hbar} \; 
\langle \, 
\rme^{-\frac{\rmi}{\hbar} \epsilon \tau}\, 
\, \rangle_{\epsilon}
\; 
\Theta(-\tau) \;,
\end{array}
\end{eqnarray}
where we introduced a short-hand notation for the integration over energy
levels. In terms of the Fourier transforms of $\rho(\epsilon)$ it reads
\begin{eqnarray}
G^R(\tau) =
\frac{\rmi}{\hbar} \; 2 \pi \rho(\tau/\hbar) \Theta(\tau) \;,
\qquad
G^A(\tau) =
\frac{\rmi}{\hbar} \; 2 \pi \rho(\tau/\hbar) \Theta(-\tau) \;.
\end{eqnarray}

\subsubsection{Fourier transforms}
\begin{eqnarray}
\begin{array}{l}
G^K(\omega) =  \displaystyle \pi \hbar \tanh\left(\beta\frac{\hbar\omega-\mu}{2}\right)
\rho(\hbar\omega) \in  \mathbb{R} \;, \\
G^R(\omega) +\displaystyle  G^A(\omega) =   2 \rmi \pi \rho(\hbar\omega) \in \rmi \mathbb{R}
\;.
\end{array}
\end{eqnarray}
Since $\rho(\epsilon)$ is real, one computes
\begin{eqnarray}
\rm{Im} G^R(\omega) =  \pi \rho(\hbar\omega) \;.
\end{eqnarray}
Thus we get, as a check, the grand-canonical fermionic fluctuation-dissipation
theorem that is established generally in Sect.~\ref{sec:FDTProof}:
\begin{equation}
 G^K(\omega) = \hbar \tanh\left(\beta\frac{\hbar\omega-\mu}{2}\right) \mbox{Im }
G^R(\omega) \;.
\end{equation}

\section{Fluctuation-Dissipation Theorem}\label{sec:FDTProof}
In this Section we give a proof of the fluctuation-dissipation theorem both in
its bosonic and fermionic versions. This theorem only holds in equilibrium and
gives a relation between the Green's functions. In the grand-canonical ensemble,
the initial density operator reads $\varrho(0) \propto \exp\left( -\beta (H -
\mu N) \right)$, where $N$ is the number operator commuting with $H$ (in
non-relativistic quantum mechanics), $\mu$ is the chemical potential fixing the
average number of particles. One can obtain the 
theorem for the canonical ensemble by formally setting $\mu=0$. Let us consider
a pair of either bosonic or fermionic operators, for instance creation and
annihilation operators $\phi^\dagger$ and $\phi$. Let us write the following
Keldysh Green's function
\begin{equation} 
\rmi\hbar G^{+-}(t,t') =  \mbox{Tr}\left[ \mathsf{T}_{\cal C} \ {\bf \phi}_{\rm
H}(t,+) {\bf \phi}^{\dagger}_{\rm H}(t',-) \varrho(0) \right] \;.
\end{equation}
By resolving the time-ordering we get
\begin{equation}
\rmi\hbar G^{+-}(t,t') =  \zeta \ \mbox{Tr}\left[ {\bf \phi}^{\dagger}_{\rm
H}(t') {\bf \phi}_{\rm H}(t) \varrho(0)\right] \;,
\end{equation}
with $\zeta = +1$ in the bosonic case and $\zeta = -1$ in the fermionic case.
Using the \textit{analyticity} of the Green's functions and then expanding ${\bf
\phi}_{\rm H}(t+\rmi\beta\hbar) = \exp\left(-\beta H \right) {\bf \phi}_{\rm
H}(t) \exp\left(+\beta H \right) $, we get
\begin{eqnarray}
\rmi\hbar G^{+-}(t+\rmi\beta\hbar,t') &=&  \zeta \ \mbox{Tr} \left[ {\bf
\phi}^{\dagger}_{\rm H}(t') {\bf \phi}_{\rm H}(t+\rm\rmi\beta\hbar) \varrho(0)
\right]  \\
&\propto&  \zeta \ \mbox{Tr} \left[ {\bf \phi}^{\dagger}_{\rm H}(t') 
\exp\left(-\beta H \right) {\bf \phi}_{\rm H}(t)  \exp\left(\beta \mu N \right)
\right] \;.
\end{eqnarray}
Since $H$ and $N$ commute and since for any operator $f(N)$, one has $\phi f(N)
= f(N+1) \phi$, we have
\begin{eqnarray}
{\bf \phi}_{\rm H}(t)  \exp\left(\beta \mu N \right) = \exp\left(\beta \mu (N +
1)\right) {\bf \phi}_{\rm H}(t) \;,
\end{eqnarray}
and so
\begin{eqnarray}
 \rmi\hbar G^{+-}(t+\rmi\beta\hbar,t') =  \zeta \exp(\beta\mu) \ \mbox{Tr}
\left[ {\bf \phi}^{\dagger}_{\rm H}(t')  \varrho(0) {\bf \phi}_{\rm H}(t)
\right] \;.
\end{eqnarray}
Using the \textit{cyclicity} of the trace, we come to
\begin{eqnarray}
 \rmi\hbar G^{+-}(t+\rmi\beta\hbar,t') &=&  \zeta \exp(\beta\mu) \ \mbox{Tr}
\left[ {\bf \phi}_{\rm H}(t) {\bf \phi}^{\dagger}_{\rm H}(t') \varrho(0) \right]
\\
 &=&  \zeta \ \exp(\beta\mu) \ \rmi\hbar G^{-+}(t,t') \;.
\end{eqnarray}
If the system is in equilibrium, the \textit{time translational
invariance} of the previous equation gives the KMS relation:
\begin{equation}
G^{+-}(\omega) \ \exp(\beta\hbar\omega) =  \zeta\exp(\beta\mu) \ G^{-+}(\omega)
\;. \label{eq:KMS_f}
\end{equation}
Using eqs.~(\ref{eq:RelationPMBasisToLOBasis2}) and
(\ref{eq:RelationPMBasisToLOBasis3}), we have on the one hand
\begin{equation}
 G^R(\omega)+G^A(\omega) = G^{+-}(\omega) (1 - \zeta \
\exp(\beta(\hbar\omega-\mu)) \;. \label{eq:KMS2RA_f}
\end{equation}
On the other hand eq.~(\ref{eq:RelationPMBasisToLOBasis1}) implies 
\begin{eqnarray}
 G^K(\omega)  =  \frac{\rmi\hbar}{2} G^{+-}(\omega) [1 + \zeta
\exp(\beta(\hbar\omega-\mu))] \;.
\end{eqnarray}
These two last relations yield the grand-canonical quantum FDT:
\begin{eqnarray} 
G^K(\omega)  = \hbar \
\tanh\left(\beta\frac{\hbar\omega-\mu}{2}\right)^{-\zeta} \ \mbox{Im } 
G^R(\omega)  \;. \label{eq:BosonicFDT_gc} \label{eq:FermionicFDT_gc}
\end{eqnarray}

\section{Computing the self-energy}
\subsection{Derivation within the Schwinger-Keldysh formalism}\label{app:Sigma}
In the Schwinger-Keldysh path-integral representation we had (see
eq.~(\ref{eq:generating_func})) for the whole system (rotors and environment)
\begin{eqnarray}\label{eq:wholerep}
{\cal Z}[\boldsymbol{\eta}^\pm] \equiv \int_{\rm c}
\uD[\boldsymbol{n}^\pm,\boldsymbol{\psi}^\pm,\boldsymbol{\bar\psi}^\pm]
\rme^{\frac{\rmi}{\hbar}
S[\boldsymbol{n}^\pm,\boldsymbol{\psi}^\pm,\boldsymbol{\bar\psi}^\pm]  } \langle
\boldsymbol{n}^{+}(0),  \boldsymbol{\bar\psi}^{+}(0) |  \varrho(0) |
\boldsymbol{n}^{-}(0), \boldsymbol{\psi}^{-}(0) \rangle  \;,
\end{eqnarray}
At time $t=0$, just after the quench, the initial density is assumed to be
factorized: $\varrho(0) = I_S \otimes \varrho_L^{\rm free}(0) \otimes
\varrho_R^{\rm free}(0)$ (see Sect.~\ref{sec:QuenchSetUp}) yielding
\begin{eqnarray}
& & \langle \boldsymbol{n}^{+}(0), \boldsymbol{\bar\psi}^{+}(0) |  \varrho(0) |
\boldsymbol{n}^{-}(0), \boldsymbol{\bar\psi}^{-}(0) \rangle  \nonumber \\
& & \qquad = \delta(\boldsymbol{n}^{+}(0) - \boldsymbol{n}^{-}(0))
 \ \langle \boldsymbol{\bar\psi}_L^{+}(0) |  \varrho_L^{\rm free}(0) |
\boldsymbol{\psi}_L^{-}(0) \rangle
 \ \langle \boldsymbol{\bar\psi}_R^{+}(0) |  \varrho_R^{\rm free}(0) |
\boldsymbol{\psi}_R^{-}(0) \rangle \;.
\end{eqnarray}
The generating functional reads
\begin{eqnarray}
\cal Z[\boldsymbol{\eta}^\pm] = \int_{\rm c'} \uD[\boldsymbol{n}^+,
\boldsymbol{n}^-]
\rme^{\frac{\rmi}{\hbar} S_{S}[\boldsymbol{n}^+,
\boldsymbol{n}^-,\boldsymbol{\eta}] } \ \
\langle\langle \ \rme^{\frac{\rmi}{\hbar} S_{SB}[\boldsymbol{n}^+, 
\boldsymbol{\psi}^+, \boldsymbol{\bar\psi}^+, \boldsymbol{n}^-,
\boldsymbol{\psi}^-, \boldsymbol{\bar\psi}^-] } \ \rangle_{L}\rangle_{R} \;.
\end{eqnarray}
The index ${\rm c'}$ at the bottom of the integral is here to remind the constraints
on the field integration, namely ${{\bf n}_i^{+}(t)}^2 = {{\bf n}_i^{-}(t)}^2 = 1$ and ${\bf n}_i^{+}(0) = {\bf
n}_i^{-}(0)  \ \forall \ i$. We introduced the average over the free environment
composed of the two reservoirs:
\begin{eqnarray}
 \langle\langle \ \cdots \ \rangle_{L}\rangle_{R} 
\hspace{-1ex} &\equiv& \hspace{-1ex}
\int
\uD[\boldsymbol{\psi}^\pm, \boldsymbol{\bar\psi}^\pm]
\ \cdots \
\rme^{\frac{\rmi}{\hbar} S_{L}^{L} }
\rme^{\frac{\rmi}{\hbar} S_{R}^{R} } \nonumber \\
& & \qquad \qquad \times \langle \boldsymbol{\bar\psi}_L^{+}(0) | 
\varrho_L^{\rm free}(0) | \boldsymbol{\psi}_L^{-}(0) \rangle
 \ \langle \boldsymbol{\bar\psi}_R^{+}(0) |  \varrho_R^{\rm free}(0) |
\boldsymbol{\psi}_R^{-}(0) \rangle \;.
\end{eqnarray}
We now develop the coupling $\rme^{\frac{\rmi}{\hbar} S_{SB} }$ up to the second
order,
\begin{eqnarray} \label{eq:SSB2formal}
 \langle\langle \ \rme^{\frac{\rmi}{\hbar} S_{SB} }\ \rangle_{L}\rangle_{R}
\simeq 1 + \frac{\rmi}{\hbar} \langle\langle \ S_{SB} \ \rangle_{L}\rangle_{R}-
\frac{1}{2\hbar^2} \langle\langle \  S_{SB}^2 \ \rangle_{L}\rangle_{R} \;.
\end{eqnarray}
The first order term is zero. The second order term reads
\begin{eqnarray}\label{eq:SSB2}
 \langle\langle \  S_{SB}^2 \ \rangle_{L}\rangle_{R} \hspace{-0.7em} &=&
\hspace{-0.7em} 
M \left(\frac{\hbar\omega_c}{N_s}\right)^2 
\sum_{ab=\pm} ab \iint_{0}^{\infty} \Udd{t}{t'} \sum_{ij=1}^N \sum_{kk'qq'=1}^{N_s}
\sum_{\mu\nu = 1}^{M} \sum_{ll'mm'=1}^{\cal M}
n_i^{\mu a}(t) n_j^{\nu b}(t') \ \sigma_{ll'}^{\mu} \sigma_{mm''}^{\nu}
\nonumber\\
& & \hspace{-1em} \times \langle\langle \
\left[  \bar\psi^{a}_{Likl}(t)  \psi^{a}_{Rik'l'}(t) + L \leftrightarrow
R\right] 
\left[  \bar\psi^{b}_{Ljqm}(t')  \psi^{b}_{Rjq'm'}(t') + L \leftrightarrow
R\right]
\ \rangle_{L}\rangle_{R} \;.
\end{eqnarray}
Developing the term on the second line, we obtain
\begin{eqnarray}
& & \langle\langle \
\left[  \bar\psi^{a}_{Likl}(t)  \psi^{a}_{Rik'l'}(t) + L \leftrightarrow
R\right] 
\left[  \bar\psi^{b}_{Ljqm}(t')  \psi^{b}_{Rjq'm'}(t') + L \leftrightarrow
R\right]
\ \rangle_{L}\rangle_{R} \nonumber
 \nonumber\\
& & \qquad = 
 \langle\langle \
   \bar\psi^{a}_{Rikl}(t)  \psi^{a}_{Lik'l'}(t) \bar\psi^{b}_{Ljqm}(t') 
\psi^{b}_{Rjq'm'}(t') +  L \leftrightarrow R
 \ \rangle_{L}\rangle_{R}
\nonumber\\
& & \qquad = - 
 \langle  \psi^{a}_{Lik'l'}(t) \bar\psi^{b}_{Ljqm}(t') \rangle_{L} \
 \langle  \psi^{b}_{Rjq'm'}(t') \bar\psi^{a}_{Rikl}(t)\rangle_{R}
 +  L \leftrightarrow R
\nonumber\\
& & \qquad =  \delta_{ij} \delta_{k'q} \delta_{kq'} \delta_{l'm}  \delta_{lm' }
\hbar^2 \
\left[
G^{ab}_{Lk'}(t,t')
G^{ba}_{Rk}(t',t)
 +  L \leftrightarrow R
\right] \;.
\end{eqnarray}
With the free fermionic Green's functions defined on the Keldysh contour as
$\rmi\hbar G^{ab}_{\alpha k}(t,t') = \langle \psi^{a}_{k}(t)
\bar\psi^{b}_{k}(t') \rangle_{\alpha}$ for $\alpha= L,R$, $a,b=\pm$ and where
$k$ labels the electron's energy. Expression (\ref{eq:SSB2}) now reads
\begin{eqnarray}
 \langle \  S_{SB}^2 \ \rangle_{LR} 
\hspace{-1ex} &=& \hspace{-1ex}
\hbar^2 M \left(\frac{\hbar\omega_c}{N_s}\right)^2
\sum_{ab=\pm} ab \iint_{0}^{\infty} \Udd{t}{t'} \sum_{i=1}^N \sum_{kk'=1}^{N_s}
\sum_{\mu\nu = 1}^{M} \sum_{ll'=1}^{\cal M}
n_i^{\mu a}(t) n_i^{\nu b}(t') \ \sigma_{ll'}^{\mu} \sigma_{l'l}^{\nu}
\nonumber\\
& & \qquad \times
\left[
G^{ab}_{Lk'}(t,t')
G^{ba}_{Rk}(t',t)
 +  L \leftrightarrow R
\right] \;.
\end{eqnarray}
By using the property $\mbox{Tr } \sigma^\mu \sigma^\nu = \delta_{\mu\nu}$, we
get
\begin{eqnarray}
 \langle \  S_{SB}^2 \ \rangle_{LR} 
\hspace{-1ex} &=& \hspace{-1ex}
M \hbar^2 \left(\frac{\hbar\omega_c}{N_s}\right)^2
\sum_{ab=\pm} ab \iint_{0}^{\infty} \Udd{t}{t'} \sum_{i=1}^N
{\bf n}_i^{a}(t) \cdot {\bf n}_i^{b}(t') \  \nonumber\\
& & \qquad \qquad\qquad\quad \times
\sum_{kk'} 
\left[
 G^{ab}_{Lk'}(t,t')
G^{ba}_{Rk}(t',t)
 +  L \leftrightarrow R
\right] \;.
\end{eqnarray}
Finally expression (\ref{eq:SSB2formal}) can be recast into
\begin{eqnarray}
 \langle\langle \ \rme^{\frac{\rmi}{\hbar} S_{SB} }\ \rangle_{L}\rangle_{R}
\simeq \rme^{\frac{\rmi}{\hbar} S^{(2)}_{SB} } \;,
\end{eqnarray}
with
\begin{eqnarray}
   S^{(2)}_{SB}[\boldsymbol{n}^+, \boldsymbol{n}^-] \equiv  -\frac{1}{2}
M\sum_{ab=\pm} \iint_{0}^{+\infty} \Udd{t}{t'} \Sigma_B^{ab}(t,t') \sum_{i=1}^N  {\bf n}_i^{a}(t)
\cdot {\bf n}_i^{b}(t') \;,
\end{eqnarray}
where the exponent $(2)$ is here to recall that we developed until second
order and with the self-energy
\begin{eqnarray} \label{eq:SelfEnergy}
\Sigma_B^{ab}(t,t') \equiv - ab \rmi\hbar  \ (\hbar\omega_c)^2 \left[
G_{L}^{ab}(t,t') G_{R}^{ba}(t',t) + G_{R}^{ab}(t,t') G_{L}^{ba}(t',t) \right]
\;,
\end{eqnarray}
where the Keldysh Green's functions of the fermions in the $\alpha$-reservoir
($\alpha=L,R$) are given by
\begin{eqnarray}
G_{\alpha}^{ab}(t,t') \equiv \int \Ud{\epsilon_\alpha}
\rho_\alpha(\epsilon_\alpha) G_{\alpha}^{ab}(\epsilon_\alpha;t-t')  =
G_{\alpha}^{ab}(t-t') \;.
\end{eqnarray}
$\rho_{\alpha}(\epsilon)$ is the density of states in $\alpha$-reservoir and
$G_{\alpha}^{ab}(\epsilon;\tau)$ are the Keldysh Green's functions of a free
fermion with energy $\epsilon$ in equilibrium in the $\alpha$-reservoir  (see
Appendix~\ref{app:FreeFermions}):
\begin{eqnarray}
\begin{array}{rcl}
\rmi \hbar G_{\alpha}^{+-}(\epsilon;\tau) &=& - n_{\alpha}(\epsilon)
\rme^{-\frac{\rmi}{\hbar} \rmi\epsilon \tau} \;, \\
\rmi \hbar G_{\alpha}^{-+}(\epsilon;\tau) &=&  \left[ 1- n_{\alpha}(\epsilon)
\right] \rme^{-\frac{\rmi}{\hbar} \epsilon \tau} \;, \\
\rmi \hbar G_{\alpha}^{++}(\epsilon;\tau) &=&  \rmi \hbar
G_{\alpha}^{-+}(\epsilon;\tau)  \Theta(\tau) + \rmi \hbar 
G_{\alpha}^{+-}(\epsilon;\tau)  \Theta(-\tau) \;, \\
\rmi \hbar G_{\alpha}^{--}(\epsilon;\tau) &=&  \rmi \hbar
G_{\alpha}^{+-}(\epsilon;\tau)  \Theta(\tau) + \rmi \hbar 
G_{\alpha}^{-+}(\epsilon;\tau)  \Theta(-\tau) \;,
\end{array}
\end{eqnarray}
with the Fermi factor $n_{\alpha}(\epsilon) \equiv (1+\rme^{\beta_\alpha
(\epsilon - \mu_\alpha)})^{-1}$.
It is clear then that the self-energy is time translational invariant:
$\Sigma_B^{ab}(t,t') \equiv \Sigma_B^{ab}(\tau)$ with $\tau \equiv t-t'$.
Moreover $\Sigma_B^{ab}(\tau)$ is a symmetric matrix with respect to time and
Keldysh indices:
\begin{equation}
\Sigma_B^{ab}(\tau) = \Sigma_B^{ba}(-\tau) \;,
\end{equation}
Using the time reversal property eq.~(\ref{eq:G_properties}) of the Keldysh
Green's functions one also establishes
\begin{equation}
 {\Sigma_B^{ab}(\tau)}^* = -\Sigma_B^{\bar a \bar b}(\tau)\;,
\end{equation}
where we note $\bar a \equiv -a$.

After a Keldysh rotation of the rotors coordinates, it yields
\begin{eqnarray}
  \frac{\rmi}{\hbar} S^{(2)}_{SB}[\boldsymbol{n}^{{(1)}}, \boldsymbol{n}^{{(2)}}] =
\frac{1}{2} M \sum_{rs={(1)}, {(2)}} \iint_{0}^\infty \Udd{t}{t'} 
\Sigma_{B}^{rs}(t,t') \sum_{i=1}^N {\bf n}_i^r(t) {\bf n}_i^s(t') \;,
\end{eqnarray}
with
\begin{eqnarray}
\begin{array}{rcl}
\Sigma_B^{{(22)}} &=& - \rmi\hbar/2 \ \left[ \Sigma_B^{++} + 
\Sigma_B^{--}\right] \;, \\
 \Sigma_B^{{(21)}} &=& -\rmi\left[ \Sigma_B^{++} +  \Sigma_B^{+-}\right] \;, \\
\Sigma_B^{{(12)}} &=& -\rmi\left[ \Sigma_B^{++} +  \Sigma_B^{-+}\right] \;, \\ 
 \Sigma_B^{{(11)}} &=& -\rmi/\hbar \ \left[ \Sigma_B^{++} +  \Sigma_B^{+-} +
\Sigma_B^{-+} +\Sigma_B^{--} \right] =  0 \;.
\end{array}
\end{eqnarray}
which is inverted as 
\begin{eqnarray}
 \rmi\hbar\Sigma_B^{ab} = -ab \Sigma_B^{{(22)}} - \frac{\hbar}{2}
\left( a \Sigma_B^{{(21)}} + b \Sigma_B^{{(12)}} \right) \;.
\end{eqnarray}

\subsection{FDT check}\label{app:FDT2ndkind}

We checked that the fermion-reservoir self-energy satisfies the bosonic FDT.
This is only valid when the reservoirs constitute an equilibrium bath,
\textit{i.e.}, $\beta_L = \beta_R = \beta$ and $\mu_L = \mu_R = \mu_0$ ($V=0$).
Note that distribution functions $\rho_L(\omega)$ and  $\rho_R(\omega)$ can be
different although the proof given below uses $\rho_L(\omega)=\rho_R(\omega) =
\rho(\epsilon)$ for simplicity reasons. The goal is to check 
\begin{equation}\label{eq:FDTcheck}
\Sigma_B^K (\omega) 
= \hbar \; \coth\left(\beta \frac{\hbar\omega}{2}\right) \; \mbox{Im } 
\Sigma_B^R(\omega)
= \hbar \coth\left(\beta \frac{\hbar\omega}{2}\right) \; \frac{\left[\Sigma_B^R
+ \Sigma_B^A \right](\omega)}{2\rmi} \;.
\end{equation}
We first develop the term in the {\sc lhs}, then we do the same with the {\sc rhs} to prove
their equality.
\begin{eqnarray}
\Sigma_B^K(\omega) &=& \rm{TF} \ \Sigma_B^K(\tau) \nonumber\\
&=& -2 (\hbar \omega_c)^2 \ \rm{TF} \left\{ G^{K} G^{K*} -
{\hbar^2}/{4}\, \left[ G^{A} G^{A*} + G^{R} G^{R*} \right] \right\} \nonumber\\
&=& -2 (\hbar \omega_c)^2  \ \rm{TF} \left\{ G^{K} G^{K*} -
{\hbar^2}/{4}\, \left[G^{R} + G^{A}\right] \left[G^{R*} +  G^{A*}\right]\right\} \;,
\end{eqnarray}
where we used the nullity
of cross terms of the type $G^{R} G^{A}$ since $G^{R} \propto \Theta(\tau)$ and
$G^{A} \propto \Theta(-\tau)$.
\begin{eqnarray}
 \Sigma_B^K(\omega) = -2 (\hbar \omega_c)^2  \left\{ G^{K} \circ
G^{K *} - {\hbar^2}/{4}\,  \left[ G^{R} +  G^{A}\right] \circ \left[G^{R *} + G^{A *} \right] \right\}
\;, \label{eq:TFconvo} 
\end{eqnarray}
where $\circ$ is the symbol for the convolution (see
Appendix~\ref{app:Conventions}) and $G^{R*}(\omega)$ stands for the Fourier
transform of $G^{R}(\tau)^*$. Since we easily obtain
\begin{eqnarray}
\begin{array}{rcl}
 G^{R}(\omega) + G^{A}(\omega) &=&  2 \rmi \pi \rho(\hbar\omega) \;, \\
 G^{R *}(\omega) + G^{A *}(\omega) &=& - 2 \rmi \pi \rho(-\hbar\omega) \;,
\end{array}
\end{eqnarray}
and
\begin{eqnarray}
\begin{array}{rcl}
 G^{K}(\omega)  &=& \pi \hbar \rho(\hbar\omega) \tanh{\left(\beta \frac{\hbar
\omega - \mu_0}{2}\right)} \;, \\
 G^{K *}(\omega) &=& \pi \hbar \rho(-\hbar\omega) \tanh{\left(\beta \frac{-\hbar
\omega - \mu_0}{2}\right)} \;,
\end{array}
\end{eqnarray}
we get by replacing in (\ref{eq:TFconvo})
\begin{eqnarray}
\begin{array}{rcl}
  \Sigma_B^K(\omega) 
\hspace{-0.7em} &=& \hspace{-0.7em} - 2 (\hbar \omega_c)^2 (\pi \hbar)^2
 \\ & & \times \left\{
\hspace{-0.5ex}
\left[\rho(\hbar\omega) \tanh{\hspace{-0.5ex}\left(\beta \frac{\hbar \omega -
\mu_0}{2}\right)}\right]
\hspace{-0.5ex}\circ\hspace{-0.5ex}\left[\rho(-\hbar\omega)
\tanh{\hspace{-0.5ex}\left(\beta
\frac{-\hbar \omega - \mu_0}{2}\right)}\right]
- \left[\rho(\hbar\omega)\right]
\hspace{-0.5ex} \circ \hspace{-0.5ex} \left[\rho(-\hbar\omega)\right]
\hspace{-0.5ex}
\right\}
 \\
\hspace{-0.7em} &=& \hspace{-0.7em} - 2 (\hbar \omega_c)^2 (\pi \hbar) \int
\frac{{\rm d}\epsilon'}{2\pi} \rho(\epsilon')\rho(\epsilon' - \hbar\omega)
\left\{ \tanh{\left(\beta \frac{\epsilon' - \mu_0}{2}\right)}
\tanh{\left(\beta\frac{\epsilon' - \hbar\omega - \mu_0}{2}\right)} - 1 \right\}
\\
\hspace{-0.7em} &=& \hspace{-0.7em} - \pi \hbar (\hbar \omega_c)^2 
\coth{\left(\beta\frac{\hbar \omega}{2}\right)} \int \Ud{\epsilon'}
\rho(\epsilon')\rho(\epsilon' - \hbar\omega) 
\left\{ \tanh{\left( \beta \frac{\epsilon' - \hbar\omega - \mu_0}{2}\right)} -
\tanh{\left( \beta \frac{\epsilon' - \mu_0}{2}\right)}\right\} \;,
\end{array}
\label{eq:LHSdev}
\end{eqnarray}
where we used the trigonometry relation 
\begin{eqnarray} 
\tanh{(x-y)} = \frac{\tanh{x} - \tanh{y}}{ 1 - \tanh{x}\tanh{y}} \;.
\nonumber
\end{eqnarray}
Let's now calculate the {\sc rhs} of (\ref{eq:FDTcheck}).
\begin{eqnarray} 
\frac{\left[\Sigma_B^R + \Sigma_B^A \right](\omega)}{2\rmi}
&=& \rmi (\hbar \omega_c)^2 
  \ \rm{TF} \left\{G^{R} G^{K *} +  G^{A}  G^{K *} +  G^{K} G^{R*}
+   G^{K} G^{A*} \right\} \nonumber \\
&=& \rmi (\hbar \omega_c)^2
 \ \rm{TF} \left\{ (G^{R} + G^{A}) G^{K *}   +
G^{K}(G^{R *} + G^{A *})\right\} \nonumber \\
&=&  \rmi (\hbar \omega_c)^2 
 \left\{ \left[ G^{R} + G^{A} \right]  \circ \left[ G^{K *}
\right] 
+ \left[ G^{K} \right] \circ \left[ G^{R *} + G^{A *} \right] \right\}\;,
\end{eqnarray}
giving
\begin{equation}
\begin{array}{l}
  \hbar\coth\left(\beta\frac{\hbar\omega}{2}\right)
\frac{\left[\Sigma_B^R + \Sigma_B^A \right](\omega)}{2\rmi} \\ 
\qquad =  -2 (\pi\hbar)^2
(\hbar \omega_c)^2 
\coth{\left(\beta\frac{\hbar \omega}{2}\right)} \\ 
 \qquad \qquad \qquad  \times\left\{ \left[ \rho(\hbar\omega) \right] \circ
\left[\rho(-\hbar\omega) 
\tanh{\left(\beta\frac{-\hbar \omega - \mu_0}{2}\right)}  \right] 
- \left[ \rho(\hbar\omega)  \tanh{\left(\beta \frac{\hbar \omega -
\mu_0}{2}\right)} \right] \circ \left[ \rho(-\hbar\omega) \right] \right\}
\nonumber \\
  = -\pi \hbar (\hbar \omega_c)^2 (2 \pi \hbar) \coth{\left(\beta\frac{\hbar
\omega}{2}\right)} \int \ud{\epsilon'} \rho(\epsilon')\rho(\epsilon' -
\hbar\omega) 
\left\{ \tanh{\left(\beta \frac{\epsilon' - \hbar\omega - \mu_0}{2}\right)} -
\tanh{\left(\beta\frac{\epsilon' - \mu_0}{2}\right)}\right\} \;.
\end{array}
\end{equation}
We recognize here the development~(\ref{eq:LHSdev}) of $\Sigma_B^K$. We just
proved that the bosonic FDT is satisfied provided that the two fermionic
reservoirs have the same temperature and chemical potential. They can have a
different density of states.

\section{Dynamics}\label{app:Dynamics}
\subsection{Quadratic effective action}
One can render the effective action quadratic at the price of introducing new
fields.
For a given $i$ and a given pair of $(r, \mu, t)$ and $(s, \nu, t')$, the
identity
\begin{eqnarray}
 1 = \int \ud{Q_{i\mu\nu}^{\ rs}(t,t')}   \delta \left( n^{\mu r}_i(t)n^{\nu
s}_i(t') -  Q_{i\mu\nu}^{\ rs}(t,t') \right)\;, \label{eq:identityresolv}
\end{eqnarray}
becomes, after using the integral representation of the delta
distribution (see Appendix~\ref{app:Conventions}),
\begin{eqnarray}
1 \propto \int \udd{Q_{i\mu\nu}^{\ rs}(t,t')}{\lambda_{i\mu\nu}^{\ rs}(t,t')} 
\exp \left( -\rmi \frac{M}{2} \lambda_{i\mu\nu}^{\ rs}(t,t') \left( n^{\mu
r}_i(t)n^{\nu s}_i(t') -  Q_{i\mu\nu}^{\ rs}(t,t') \right)  \right) \;.
\end{eqnarray}
Introducing similar identities for all possible pairs of $(r, \mu, t)$ and $(s,
\nu, t')$, we obtain a path integral over two\footnote{There are $N(M^2K^2 +
MK)/2$ of each of these fields, where $K=2$ is the number of possible
Keldysh indices.} fields $Q_{i\mu\nu}^{\ rs}(t,t')$ and $\lambda_{i\mu\nu}^{\
rs}(t,t')$ that are symmetric in the Keldysh indices, times and rotor
components: $Q_{i\nu\mu}^{\ sr}(t',t) = Q_{i\mu\nu}^{\ rs}(t,t')$ and
$\lambda_{i\nu\mu}^{\ sr}(t',t) = \lambda_{i\mu\nu}^{\ rs}(t,t')$.
The effective action is now also a functional of $\boldsymbol{Q}$ and
$\boldsymbol{\lambda}$ and reads
\begin{eqnarray}
 \frac{\rmi}{\hbar} S_{\rm eff} 
\hspace{-1ex} &=& \hspace{-1ex}
-\frac{M}{2} \sum_{r,s = {(1)}, {(2)}} \iint \udd{t}{t'} \sum_{i} \sum_{\mu\nu} n_i^{\mu r}(t) \left[  Op_{i\mu\nu}^{\ rs}(t,t') + \rmi \lambda_{i\mu\nu}^{\ rs}(t,t') \right] n_i^{\nu s}(t') \\
& & + 
\frac{J^2 M^2}{2 N} \sum_{i,j} \iint \udd{t}{t'} \sum_{\mu,\nu} Q_{i\mu\nu}^{\ {(11)}}(t,t') Q_{j\mu\nu}^{\ {(22)}}(t,t') + Q_{i\mu\nu}^{\ {(12)}}(t,t') Q_{j\mu\nu}^{\ {(21)}}(t,t')
 \nonumber \\
& & + 
\frac{\rmi}{\hbar} \frac{M}{2}  \sum_{a} a \int \ud{t} \sum_i z_i^a(t) 
+ \rmi \frac{M}{2}  \sum_{rs} \iint \udd{t}{t'} \sum_{i} \sum_{\mu\nu} \lambda_{i\mu\nu}^{\ rs}(t,t') Q_{i\mu\nu}^{\ rs}(t,t')  \nonumber \\
& & +  \quad \mbox{boundary terms} \;, \nonumber
\end{eqnarray}
where we introduced the operator $Op_{i\mu\nu}^{\ rs}(t,t')$ defined as
\begin{eqnarray} \label{eq:defOprs}
\begin{array}{rcl}
Op_{i\mu\nu}^{\ {(12)}}(t,t') &\equiv& \displaystyle \rmi\delta_{\mu\nu} \delta(t-t') \left[  \frac{1}{\Gamma} \partial^2_{t'} + \frac{1}{2} \sum_{a=\pm} z_i^a(t)  \right]
		-\rmi  \delta_{\mu\nu} \Sigma_B^R(t',t) \;,
\\
Op_{i\mu\nu}^{\ {(21)}}(t,t') &\equiv& \displaystyle Op_{i\nu\mu}^{\ {(12)}}(t',t) \;, \\
Op_{i\mu\nu}^{\ {(22)}}(t,t') &\equiv& \displaystyle \frac{\rmi\hbar}{4} \delta_{\mu\nu} \delta(t-t') \sum_{a=\pm} a z_i^a(t) 
		+ \delta_{\mu\nu} \Sigma_B^K(t,t')  \;, \\
 Op_{i\mu\nu}^{\ {(11)}}(t,t') &\equiv& \displaystyle \frac{\rmi}{2\hbar} \delta_{\mu\nu} \delta(t-t') \sum_{a=\pm} a  z_i^a(t)\;.
\end{array}
\end{eqnarray}
$Op_{i\mu\nu}^{\ rs}(t,t')$ is symmetric in the Keldysh indices, times and rotor components: $Op_{i\nu\mu}^{\ sr}(t',t) = Op_{i\mu\nu}^{\ rs}(t,t')$.
The functional integration over $n_i^{\mu r}$ is now quadratic and can be performed, leading to
\begin{eqnarray}
\frac{\rmi}{\hbar}  S_{\rm eff} 
\hspace{-1ex} &=& \hspace{-1ex}
 -\frac{1}{2} \mbox{Tr } \ln  M \left( Op +  \rmi \lambda  \right) \\
& & - \frac{J^2 M^2}{2 N} \sum_{i,j} \iint \udd{t}{t'} \sum_{\mu,\nu} Q_{i\mu\nu}^{\ {(11)}}(t,t') Q_{j\mu\nu}^{\ {(22)}}(t,t') + Q_{i\mu\nu}^{\ {(12)}}(t,t') Q_{j\mu\nu}^{\ {(21)}}(t,t')
 \nonumber \\
 & & + \frac{\rmi}{\hbar}  \frac{M}{2} \sum_{a} a \int \ud{t} \sum_i z_i^a(t) 
+ \rmi \frac{M}{2}  \sum_{rs} \iint \udd{t}{t'} \sum_{i} \sum_{\mu\nu} \lambda_{i\mu\nu}^{\ rs}(t,t') Q_{i\mu\nu}^{\ rs}(t,t') \nonumber
\end{eqnarray}
where the trace in the first term is spanning the whole space of indices, namely rotor sites, Keldysh indices, times and rotor components.

\subsection{Saddle-point evaluation}
In this subsection, we evaluate in the limit $M\to \infty$ the saddle-point equations with respect to the dummy fields we introduced previously, namely $\lambda_{i\mu\nu}^{\ rs}(t,t')$, $Q_{i\mu\nu}^{\ rs}(t,t')$ and $z_{i}^{a}(t)$. 
The fluctuations around the saddle are neglected. In particular, using eq.~(\ref{eq:identityresolv}) we have the identity (see the definition of Green's functions in Sect.~\ref{sec:KeldyshRotaBos})
\begin{eqnarray}
 Q_{i\mu\nu}^{\ rs}(t,t') = \rmi\hbar G_{ii\mu\nu}^{\ rs}(t,t') \;. \label{eq:QeqGrs}
\end{eqnarray}
Along the lines we prove that the solution in the saddle is $O(NM)$, like the starting Hamiltonian.

The saddle-point with respect to $\lambda_{i\mu\nu}^{\ rs}(t,t')$ yields
\begin{eqnarray}
 \frac{\delta  S_{\rm eff}}{\delta \lambda_{i\mu\nu}^{\ rs}(t,t')} =  -\frac{1}{2} \ \mbox{Tr } \frac{\delta }{\delta \lambda_{i\mu\nu}^{\ rs}(t,t')} \ln  M \left( Op + \rmi \lambda  \right) 
+ \rmi \frac{M}{2} Q_{i\mu\nu}^{rs}(t,t') = 0 \;,
\end{eqnarray}
giving in matrix notations
\begin{eqnarray}
  {}^t \!\left( Op + \rmi \lambda \right)^{-1} =  M  Q \;,
\end{eqnarray}
where the symbol ${}^t$ represents the transposition. Since all operators in the last equation are symmetric by definition, we get 
\begin{eqnarray}\label{eq:saddle1rs}
  Op + \rmi \lambda = \frac{1}{M} Q^{-1} \;.
\end{eqnarray}

The saddle-point equation with respect to $Q_{i\mu\nu}^{\ rs}(t,t')$ yields
\begin{equation}\label{eq:saddle2rs}
 \rmi\lambda_{i\mu\nu}^{\ rs}(t,t') = \frac{J^2 M}{N} \sum_{j} Q_{j\mu\nu}^{\ \bar r \bar s}(t,t') \quad \forall \ i\;,
\end{equation}
where $\overline{(2)} \equiv {(1)}$ and $\overline{(1)} \equiv {(2)}$.
The {\sc rhs} of this last equation being site-independent, $\lambda_{i\mu\nu}^{\ rs}(t,t')$ does not depend on $i$:  $\lambda_{i\mu\nu}^{\ rs}(t,t') = \lambda_{\mu\nu}^{rs}(t,t')$. Equations (\ref{eq:saddle1rs}) and (\ref{eq:saddle2rs}) imply
\begin{equation}\label{eq:saddle1+2rs}
  Op_{i}^{\ rs} + \frac{J^2 M}{N} \sum_{j} Q_{j}^{\ \bar r \bar s} - \frac{1}{M} {Q^{-1}}_i^{\ rs} =0 \;.
\end{equation}
The saddle-point equation with respect to $z_{i}^{a}(t)$ yields to the two equations:
\begin{eqnarray}
\begin{array}{rcl}
\displaystyle \sum_\mu  \left(\left[ Op + \rmi \lambda \right]^{-1} \right)_{i\mu\mu}^{\ {(12)}}(t,t) + \left(\left[ Op + \rmi \lambda \right]^{-1} \right)_{i\mu\mu}^{\ {(21)}}(t,t) =  0 \;, \\
\displaystyle \sum_\mu  \left(\left[ Op + \rmi \lambda \right]^{-1} \right)_{i\mu\mu}^{\ {(11)}}(t,t) +  \frac{\hbar^2}{4} \left(\left[ Op + \rmi \lambda \right]^{-1} \right)_{i\mu\mu}^{\ {(22)}}(t,t) =   M \;.
\end{array}
\end{eqnarray}
This is nothing more than the constraint that rotors should have a unit length. However, $\lambda$ being site-independent, it is clear from these equations that it has to be the same for $Op$. Finally at the saddle, $Op$, $Q$ and $z$ are site-independent (homogeneous) so we can get rid of the sites indices: $Op_{i\mu\nu}^{\ rs}(t,t') = Op_{\mu\nu}^{rs}(t,t')$, $Q_{i\mu\nu}^{\ rs}(t,t') = Q_{\mu\nu}^{rs}(t,t')$ and $z_i^a(t) = z^a(t)$.
Equation~(\ref{eq:saddle1+2rs}) becomes
\begin{equation}
   Op^{rs} + J^2 M Q^{\bar r \bar s} - \frac{1}{M} {Q^{-1}}^{rs} = 0\;.
\end{equation}
Since from its definition (\ref{eq:defOprs}) $Op_{\mu\nu}^{rs}(t,t')\propto\delta_{\mu\nu}$, the previous equation tells us that it has to be the same for $Q_{\mu\nu}^{rs}(t,t') $ so we can get rid of all the rotor component indices. Multiplying  by $Q^{sv}(t',t'')$, and summing over $s$ and $t'$, we get
\begin{equation}
  \int \ud{t'} \sum_s Op^{rs}(t,t') Q^{sv}(t',t'')   +  J^2 M Q^{\bar r \bar s}(t,t') Q^{sv}(t',t'')  - \frac{1}{M} \delta_{rv} \delta(t-t'') = 0 \;.
\end{equation}
 The macroscopic Green's function reading $\rmi \hbar G^{rs}(t,t') = M Q^{rs}(t,t')$ we obtain
\begin{equation}\label{eq:EQDyna_00}
  \int \ud{t'} \sum_s Op^{rs}(t,t') \rmi \hbar G^{sv}(t',t'')   + J^2  \rmi \hbar G^{\bar r \bar s}(t,t') \rmi \hbar G^{sv}(t',t'') - \delta_{rv} \delta(t-t'') = 0 \;.
\end{equation}

\subsection{Schwinger-Dyson equations}
The $(r={(2)}, v={(1)})$ component of eq.~(\ref{eq:EQDyna_00}) gives a complex equation the real part of which yields
\begin{eqnarray}
 z^+(t) = z^-(t) \equiv z(t) \ \forall \ t \;,
\end{eqnarray}
and the imaginary part of which is the dynamic equation for the self-correlation:
\begin{eqnarray} \label{eq:Dyson1app}
\left[\frac{1}{\Gamma}\frac{\partial^2}{\partial t^2} + z(t)\right] C(t,t') 
         = \int_{0}^{t'} \ud{t''}  \Sigma^K (t,t'')R(t',t'') + \int_{0}^{t} \ud{t''}  \Sigma^R (t,t'')C(t'',t') \;, 
\end{eqnarray}
where we introduced
\begin{eqnarray}
  \Sigma^K  \equiv  J^2 C + \Sigma_B^K  \;, \qquad   \Sigma^R   \equiv J^2 R + \Sigma_B^R  \;.
\end{eqnarray}
Similarly, the $(r={(2)}, v={(2)})$ component of eq.~(\ref{eq:EQDyna_00}) yields the equation of motion for the self-response:
\begin{eqnarray}
\left[\frac{1}{\Gamma}\frac{\partial^2}{\partial t^2} + z(t)\right] R(t,t') 
         = \delta(t-t') + \int_{t'}^{t} \ud{t''}  \Sigma^R (t,t'')R(t'',t') \;. \label{eq:Dyson2app}
\end{eqnarray}
The $(r={(1)}, v={(1)})$ component of eq.~(\ref{eq:EQDyna_00}) leads to the same equation and the $(r={(1)}, v={(2)})$ component expresses $0=0$. Setting $t'=t$ in eq.~(\ref{eq:Dyson1app}) we obtain the expression for the Lagrange multiplier
\begin{eqnarray}\label{eq:Dyson3app}
  z(t) =  \int_{0}^{t} \ud{t''}  \Sigma^K (t,t'')R(t,t'') +  \Sigma^R (t,t'')C(t,t'') - \frac{1}{\Gamma} \frac{\partial^2 C}{\partial t^2}(t,t'\to t^-)\;. \label{eq:zapp}
\end{eqnarray}
Equations (\ref{eq:Dyson1app}) and (\ref{eq:Dyson2app}) together with eq.~(\ref{eq:zapp}) constitute the Schwinger-Dyson equations that fully determine the dynamics of the interacting system.


\begin{thebibliography}{99}

\bibitem{Leggett}
A. J. Leggett, S. Chakravarty, A. T. Dorsey, M. P. A. Fisher, A. Garg, W. Zwerger, Rev. Mod. Phys. \textbf{59}, 1 (1987).

\bibitem{Barbara} 
I.S. Tupitsyn and B. Barbara, {\it Quantum tunneling of magnetization in molecular 
complexes with large spins. Effect of the environment.} Magnetism: Molecules to Materials III, 109 (Wiley, New York, 2001).

\bibitem{Calabrese}
See \textit{e.g} P. Calabrese and J. Cardy, 
J. Phys. A {\bf 42}, 504005 (2009).

\bibitem{cut-offReichman} 
D. Segal, D. R. Reichman and A. J. Millis, Phys. Rev. B {\bf 76}, 195316 (2007).

\bibitem{Seiji}
S. Miyashita, S. Tanaka, H. De Raedt, and B. Barbara,
J. Phys.: Conf. Ser. 143 012005 (2009).

\bibitem{quantum-annealing}
T. Kadowaki and H. Nishimori, 
Phys. Rev. E {\bf 58}, 5355 (1998).\\
E. Farhi, J. Goldstone, S. Gutmann, J. Lapan, A. Lundgren, and D. Preda,
Science \textbf{292}, 472 (2001). \\
G. E. Santoro and E. Tosatti, J. Phys. A {\bf 39}, R393 (2006). 


\bibitem{Zurek}
A. Polkovnikov, Phys. Rev. B {\bf 72}, 161201 (2005). \\
W. H. Zurek, U. Dorner, and P. Zoller, 
Phys. Rev. Lett. {\bf 95}, 105701 (2005). 

\bibitem{Greg}
L. F. Cugliandolo, G. S. Lozano, and H. Lozza, 
Phys. Rev. B {\bf 71}, 224421 (2005). \\
 G. Schehr and  H. Rieger
Phys. Rev. Lett. {\bf 96}, 227201 (2006).
J. Stat. Mech. (2008) P04012.

\bibitem{Lili}
L. Arrachea, Phys. Rev. B {\bf 70}, 155407 (2004). \\
A. Caso, L. Arrachea, and G. S. Lozano, Phys. Rev. B {\bf 81}, 041301 (2010)

\bibitem{LiliCu}
L. Arrachea and L. F. Cugliandolo
Europhys. Lett. {\bf 70}, 642 (2005).

\bibitem{OnukiKawasaki} A. Onuki and K. Kawasaki, 
Ann. Phys. (N. Y.) {\bf 121}, 456 (1979).

\bibitem{Hinrichsen}
H. Hinrichsen,  Adv. Phys. \textbf{49} 815 (2000).

\bibitem{DombGreen} B. Schmittmann and R. K. P. Zia, 
Vol. 17 of Phase Transitions and Critical  Phenomena,
eds. C. Domb and J. L. Lebowitz (Academic Press, London, 1995).

\bibitem{Tauber}
U. C. T\"{a}uber, Chap. 7 of Ageing and the Glass Transition in Lecture Notes in Physics (Springer, Berlin, 2007).

\bibitem{Struick} L. C. E. Struick, {\it Physical Aging 
in Amorphous Polymers and Other Materials}, (Elsevier, Amsterdam, 1978).

\bibitem{LeticiaLesHouches} L. F. Cugliandolo,
in Les Houches Session LXXVII, J-L. Barrat {\it et al}
eds. (Springer-EDP Sciences, Berlin-Les Ulis, 2003).

\bibitem{DalidovichPhillips} D. Dalidovich and P. Phillips,
  Phys. Rev. Lett. {\bf 93}, 027004 (2004).

\bibitem{GreenSondhi} A. G. Green and S. L. Sondhi,
  Phys. Rev. Lett. {\bf 95}, 267001 (2005).

\bibitem{HoganGreen} P. M. Hogan and A. G. Green, arXiv:cond-mat/0607522.

\bibitem{MitraKimMillis} 
 A. Mitra, S. Takei, Y. B. Kim, and A. J. Millis, Phys. Rev. Lett. \textbf{97}, 236808 (2006).

\bibitem{Feldman}  D. E. Feldman, Phys. Rev. Lett. {\bf 95}, 177201 (2005).

\bibitem{MitraMillis}
A. Mitra and A. J. Millis, Phys. Rev. B \textbf{77}, 220404 (2008).

\bibitem{AroBiroCuglio} C. Aron, G. Biroli, and L. F. Cugliandolo,
Phys. Rev. Lett. {\bf 102}, 050404 (2009).

\bibitem{SachdevBook} S. Sachdev, {\it Quantum Phase Transitions}, 
(Cambridge Univ. Press, 1999).

\bibitem{SchwingerKeldysh} 
U. Weiss, {\it Quantum dissipative systems} (World Scientific, Singapore, 1993).

\bibitem{Kamenev}
A. Kamenev, arXiv:cond-mat/0412296, arXiv:cond-mat/0109316.

\bibitem{SachdevYe} J. Ye, S. Sachdev, and N. Read, Phys. Rev. Lett. {\bf 70}, 4011 (1993). \\
T. K. Kope\'c, Phys. Rev. B {\bf 50}, 9963 (1994).

\bibitem{Shukla}
P. Shukla and S. Singh, Phys. Lett. A \textbf{81}, 477 (1981).

\bibitem{Rokni}
M. Rokni and P. Chandra, Phys. Rev. B {\bf 69},  094403 (2004).
 
\bibitem{CugliandoDean}
L. F. Cugliandolo and D. S. Dean, J. Phys. A {\bf 28}, 4213 (1995). \\
L. F. Cugliandolo and D. S. Dean, J. Phys. A {\bf 28}, L453 (1995).

\bibitem{CugliandoloLozano}
L. F. Cugliandolo and G. S. Lozano, Phys. Rev. B {\bf 59}, 915 (1999). \\
L. F. Cugliandolo and G. S. Lozano, Phys. Rev. Lett. {\bf 80}, 4979 (1998).

\bibitem{Ingold}
A. Schmid, J. Low Temp. {\bf 49}, 609 (1982). \\
A. O. Caldeira and A. J. Leggett, Physica A {\bf 121}, 587 (1983). \\
H. Grabert, P. Schramm and G.-L. Ingold, Phys. Rep. {\bf 168}, 115 (1988). \\
C. Greiner and S. Leupold, Ann. Phys. {\bf 270}, 328 (1998). 

\bibitem{NunezDuine}  A. S. N\'u\~nez and R. A. Duine, Phys. Rev. B \textbf{77}, 054401 (2008).

\bibitem{Basko} D. M. Basko and M.G. Vavilov, Phys. Rev. B \textbf{79}, 064418 (2009).

\bibitem{effect-bath}
L. F. Cugliandolo, D. R. Grempel, G. S. Lozano, H. Lozza, and C. A. da Silva Santos,
Phys. Rev. B {\bf 66}, 014444 (2002). \\
L. F. Cugliandolo, D. R. Grempel, G. S. Lozano, and H. Lozza,
Phys. Rev. B {\bf 70}, 024422 (2004).

\bibitem{Cukupe} L. F. Cugliandolo, J. Kurchan, and L. Peliti, Phys. Rev. E \textbf{55}, 3898 (1997).

\bibitem{other-quantum} 
M. P. Kennett and C. Chamon, Phys. Rev. Lett. {\bf 86}, 1622 (2001). \\
M. P. Kennett, C. Chamon, and J. Ye, Phys. Rev. B \textbf{64}, 224408 (2001). \\
G. Biroli and O. Parcollet, Phys. Rev. B {\bf 65}, 094414 (2002).\\
H. Westfahl, J. Schmalian, P. G. Wolynes, Phys. Rev. B {\bf 68}, 134203 (2003). 
G. Busiello, E. V. Gazeeva, R. V. Saburova, 
I. R. Khaibutdinova, G. P. Chugunova, 
Physics of metals and metallography {\bf 97}, 552 (2004);
 {\it ibid}  {\bf 101}, 109 (2006); 
{\it ibid} {\bf 102}, 244 (2006).
L. F. Cugliandolo, T. Giamarchi, and P. Le Doussal, 
Phys. Rev. Lett. {\bf 96}, 217203 (2006).

\bibitem{Kubo} R. Kubo, M. Toda, and N. Hashitsume,
\textit{Statistical physics II: Nonequilibrium stastical mechanics},
(Springer, New York 1985).

\bibitem{Zinn-Justin} J. Zinn-Justin, 
{\it Quantum Field Theory and Critical Phenomena} 
(Oxford Univ. Press, 2002).

\bibitem{CuBi} G. Biroli and L. F. Cugliandolo,
Phys. Rev. B \textbf{64}, 014206 (2001).

\bibitem{KondoExp}
See, for example, D. Goldhaber-Gordon et al., Nature (London) {\bf 391}, 156 (1998). \\
D. Goldhaber-Gordon et al., Phys. Rev. Lett. {\bf 81}, 5225 (1998). \\
W. Liang et al., Nature (London) {\bf 417}, 725 (2002).

\bibitem{Kondo}
P. Mehta and N. Andrei, Phys. Rev. Lett. {\bf 96}, 216802 (2006).\\
S. Kehrein, Phys. Rev. Lett. {\bf 95}, 056602 (2005). \\
E. Boulat, H. Saleur, and P. Schmitteckert, Phys. Rev. Lett. {\bf 101}, 140601 (2008).\\
B. Doyon, N. Andrei, Phys. Rev. B {\bf 73},  245326 (2006).

\bibitem{Landauer}
R. Landauer, IBM, J. Res. Dev. {\bf 1}, 223 (1957); Philos. Mag. {\bf 21}, 863 (1970). \\
M. B\"uttiker, Phys. Rev. Lett. {\bf 57}, 1761 (1986).


\end{thebibliography}
\end{document}